\documentclass[twocolumn,twocolappendix]{aastex701}
\usepackage{lmodern}
\usepackage{amsmath}
\usepackage{mathtools}
\usepackage{multirow}
\usepackage[capitalise]{cleveref}
\usepackage{tabularx}
\usepackage{booktabs}
\usepackage{physics}
\usepackage{xspace}
\usepackage{dsfont}
\usepackage{xcolor}
\usepackage[squaren]{SIunits}

\addunit{\massh}{\mathit{h}^{-1}M_\odot}
\addunit{\Mpch}{\mathit{h}^{-1}Mpc}
\addunit{\Gpch}{\mathit{h}^{-1}Gpc}
\addunit{\kpch}{\mathit{h}^{-1}kpc}
\addunit{\kms}{km\ s^{-1}}
\addunit{\invMpch}{\mathit{h}\ Mpc^{-1}}
\addunit{\GHz}{GHz}
\addunit{\sqamin}{arcmin^2}
\addunit{\sqdeg}{deg^2}
\addunit{\Myr}{Myr}

\usepackage{scalefnt}
\newcommand\smaller[2][0.85]{{\scalefont{#1}#2}}

\defcitealias{springel2003cosmological}{SH03} %
\defcitealias{frontiere2022simulating}{FRO23} %
\defcitealias{frontiere2017}{FRO17} %
\defcitealias{hopkins2023fire}{FIRE3} %
\defcitealias{haardt2012}{HM12} %
\defcitealias{faucher2020}{FG20}
\newcommand{\FRONTIERE}{\mbox{Frontier-E}\xspace}
\newcommand{\CRKHACC}{\smaller{CRK-HACC}\xspace}
\newcommand{\COSMOTOOLS}{\smaller{CRK-HACC}\xspace}
\newcommand{\HACC}{\smaller{HACC}\xspace}
\newcommand{\HAVOCC}{\smaller{HAvoCC}\xspace} 
\newcommand{\CRKSPH}{\smaller{CRKSPH}\xspace}
\newcommand{\SWFFT}{\smaller{SWFFT}\xspace}
\newcommand{\CLOUDY}{\smaller{CLOUDY}\xspace}
\newcommand{\FIRE}{\smaller{FIRE}\xspace}
\newcommand{\FIIRE}{\smaller{FIRE-2}\xspace}
\newcommand{\FIIIRE}{\smaller{FIRE-3}\xspace}
\newcommand{\TNG}{\smaller{Illustris-TNG}\xspace}
\newcommand{\TNGCLUSTER}{\smaller{TNG-Cluster}\xspace}
\newcommand{\TNGONE}{TNG100\xspace}
\newcommand{\GX}{\smaller{GADGET-X}\xspace}
\newcommand{\TNGTHREE}{TNG300\xspace}
\newcommand{\MTNG}{\smaller{MillenniumTNG}\xspace}
\newcommand{\MTNGNUM}{MTNG740\xspace}
\newcommand{\HORIZON}{\smaller{Horizon-AGN}\xspace}
\newcommand{\EAGLE}{\smaller{EAGLE}\xspace}

\newcommand{\FLAMINGO}{\smaller{FLAMINGO}\xspace}
\newcommand{\COLIBRE}{\smaller{COLIBRE}\xspace}
\newcommand{\SIMBA}{\smaller{SIMBA}\xspace}
\newcommand{\UM}{\smaller{UniverseMachine}\xspace}
\newcommand{\EMERGE}{\smaller{EMERGE}\xspace}

\newcommand{\RAMACHANDRA}{Ramachandra et al. (2025, in prep.)}

\newcommand{\Wrij}{\ensuremath{\mathcal{W}_{ij}^R}}

\newcommand{\pluscup}{\mathrel{\raisebox{-0.3ex}{$\uplus$}}}

\makeatletter
\newcommand{\showfontsize}{The current font size is \f@size pt.}
\makeatother 
\usepackage{hyperref}
\begin{document}

\title{Modeling Galaxy Formation in Cosmological Simulations with \CRKHACC}

\author[0009-0005-8598-4292]{Nicholas~Frontiere}
\affiliation{CPS Division,  Argonne National Laboratory, Lemont, IL 60439, USA}
\affiliation{HEP Division, Argonne National Laboratory, Lemont, IL 60439, USA}
\email{nfrontiere@anl.gov}
\author[0000-0003-1406-0744]{J.D.~Emberson}
\affiliation{CPS Division, Argonne National Laboratory, Lemont, IL 60439, USA}
\email{jemberson@anl.gov}
\author[0000-0002-8469-4534]{Michael~Buehlmann}
\affiliation{CPS Division, Argonne National Laboratory, Lemont, IL 60439, USA}
\email{mbuehlmann@anl.gov}
\author[0000-0002-7832-0771]{Salman~Habib}
\affiliation{CPS Division, Argonne National Laboratory, Lemont, IL 60439, USA}
\affiliation{HEP Division, Argonne National Laboratory, Lemont, IL 60439, USA}
\email{habib@anl.gov}
\author[0000-0003-1468-8232]{Katrin~Heitmann}
\affiliation{HEP Division, Argonne National Laboratory, Lemont, IL 60439, USA}
\email{heitmann@anl.gov}
\author[0000-0001-7772-0346]{Nesar~Ramachandra}
\affiliation{CPS Division, Argonne National Laboratory, Lemont, IL 60439, USA}
\email{nramachandra@anl.gov}
\author[0000-0002-4900-6628]{Claude--Andr\'e Faucher--Gigu\`ere}
\affiliation{Department of Physics and Astronomy and Center for Interdisciplinary Exploration and Research in Astrophysics (CIERA), Northwestern University, Evanston, IL 60201, USA}
\email{cgiguere@northwestern.edu}

\begin{abstract}

Self-consistently modeling baryonic effects in survey-scale cosmological simulations has become increasingly important as the diversity, precision, and statistical reach of modern observations continue to improve. The advent of exascale computing now enables a new generation of simulations that couple these physical processes across full-sky volumes with excellent statistical sampling of large-scale structure tracers such as galaxies, groups, and clusters. To support these efforts, we extend the \CRKHACC framework, a GPU-accelerated cosmological hydrodynamics code, with a suite of astrophysical subgrid models that simulate radiative cooling, star formation, stellar evolution, and AGN feedback within a numerically robust formulation optimized for scalability on modern exascale architectures. The models were selected and calibrated to reproduce observed galaxy stellar mass functions over the redshift range $0 < z < 2$ and cluster populations probed by cosmological surveys, capturing the large-scale baryonic evolution relevant for multi-wavelength, cross-correlated analyses. We describe the implementation and calibration of these models and demonstrate their consistency with observed galaxy population statistics and modern hydrodynamic simulations, establishing the baseline for exascale efforts that extend this framework to survey-scale volumes.

\end{abstract}

\keywords{methods: numerical -- cosmology: theory -- galaxies: formation -- galaxies: evolution}

\section{Introduction} \label{sec:intro}

Simulating the formation of large-scale structure is a central tool in modern cosmology, providing a major component of the theoretical and modeling foundation for interpreting current and upcoming observational surveys. By solving for the gravitational evolution of matter from primordial fluctuations to the present-day cosmic web, large-volume gravity-only simulations link the initial conditions of the early universe to the complex distribution of galaxies, halos, and clusters observed today. Such simulation efforts offer computational efficiency and have yielded many foundational predictions and insights (see \citet{angulo2021} for a recent review). However, as multi-wavelength surveys have grown increasingly informative, it has become necessary to incorporate baryonic processes through hydrodynamic simulations to produce mock observables and statistical predictions of sufficient fidelity for survey interpretation \citep[e.g.,][]{vanDaalen2011,harnois2015, copeland2018, chisari2019}.

Accurately modeling baryonic effects requires incorporating astrophysical processes such as radiative cooling, star formation, chemical enrichment, and feedback from supernovae and active galactic nuclei (AGN). These mechanisms operate below the resolution scale of simulations addressing cosmological volumes and must therefore be represented using sub-resolution (``subgrid'') models. As subgrid models are necessarily coarsened and phenomenological representations of the relevant physical processes, they often require calibration against observational constraints, balancing physical motivation with empirical tuning to achieve predictive accuracy. 

Incorporating these baryonic processes at cosmological scales is computationally demanding. While many current hydrodynamic simulations are limited to modest box sizes by numerical cost, modern observational surveys increasingly require much larger volumes to capture rare objects, long-range correlations, and the full spatial extent needed for realistic mock survey realizations. This demand has motivated recent large-volume efforts (e.g., \MTNG, \citealt{pakmor2023}; \FLAMINGO, \citealt{schaye2023flamingo}), which trade spatial and mass resolution for cosmological volume. 

With the advent of exascale computing and GPU-accelerated architectures, the effective dynamic range can now be expanded by more than an order of magnitude in total particle count and simulated volume, enabling simulations that combine the physical complexity of baryonic processes with the statistical power required for survey-scale analysis. An example is the \FRONTIERE exascale cosmological simulation~\citep{frontiere2025GB}, the first trillion-particle-class hydrodynamic run executed on the Frontier exascale system.\footnote{\url{https://www.olcf.ornl.gov/frontier/}} \FRONTIERE evolved four trillion dark matter and baryonic particles to redshift zero within a $\sim\!100~\mathrm{Gpc}^3$ volume, achieving a scale comparable to state-of-the-art gravity-only simulations while self-consistently modeling gas dynamics and astrophysical subgrid processes.

The \FRONTIERE\ simulation was performed with the \CRKHACC\ framework. Originally described by \citet{frontiere2022simulating} (hereafter \citetalias{frontiere2022simulating}), \CRKHACC extends the gravity-only \HACC framework developed by \citet{habib2016hacc}, which was designed for extreme scalability on modern supercomputers. It incorporates a high-order Lagrangian smoothed particle hydrodynamics (SPH) scheme to simulate baryonic evolution with high accuracy. In this work, we describe the astrophysical subgrid models implemented in \CRKHACC, which provide the foundation for self-consistent simulations of galaxy formation and feedback in large cosmological volumes, such as those used for \FRONTIERE.

This study is optimized for the simulation scale required by mock full-sky surveys, where achieving extreme volumes implies a comparatively coarse baryonic mass resolution (on the order of $10^8\,M_\odot$ per particle). Within this regime, we prioritize subgrid formulations that are numerically robust while also scalable and GPU-efficient for exascale deployment. This focus contrasts with higher-resolution galaxy-formation and zoom-in studies that aim to resolve the internal structure and morphology of individual galaxies.

Given this scope, we set expectations accordingly. We do not aim to resolve the faintest galaxies, the earliest epochs where our particle masses become limiting, or detailed morphologies and chemical substructure. Instead, our goal is to deliver physically realistic, survey-scale predictions suited for mock all-sky analyses and instrument modeling. These include converged clustering and halo population statistics, realistic galaxy and cluster samples for synthetic survey generation, and large-scale structure fields suitable for cross-probe predictions across optical, X-ray, and millimeter-band observables. In this context, we focus on reproducing galaxy stellar mass functions consistent with observations over the redshift range $0<z<2$ and on modeling cluster density profiles that remain realistic at our mass resolution, establishing the basis for the mock-sky and cosmological analyses enabled by \CRKHACC.

While a number of cosmological simulation codes implement subgrid physics, they differ in their modeling assumptions, numerical coupling strategies, and calibration approaches. Rather than attempting to survey this landscape (see, e.g.,~\citealt{somerville2015physical,naab2017,vogelsberger2020,crain2023} for detailed reviews), we focus here on the specific implementation adopted in \CRKHACC. Our fiducial model was selected after testing a range of candidate strategies from the literature, with final choices and modifications guided by the performance and fidelity requirements of simulating large-scale observational campaigns. A companion paper detailing the Bayesian model calibration procedure is presented in \RAMACHANDRA.

The manuscript is organized as follows. Section~\ref{sec:solvers} describes the numerical framework and astrophysical subgrid models implemented in \CRKHACC, including radiative cooling, star formation, feedback, and chemical enrichment, all built to support GPU acceleration and parallel scalability. Section~\ref{subsec:calib} outlines the calibration procedure used to tune these models against selected observational benchmarks. Section~\ref{sec:results} presents comparisons between the fiducial model and additional, non-calibrated observational measurements to evaluate its broader physical fidelity and overall performance. Finally, Section~\ref{sec:conclude} summarizes the main findings and discusses future directions.

The appendices provide supplementary material, including detailed \CRKHACC solver updates required for subgrid physics (Appendix~\ref{app:CRKUpdate}), self-similar radiative cooling validation tests (Appendix~\ref{sec:selfsimilarcool}), a cluster code comparison study (Appendix~\ref{sec:nifty}), the construction of the cooling and emissivity tables (Appendix~\ref{sec:cloudy_tab}), and the integrated enrichment model used in our stellar-feedback implementation (Appendix~\ref{app:Enrich}).

\section{Solvers and Methods} \label{sec:solvers}

\begin{deluxetable*}{@{}p{0.25\textwidth} p{0.305\textwidth} p{0.40\textwidth}@{}}
\tabletypesize{\footnotesize}
\tablecaption{Overview of the physics modeled in \CRKHACC.\label{tab:subgrid-summary}}
\tablehead{
  \colhead{Physical Process} &
  \colhead{Base Model(s)} &
  \colhead{Implementation in \CRKHACC}}
\startdata
 \centering 
{\textbf{Gravity \& Hydrodynamics} \\ Section~\ref{subsec:GH}}  &
 TreePM + CRKSPH \citep{habib2016hacc, frontiere2017, frontiere2022simulating}. &
GPU-accelerated, high-order hydrodynamics and gravity solver using a hybrid particle-mesh and tree-based approach optimized for exascale performance. 
\\[6pt]
 \centering 
\textbf{Radiative Cooling \& Heating} \\ Section~\ref{subsec:radcool} &
Optically thin, ionization-equilibrium gas with a UV background from \citet{faucher2020} and a metal-line cooling treatment similar to \citet{wiersma2009effect}. &
Modeled consistently using metallicity-scaled, self-shielding--attenuated \CLOUDY\ \citep{ferland2017} rates assuming relative solar abundances, applied via a modified exact-integration scheme \citep{townsend2009exact}. \\[6pt]
 \centering 
\textbf{Star Formation} \\  Section~\ref{subsec:SF}&
Two-phase ISM model \citep{springel2003cosmological} with a softened equation of state. &
Gas transitions smoothly into the ISM above a density threshold, with star formation integrated stochastically. ISM metallicity and helium floors model unresolved early enrichment. 
\\[6pt]
 \centering 
\textbf{Galactic Winds} \\ Section~\ref{subsec:wind} &
Modified kinetic wind model from \citet{vogelsberger2013model, pillepich2018}. &
Decoupled stochastic winds with energy-based mass loading and dark matter velocity-dispersion scaling \citep{oppenheimer2006}. Metal loading regulates ISM entrainment, with density/time-based recoupling. 
\\[6pt]
 \centering 
\textbf{Chemical Enrichment} \\ Section~\ref{subsec:chem} &
Supernovae~II/Ia and stellar-wind yields and rates from \citet{hopkins2023fire}. &
Stellar evolution modeled with time- and metallicity-dependent fits to integrated instantaneous rates, enabling exact, continuous tracking of metal and helium enrichment without time-discretization errors. 
\\[6pt]
 \centering 
\textbf{Black Hole Evolution} \\ Section~\ref{subsec:bhseed}--\ref{subsec:bhaccretion} &
Conceptual framework from \citet{springel2005AGN}, modeling BHs as collisionless sink particles with separate internal-mass evolution. &
Custom implementation with resolution-adaptive seeding at newly formed galaxy centers and stable repositioning--merging procedures that prevent spurious displacements. Unboosted Bondi accretion. Continuous mass growth as in \citet{bahe2022}. 
\\[6pt]
 \centering 
\textbf{AGN Feedback} \\ Section~\ref{subsec:bhfeedback} &
Two-mode thermal and kinetic AGN feedback model following \citet{weinberger2016simulating}. &
Synchronized energy injection with mode set by the Eddington ratio. Thermal feedback deposits heat continuously. Randomly oriented kinetic feedback is triggered when accumulated energy reaches a constant threshold defined by a simplified jet-velocity parameter. \\[6pt]
\enddata
\end{deluxetable*}

We begin by describing the enhancements to the \CRKHACC framework introduced in \citetalias{frontiere2022simulating}, which extend its capabilities by incorporating astrophysical subgrid models that simulate galactic feedback processes within the context of large-scale structure formation. We then detail how these processes are implemented to ensure parallel reproducibility and introduce a GPU-accelerated in situ analysis pipeline for galaxy finding, used for efficient black-hole seeding and simulation analysis. Table~\ref{tab:subgrid-summary} summarizes the modeled physics, with detailed explanations provided in the sections below. In addition to expanding the scientific capabilities, several modifications to the \CRKHACC solver were required to accommodate subgrid source terms. These algorithmic updates are summarized separately in Appendix~\ref{app:CRKUpdate}.

Throughout this work, particle positions $\vb{x}_i$ and smoothing lengths $h_i$ are expressed in comoving coordinates. Consistent with \citetalias{frontiere2022simulating}, SPH interpolation employs a \citet{Wendland1995} $C^4$ kernel, $W_{C^4}(|\vb{x}_{ij}|, h_i)$, which we denote compactly as $W_{ij}(h_i)$. Here, $|\vb{x}_{ij}| \equiv |\vb{x}_i - \vb{x}_j|$ is the separation between particles $i$ and $j$. Unless otherwise noted, subgrid relations are formulated in proper units to remain consistent with their physical calibrations. Spatial quantities expressed in proper units are labeled accordingly (e.g., pkpc).

\subsection{Gravity and Hydrodynamics}\label{subsec:GH}

The \CRKHACC framework combines a highly scalable gravitational $N$-body solver with a modern SPH scheme to model the coevolution of dark matter and baryonic gas during cosmological structure formation. The gravitational potential is decomposed into slow- and fast-varying components, which define the long- and short-range force operators, respectively.

The slow component is computed using a particle-mesh (PM) method with high-order spectral filtering and a distributed FFT implementation, \SWFFT,\footnote{\url{https://git.cels.anl.gov/hacc/SWFFT}} optimized for large-scale parallel performance. The fast component is handled with a tree-based method designed for GPU acceleration. The combined TreePM solver ensures low-noise force estimates and a compact handover scale, achieving accurate modeling of gravitational interactions across a wide dynamic range while maintaining excellent performance and portability on modern supercomputing architectures \citep{habib2016hacc}. 

\CRKHACC models gas dynamics using Conservative Reproducing Kernel Smoothed Particle Hydrodynamics (\CRKSPH), a Lagrangian method that improves upon traditional SPH by exactly reproducing linear fields while conserving mass, momentum, and energy (see \cite{frontiere2017}; henceforth \citetalias{frontiere2017}). The reproducing kernel corrections enhance the accuracy of fluid interpolation and derivative estimates, particularly in shearing and mixing flows. To handle shocks, \CRKSPH employs a modified artificial viscosity scheme with a limiter function that suppresses dissipation in smooth regions. The thermal evolution follows a ``compatible energy'' formalism that ties the energy update directly to the hydrodynamic work, thereby preserving entropy in adiabatic flows. In practice, local hydrodynamic interactions, including neighbor searches and SPH summations, use the same tree structure as the short-range gravity solver, resolving gas and dark matter dynamics at small scales.

An important feature of the combined gravity and hydrodynamics solver is its domain decomposition strategy, which uses ``overloading'' to duplicate particles near rank boundaries in adjacent domains (summarized in Section~3.4 of \citetalias{frontiere2022simulating}). These overlapping regions, analogous to ghost zones, allow particles to be integrated independently without inter-node communication until they are refreshed and synchronized across ranks at the coarser particle-mesh timescale. This approach maintains scalability and minimizes MPI overhead in large-scale runs, and it is especially effective for the astrophysical models discussed here, which require integration at finer timesteps and would otherwise demand more frequent communication.

\subsection{Integration of Subgrid Operators}\label{subsec:SGInter}
The gravitational and hydrodynamic solvers are coupled through a symplectic time-integration scheme that supports hierarchical timestepping: long-range gravitational updates are applied at coarser intervals (PM steps) with equal spacing in scale factor, while short-range forces are integrated within smaller, fixed ``subcycles.’’ Hydrodynamic interactions are integrated on progressively finer timescales according to local dynamical CFL conditions, using power-of-two integration bins (see Section~3.1 in \citetalias{frontiere2022simulating} for a detailed description). Note that we refer to particles as \emph{active} when they are updated on a given timestep and \emph{passive} when they reside on coarser levels of the timestepping hierarchy; this terminology is used throughout the manuscript.

The addition of subgrid operators introduces sources of momentum, energy, and acceleration, along with mechanisms for mass exchange and species tracking, to model complex astrophysical processes such as star formation and galactic feedback. These subgrid operators are coupled at the appropriate timestepping hierarchies (i.e., PM, subcycling, and hydrodynamic timesteps) and are integrated using first-order Strang splitting \citep{strang1968}.

Specifically, the discretized time propagators for the particle-mesh, short-range, and hydrodynamics operators, $\hat{U}(\Delta t)$, $\hat{U}_{\text{SR}}(\Delta t')$, and $\hat{U}_H(\Delta t'')$, as described by Eqs. (39), (40), and (42) in \citetalias{frontiere2022simulating}, are modified to include subgrid sources as
\begin{align}
\hat{U}(\Delta t) &\rightarrow \hat{U}(\Delta t)\,\hat{U}^{S}_{\text{PM}}(\Delta t) \\
\hat{U}_{\text{SR}}(\Delta t') &\rightarrow \hat{U}_{\text{SR}}(\Delta t')\,\hat{U}^{S}_{\text{SR}}(\Delta t') \\
\hat{U}_H(\Delta t'') &\rightarrow \hat{U}_H(\Delta t'')\,\hat{U}^{S}_H(\Delta t'')
\label{eq:SubgridOperators}
\end{align}
where each source integrator $\hat{U}^{S}_{\{\text{PM}, \,\text{SR}, \,H\}}$ represents the composite of subgrid operators acting at the corresponding timestepping hierarchy.\footnote{$\hat{U}^{S}_H$ follows the definition in Eq.~(38) of \citetalias{frontiere2022simulating}; here, we extend the notation to include analogous source operators $\hat{U}^{S}_{\text{PM}}$ and $\hat{U}^{S}_{\text{SR}}$ for the PM and subcycle (short-range) levels, respectively.} Here, $\Delta t$, $\Delta t'$, and $\Delta t''$ denote the progressively finer timestep sizes of each timestepping hierarchy.

The primary subgrid operators are composed as follows:
the smallest hydrodynamic time interval operators are
\begin{equation}
    \hat{U}^{S}_H = \hat{U}^{\rm Z-load}_H\circ\hat{U}^{\rm recouple}_H\circ\hat{U}^{\rm conv}_H\circ\hat{U}^{\rm cool}_H, 
\end{equation}
where $\hat{U}^{\rm cool}_H$ is radiative cooling (Section~\ref{subsec:exact}), $\hat{U}^{\rm conv}_H$ is the conversion of ISM gas to stars and wind (Section~\ref{subsec:wind_algo}), and $\hat{U}^{\rm recouple}_H$ and $\hat{U}^{\rm Z-load}_H$ are wind recoupling and metal loading (Sections~\ref{subsec:windR}~\&~\ref{subsec:windZ}), respectively. The intermediate scale operators include 
\begin{equation}
    \hat{U}^{S}_{\text{SR}} = \hat{U}^{\rm AGN-feed}_{\text{SR}} \circ \hat{U}^{\rm AGN-acc}_{\text{SR}}\circ \hat{U}^{\rm enrich}_{\text{SR}},
\end{equation}
modeling stellar enrichment $\hat{U}^{\rm enrich}_{\text{SR}}$ (Section~\ref{subsec:chem}) as well as AGN accretion $\hat{U}^{\rm AGN-acc}_{\text{SR}}$ and feedback $\hat{U}^{\rm AGN-feed}_{\text{SR}}$ (Sections~\ref{subsec:bhaccretion}~\&~\ref{subsec:bhfeedback}). Lastly, the PM subgrid operators consist of
\begin{equation}
    \hat{U}^{S}_{\text{PM}} = \hat{U}^{\rm AGN-merge}_{\text{PM}}\circ \hat{U}^{\rm AGN-repos}_{\text{PM}}\circ\hat{U}^{\rm AGN-seed}_{\text{PM}},
\end{equation}
which currently include AGN seeding $\hat{U}^{\rm AGN-seed}_{\text{PM}}$ (Section~\ref{subsec:bhseed}), as well as merging $\hat{U}^{\rm AGN-merge}_{\text{PM}}$ and repositioning $\hat{U}^{\rm AGN-repos}_{\text{PM}}$ (Section~\ref{subsec:bhrepos}). 

As PM and subcycle timesteps are synchronized across all ranks, subgrid modules introduced at that cadence can be integrated consistently, which is important for particles in shared overloaded regions. For operators applied at the hydrodynamic level, where timestepping can vary between nodes, additional care is required to maintain consistent results (see Section~\ref{subsec:stochastic} for further discussion). 

The following sections describe the details of the subgrid implementations and the primary code modifications needed to support them. 

\subsection{Radiative Cooling and Heating}\label{subsec:radcool}

Radiative cooling and photoheating are fundamental processes in cosmological simulations, governing the formation of stars and galaxies and the thermal evolution of the intergalactic medium (IGM). As is standard in hydrodynamic frameworks (e.g., \TNG, \citealt{vogelsberger2013model,pillepich2018}; \HORIZON, \citealt{dubois2014}; \EAGLE, \citealt{schaye2015eagle}; \SIMBA, \citealt{dave2019}), we model the gas as optically thin and in ionization equilibrium, subject to a spatially uniform, time-dependent ultraviolet background (UVB) radiation field adopted from \citet{faucher2020}. This approximation avoids the computational cost of full radiative transfer and reionization calculations while capturing the dominant thermal influence of the UVB on diffuse gas in the IGM and circumgalactic medium (CGM). 

Additional contributions include inverse Compton cooling off the CMB \citep[e.g.,][]{ikeuchi1986}, metal-line cooling dependent on total metallicity (analogous to the metallicity-scaled formulation in \citealt{wiersma2009effect}), and a density- and redshift-dependent correction for self-shielding attenuation (Eq.~A1 in \citealt{rahmati2013}). In combination, these terms define a total radiative source function, $\Lambda(u,\rho,Z,Y,z)$, expressed as a function of gas internal energy $u$, density $\rho$, metallicity $Z$, helium fraction $Y$, and redshift $z$.

We next describe the metal-line cooling implementation in \CRKHACC, which employs \CLOUDY-based rate tables under a scaled-solar abundance approximation. We then outline the exact integration scheme used to apply these rates during the hydrodynamic update \citep{townsend2009exact}, eliminating the stiffness and timestep constraints inherent to conventional explicit or implicit solvers.

\subsubsection{Metal-Line Cooling}\label{subsec:mlc}

To generate the tabulated cooling and heating source function $\Lambda$, we use the photoionization code \CLOUDY version~17.02 \citep{ferland2017}\footnote{\url{https://gitlab.nublado.org/cloudy/cloudy/-/wikis/home}}. Unlike early treatments that evaluate radiative cooling for a primordial hydrogen--helium mixture (e.g., \citealt{katz1996}), \CLOUDY self-consistently models the contributions from metal-line cooling. A detailed description of the \CLOUDY table generation procedure, including parameter grids, UV background models, and self-shielding corrections, is provided in Appendix~\ref{sec:cloudy_tab}.

Modern hydrodynamic codes (\TNG, \EAGLE, \FLAMINGO, \HORIZON, etc.) commonly track the abundances of individual gas elements such as carbon, oxygen, and nitrogen produced by chemical enrichment models. Previous studies of metal-line cooling \citep[e.g.,][]{smith2008, wiersma2009effect} have shown that the cooling contributions of these elements can be tabulated with \CLOUDY (or comparable codes) and applied directly to simulated particles. However, such precomputed cooling tables can become prohibitively large, as their dimensionality scales with the number of tracked elements. To mitigate this cost, the contribution of individual elements, assuming solar abundances, can be rescaled to approximate gas cooling for arbitrary compositions, as described by \citealt{wiersma2009effect}.

Similar to the cooling treatment adopted in \TNG and in simulation codes that incorporate ionization and chemical processes with \smaller{GRACKLE} \citep{smith2017}, \CRKHACC does not include the cooling contribution of individual metal elements. Instead, cooling rates are computed based on the hydrogen ($X$), helium ($Y$), and total metal ($Z$) mass fractions, assuming that the relative abundances of metals follow the solar pattern.

Specifically, the abundance of each metal element ($Z_i$) relative to hydrogen is scaled as
\begin{equation}
\frac{n_{Z_i}}{n_{\rm H}} = R \left( \frac{n_{Z_i}}{n_{\rm H}} \right)_\odot,
\label{eqn:nzi}
\end{equation}
where the scaling factor $R \geq 0$ is applied uniformly to all metal species with number density $n_{Z_i}$.
Combining \cref{eqn:nzi} with the definitions of the mass fractions,
\begin{align}
X &= \frac{n_{\rm H} m_{\rm H}}{n_{\rm H} m_{\rm H} + n_{\rm He} m_{\rm He} + \sum_i n_{Z_i}m_{Z_i}}, \label{eqn:Xfrac} \\
Y &= X \frac{n_{\rm He}m_{\rm He}}{n_{\rm H}m_{\rm H}}, \label{eqn:Yfrac}\\
Z &= X \sum_i \frac{n_{Z_i}m_{Z_i}}{n_{\rm H}m_{\rm H}}  = X R\sum_i \left(\frac{n_{Z_i}}{n_{\rm H}}\right)_\odot \frac{m_{Z_i}}{m_{\rm H}}, \label{eqn:Zfrac}
\end{align}
the relationship between $R$ and the total metallicity is given by
\begin{equation}
R = \frac{Z}{X} \left( \frac{X}{Z} \right)_\odot,
\label{eqn:R}
\end{equation}
where $(X/Z)_\odot$ denotes the solar hydrogen-to-metal mass fraction. In our cooling implementation, the index $i$ runs over all 28 metal elements tracked in \CLOUDY, each with atomic mass $m_{Z_i}$. Accordingly, we adopt solar mass fractions of $X_\odot = 0.706498$, $Y_\odot = 0.280555$, and $Z_\odot = 0.012947$, based on the \CLOUDY solar composition from \citet{grevesse1998,holweger2001,prieto2001,prieto2002}.

As described in Appendix~\ref{sec:cloudy_tab}, the tabulated \CLOUDY\ cooling rates $\Lambda$ are sampled as a function of the metallicity scaling factor $R$. When particles are radiatively cooled (see Section~\ref{subsec:exact}), the local metallicity $Z$ is converted to $R$ for table lookup, thereby modeling metal-line cooling self-consistently with total metallicity under the assumption of relative solar abundances. 

\CRKHACC\ forgoes individually tracking heavy elements for several practical reasons: 1) the elemental yields of enrichment models remain highly uncertain, 2) the baryon mass resolutions typical of large-volume simulations ($\sim10^7$--$10^8~\massh$) are too coarse to reliably evolve detailed chemical abundances, 3) storing individual elemental state variables increases the memory overhead of large simulation campaigns, and 4) the differences in cooling rates between element-by-element and scaled-solar abundance treatments are modest \citep{wiersma2009effect}.

\begin{figure}
    \centering
    \includegraphics[width=\linewidth]{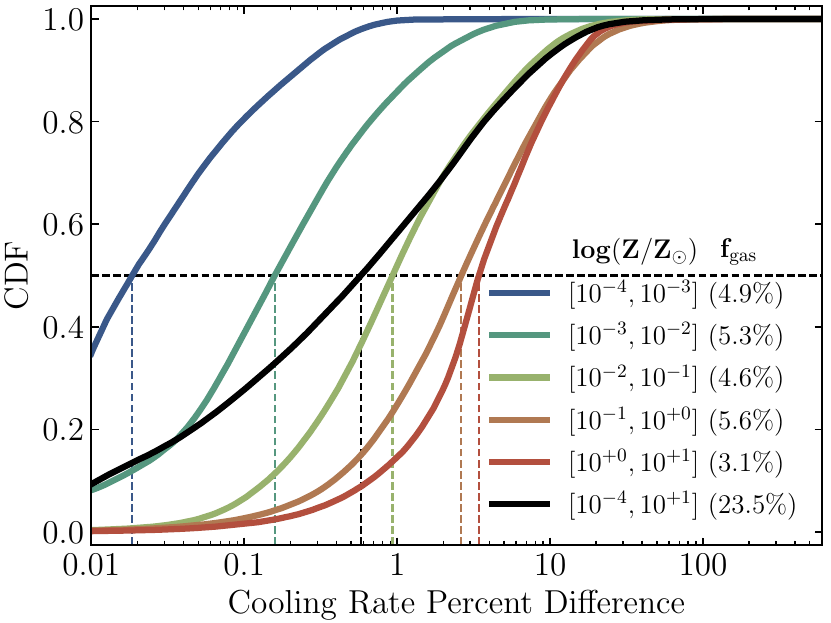}
    \caption{Cumulative distribution function of the relative difference in cooling rates computed with \CLOUDY\ when using individual versus solar-scaled metal abundances. The individual-abundance cooling rates are calculated for one million randomly selected gas particles from the m12a \FIIIRE\ simulation \citep{hopkins2023fire,sultan2025} that tracks nine metal elements. These are compared to rates derived assuming solar-scaled abundances. Square brackets in the legend indicate the metallicity bin of each curve with the round brackets denoting the fraction of \CRKHACC\ gas particles in that bin. Vertical dashed lines show the median error of each bin. Among metal-enriched gas particles with $Z/Z_\odot \geq 10^{-4}$, we find that $93\%$ ($58\%$) have relative errors below $10\%$ ($1\%$). This translates into $98\%$ ($90\%$) of all simulation particles, confirming that nearly all gas in \CRKHACC cools at rates consistent with element-by-element tracking.}
    \label{fig:cloudy}
\end{figure}

To verify this last point, we computed cooling rates with \CLOUDY using the individually tracked metal abundances of a random sample of one million particles drawn from the $z = 0$ snapshot of the m12a \FIIIRE\ simulation \citep{hopkins2023fire,sultan2025}. This simulation models a $\unit{2\times10^{12}}{M_\odot}$ halo at $\unit{6\times10^4}{M_\odot}$ baryon mass resolution while tracking nine metal element abundances (C, N, O, Ne, Mg, Si, S, Ca, and Fe). The particles were randomly selected with weights chosen so that their metallicity distribution reproduces that of calibrated \CRKHACC runs, ensuring that the test sample reflects the range of values encountered in our simulations. For comparison, we calculated a second set of cooling rates assuming scaled-solar abundances, where the scaling factor $R$ was computed from \cref{eqn:R} using the $X$ and $Z$ values of each randomly drawn \FIIIRE\ particle. We restrict attention to metal-enriched ($Z/Z_\odot \geq 10^{-4}$) gas particles to avoid cases where metals have negligible impact on cooling. 

Figure~\ref{fig:cloudy} shows the cumulative distribution function of the relative difference between the two methods for various bins in metallicity. In general, the relative error increases with metallicity but remains modest, with the median error of super-solar metallicity particles sitting in the few-percent range. Overall, we find that $93\%$ ($58\%)$ of the sampled particles have relative errors below $10\%$ ($1\%)$. Accounting for the majority of metal-poor ($Z/Z_\odot < 10^{-4}$) gas with minimal metal cooling, this corresponds to roughly $98\%$ ($90\%)$ of all gas particles, confirming that the vast majority of particles in \CRKHACC\ exhibit cooling rates consistent to those obtained from fully element-by-element tracking.

In agreement with \MTNG \citep{pakmor2023}, we also find that evolving only the total metallicity has minimal impact on the coarse-scale galaxy properties resolved at our simulation resolution, though it can influence the calibration of subgrid parameters against observations. Given the mass resolution limits of large-volume simulations, a more reliable approach for reproducing observable properties is to apply post-processing models for galaxy spectral energy distributions (SEDs), including dust treatment, using observation-driven methods. Our framework employs a stellar population synthesis (SPS) model following \citet{conroy2009,conroy2010}, applied to the star formation and metallicity histories of each galaxy. In addition, we incorporate Bayesian dust attenuation models from \citet{Nagaraj2022} to generate narrow-band spectra for individual galaxies. Details of this methodology and the resulting data products—including spectra and photometric catalogs—will be presented in a forthcoming publication.

\subsubsection{Exact Integration}\label{subsec:exact}

Simulated gas is subject to radiative cooling and heating subgrid source terms. The internal energy evolution due to these processes is described by the following ODE:
\begin{equation}\label{eqn:cool}
    \dv{u}{t} = - s\frac{\Lambda}{\rho}, 
\end{equation}
where $\Lambda$ is the energy change per unit time and volume, $\rho$ is the gas density, and $s = \pm 1$ distinguishes cooling ($s = +1$) from heating ($s = -1$).

\cref{eqn:cool} can be integrated exactly, following \citet{townsend2009exact}, under the assumption that the source function $\Lambda$ depends only on internal energy, while the gas density and chemical composition remain fixed. As the subgrid models in \CRKHACC are implemented using operator splitting (see Section~\ref{subsec:SGInter} and \citetalias{frontiere2022simulating}), this condition is explicitly maintained, with the gas composition updated only when the density and chemical enrichment operators are performed.

Applying separation of variables to \cref{eqn:cool} and integrating across a timestep of length $\Delta t$ yields
\begin{equation}\label{eqn:intcool}
\int_{u^0}^{u^1} \frac{\text{d}u}{\Lambda(u)} = - \frac{s}{\rho}  \Delta t,
\end{equation}
where $u^0$ and $u^1$ denote the internal energy of the particle at the beginning and end of the step, respectively.
Defining a dimensionless integration function $I(u)$, anchored to a reference energy $u_\text{a}$, as
\begin{equation}\label{eqn:I}
I(u)\equiv s\frac{\Lambda_\text{a}}{u_{\text{a}}} \int^{u_\text{a}}_u \frac{\text{d}u'}{\Lambda(u')},
\end{equation}
\cref{eqn:intcool} becomes:
\begin{equation}
    \left[I(u^0)-I(u^1)\right]\frac{u_\text{a}}{s\Lambda_\text{a}} = -\frac{s}{\rho}  \Delta t,
\end{equation}
which can be inverted to solve for the internal energy at the end of the step, 
\begin{equation}\label{eqn:u1}
    u^1 = I^{-1}\left[I(u^0)+\frac{\Lambda_\text{a}}{u_\text{a}\rho}\Delta t\right] = I^{-1}[I(u^1)],
\end{equation}
using $s^2 = 1$, $\Lambda_\text{a} \equiv \Lambda(u_{\text{a}})$, and $I(u^1) \equiv I(u^0)+\frac{\Lambda_\text{a}}{u_\text{a}\rho}\Delta t$. 

The evaluation of $I(u)$ and its inverse is straightforward when the source function is tabulated for discrete values $\Lambda_k = \Lambda(u_k)$ at logarithmically spaced energies $u_k$, and interpolated with piecewise power laws,
\begin{equation}\label{eqn:L}
    \Lambda(u) = \Lambda_k \left(\frac{u}{u_k}\right)^{\alpha_k}\!\!\!\!\!\!\!, \quad u_{k} \leq u \leq u_{k+1},
\end{equation}
where $\alpha_k \equiv \log(\Lambda_{k+1}/\Lambda_k) / \log(u_{k+1}/u_k)$ ensures continuity across bin edges.
Substituting this form into \cref{eqn:I} yields the recursive relation
\begin{equation}\label{eqn:Iu}
    I(u) = I_k+s\frac{\Lambda_\text{a} u_k}{\Lambda_k u_\text{a}}\scalebox{0.86}{$ \begin{cases}
        \frac{1}{1-\alpha_k}\left[1-\left(\frac{u}{u_k}\right)^{1-\alpha_k}\right]\!, & \alpha_k \ne 1 \\
        -\ln(\frac{u}{u_k}), & \alpha_k  =  1 
    \end{cases}$}
\end{equation}
with $I_k = I(u_k)$. The inverse function follows similarly as 
\begin{equation}\label{eqn:Iinv}
    I^{-1} = u_k\,\scalebox{0.86}{$
    \begin{cases}
        \left[ 1 - (I - I_k)(1 - \alpha_k)s \dfrac{\Lambda_k u_\text{a}}{\Lambda_\text{a} u_k} \right]^{\frac{1}{1 - \alpha_k}}\!\!\!\!\!\!\!\!\!\!\!\!\!, & \;\;\;\;  \alpha_k \ne 1 \\
        \exp \left[ -s \dfrac{\Lambda_k u_\text{a}}{\Lambda_\text{a} u_k} (I - I_k) \right], & \;\;\;\; \alpha_k = 1
    \end{cases}
    $}
\end{equation}
valid for $I_k \leq I \leq I_{k+1}$.

In practice, each particle is integrated independently using an anchor value chosen as its initial internal energy, $u_\text{a} = u^0$, which sets $I(u^0) = 0$ and $I(u^1) = \frac{\Lambda_\text{a}}{u_\text{a}\rho}\Delta t$. For heating (cooling) source functions, \cref{eqn:Iu} is recursively evaluated over successively higher (lower) energy bins starting from the anchor energy. In both cases, $I(u)$ increases monotonically across each bin. The power-law slopes in \cref{eqn:L} are obtained by log-linear interpolation of the tabulated source function at the bin edges, using the fixed particle composition values for density, metallicity, and helium fraction. 

The integration proceeds until $I(u)$ exceeds $I(u^1)$, at which point the enclosing bin is identified such that \mbox{$I_k \leq I(u^1) \leq I_{k+1}$}. The corresponding bin edge $u_k$, together with its stored $I_k$, is then used to compute the final internal energy $u^1$ by evaluating \cref{eqn:Iinv} with $I = I(u^1)$.

The integration procedure can employ any bin width, with narrower bins providing a more accurate piecewise power-law representation of $\Lambda$. A special case arises when the equilibrium energy is reached, corresponding to the bin where the source function $\Lambda$ changes sign. Care must be taken in this regime, as the integral diverges when $\Lambda \rightarrow 0$. To ensure numerical stability, we instead integrate that bin using a constant source function defined as the maximum of the two bin-edge values, $\Lambda = \max(\Lambda_k, \Lambda_{k+1})$, and constrain the energy not to exceed the equilibrium value determined by linearly interpolating the zero crossing. Using this approach, the integration converges smoothly in all cases when adopting a conservative bin width of 0.004 dex.


As described in \citet{townsend2009exact}, the exact integration method offers several advantages over traditional explicit or implicit solvers for internal energy evolution. In particular, it eliminates the need for the small timesteps required by explicit schemes to resolve short cooling times, while also avoiding the multiple roots and discontinuous solutions that can arise in implicit formulations. The result is a stable and continuous evolution of internal energy across a wide range of thermodynamic conditions.


Arguably, the main drawback of the exact integration method is the potential computational cost associated with frequent tabular lookups of $\Lambda$, particularly during the evaluation and inversion of $I(u)$. However, on modern GPUs these table accesses are highly efficient, rendering the cooling kernel subdominant in the overall \CRKHACC timestepping cycle. Consequently, the solver retains the full accuracy of the exact integration scheme while incurring only minimal computational overhead.

A dedicated validation of the radiative cooling module is presented in Appendix~\ref{sec:selfsimilarcool}, where we demonstrate that the \CRKSPH hydrodynamics solver, combined with the exact integration scheme, reproduces self-similar evolution in idealized, scale-free cosmological simulations.

\subsection{Star Formation}\label{subsec:SF}

Detailed modeling of star formation and the interstellar medium (ISM) is precluded by the limited resolution of cosmological simulations. Consequently, we adopt a coarse-grained subgrid model following \citet{springel2003cosmological} (henceforth \citetalias{springel2003cosmological}), which represents the ISM as a two-phase medium in which cold, dense clouds are embedded within a hot ambient gas in pressure equilibrium. We summarize the numerical implementation and adopted star formation parameters below and refer the reader to \citetalias{springel2003cosmological} for a detailed derivation. We also incorporate algorithmic refinements and threshold criteria similar to those used in \TNG\ \citep{vogelsberger2013model}.

While the \citetalias{springel2003cosmological} model captures the local regulation of star formation within the ISM, it does not include galactic-scale winds that expel gas from galaxies and enrich the surrounding medium. These outflows are modeled separately, with the \CRKHACC implementation described in Section~\ref{subsec:wind}.

\subsubsection{ISM Equation of State}

In the \citetalias{springel2003cosmological} ISM model, the evolution of the two-phase medium is governed by mass and energy exchange between the cold and hot gas components, which occurs through three main channels: 1) cold gas with density $\rho_c$ and internal energy $u_c$ is converted into stars on a characteristic timescale $t_{\rm sfr}(\rho) \propto \rho^{-1/2}$, consistent with gravitational free-fall and the Schmidt law. A short-lived fraction $\beta$ of this newly formed stellar population consists of massive stars that promptly explode as core-collapse supernovae, injecting mass and energy into the hot gas phase at a fiducial temperature of $T_{\rm SN} = 10^8\,\mathrm{K}$; 2) the supernova energy evaporates additional cold clouds with an efficiency parameter $A_{\rm ev}(\rho) \propto \rho^{-4/5}$ \citep{mckee1977}; and 3) radiative cooling at a rate $\Lambda$ condenses hot gas of density $\rho_h$ and internal energy $u_h$ back into the cold phase, establishing a self-regulated cycle of star formation.

The coupled processes yield an effective equation of state (EOS) that provides pressure support and mitigates runaway gravitational collapse in dense star-forming regions:
\begin{equation}\label{eq:EOS_SH03}
    P_{\rm SH03} = (\gamma-1)(\rho_h u_h + \rho_c u_c) = (\gamma-1)\rho u_{\rm eff},
\end{equation}
where $\gamma = 5/3$ is the adiabatic index of a monatomic ideal gas, and the effective specific energy is given by \mbox{$u_{\rm eff}=(1 - x_c)u_h + x_c u_c$}, with \mbox{$x_c = \rho_c / \rho$} defining the mass fraction of gas in cold clouds.

The thermal energy of the cold clouds is fixed at \mbox{$T_c = 10^3\,\mathrm{K}$}. The remaining quantities, $u_h$ and $x_c$, follow from the equilibrium solution derived in \citetalias{springel2003cosmological}:
\begin{align}
    u_h &= \frac{u_{\rm SN}}{(A_{\rm ev}+1) + u_c}, \\
    x_c &= \frac{2y}{1 + 2y + \sqrt{1 + 4y}}, \\
    y &= \frac{t_{\rm sfr}\Lambda(u_h, \,\rho,\, Z,\, Y,\, z)}{\rho\left[\beta u_{\rm SN} - (1 - \beta) u_c\right]},
\end{align}
where the ambient medium approaches equilibrium on a relaxation timescale
\begin{equation}
    \tau_{\rm relax} = t_{\rm sfr} \frac{(1 - x_c)}{\beta\,(A_{\rm ev} + 1)\,x_c}.
\end{equation}

To evaluate the core-collapse mass fraction $\beta$, we adopt the \citet{kroupa2001} initial mass function (IMF), a three-part broken power law with slopes \mbox{$\alpha = \{0.3, 1.3, 2.3\}$} over mass ranges $\{0.01\!-\!0.08,\, 0.08\!-\!0.5,\, 0.5\!-\!100\}\,M_\odot$.\footnote{The Kroupa IMF is chosen to maintain consistency with the \FIIIRE stellar enrichment fits described in Section~\ref{subsec:chem}.} Integrating the IMF using a supernova progenitor cutoff of $8\,M_\odot$ yields a massive star fraction of $\beta = 0.2$.

\Cref{eq:EOS_SH03} assumes maximal coupling of supernova feedback energy, which can lead to an overpressurized ISM. Following \citet{springel2005AGN} and \citet{hopkins2010}, we soften the ISM temperature using a dimensionless parameter $q_{\rm eos}$ that interpolates between the effective \citetalias{springel2003cosmological} model and an isothermal equation of state at $T_{\rm iso} = 10^4\,\mathrm{K}$, 
\begin{equation}\label{eq:SF_EOS}
    u_{\rm ISM} = q_{\rm eos}\, u_{\rm eff} + (1 - q_{\rm eos})\, u_{\rm iso}.
\end{equation}
We adopt $q_{\rm eos}=0.3$, consistent with the observationally motivated range $q_{\rm eos} \in [0.1, 0.3]$ from \citet{hopkins2010} and used in \citet{vogelsberger2013model}.

\subsubsection{Numerical Implementation}\label{subsec:SFImplementation}
 
Gas becomes eligible for star formation (SF) only if its hydrogen number density exceeds a fixed threshold, \mbox{$n_{\rm Hp} > n_{\rm sfr} = 0.13 \; {\rm cm^{-3}}$}. Here $n_{\rm Hp} = \rho X_{\rm p} / m_{\rm H}$ is the hydrogen number density evaluated using a fixed primordial composition with $X_{\rm p} = 0.75$ and $Y_{\rm p} = 0.25$ in \CRKHACC.\footnote{In contrast, the total hydrogen number density $n_{\rm H}$ is computed using the evolved hydrogen mass fraction $X$ from \cref{eqn:Xfrac}.} This convention ensures that the star formation threshold remains independent of local gas metallicity.

The density-dependent ISM parameters $A_{\rm ev}$ and $t_{\rm sfr}$ are evaluated using the same hydrogen number density convention:
\[A_{\rm ev}(\rho) = A_0 \left(\frac{n_{\rm Hp}}{n_{\rm sfr}}\right)^{-4/5} \!\!\text{and} \quad 
t_{\rm sfr}(\rho) = t_0 \left(\frac{n_{\rm Hp}}{n_{\rm sfr}}\right)^{-1/2}\!\!\!\!\!\!\!\!\!\!\!,
\]
where the evaporation efficiency and star formation timescale factors are set to $A_0 = T_{\rm SN} / 10^5\,{\rm K}$ and $t_0 = 2.2\,{\rm Gyr}$, respectively.

For particles exceeding the $n_{\rm sfr}$ threshold, the local ISM energy $u_{\rm ISM}$ is computed from \cref{eq:SF_EOS}. To suppress spurious star formation in hot, dense gas, particles with $u > u_{\rm ISM}$ are excluded from star formation eligibility. These particles continue to cool radiatively, but their thermal energy is prevented from dropping below $u_{\rm ISM}$. Gas with $u \leq u_{\rm ISM}$ constitutes the ISM phase and evolves thermally toward equilibrium over a relaxation timescale $\tau_{\rm relax}$ as follows,
\begin{equation}\label{eq:relax}
    u_{\rm sfgas}^1 = u_{\rm ISM} + (u_{\rm sfgas}^0 - u_{\rm ISM}) e^{-\Delta t / \tau_{\rm relax}},
\end{equation}
where the energy is updated from $u_{\rm sfgas}^0$ to $u_{\rm sfgas}^1$ over a timestep $\Delta t$.

When a gas particle becomes eligible for star formation, we impose a minimum ISM metallicity floor, $Z_{\rm ISM,\,min}$, to model unresolved early enrichment.\footnote{Regular gas is initialized with zero metallicity.} This floor is set to match the typical ISM metallicity of galaxies with baryonic masses comparable to the simulation mass resolution, which defines the scale of the first resolved stellar populations. Without this floor, newly formed low-mass galaxies would remain artificially metal-poor, requiring delayed self-enrichment to reach the observed mass--metallicity relation (MZR). As shown in Figure~\ref{fig:galmet} and discussed in Section~\ref{subsec:MZR}, the \CRKHACC\ galaxy MZR at $z = 0$ agrees well with observational data, maintaining consistency even for the smallest resolved systems.

For the simulations presented here, we adopt a metallicity floor of $Z_{\rm ISM,\,min} = 0.25\, Z_\odot$, consistent with the MZR of galaxies with stellar mass $M_* \sim 10^8\,\massh$ measured in the FIREBox simulations \citep{bassini2024}. At this mass scale, the gas-phase metallicity varies by only $\lesssim 0.2$--$0.3$~dex over $0 < z < 3$, making a fixed floor adequate for the selected mass resolution. Nonetheless, a redshift-dependent prescription, such as those inferred from observations (e.g., \citealt{zahid2014,curti2020,sanders2021}), can be readily incorporated into the imposed $Z_{\rm ISM,\,min}$.

We likewise impose a floor on the ISM helium fraction by linearly interpolating between the primordial and solar helium abundances using the same fractional factor, $Y_{\rm ISM,\,min} = Y_{\rm p} (1 - f) + f\, Y_\odot$, where $f \equiv Z_{\rm ISM,\,min}/Z_\odot$. As individual elemental abundances are not tracked in this work, disentangling contributions from specific enrichment channels is unnecessary, though such effects could be incorporated with more detailed MZR-based prescriptions.

The relaxation time in \cref{eq:relax} ensures smooth thermal evolution of ISM particles.
To maintain thermal energy continuity across the star formation threshold ($n_{\rm Hp}=n_{\rm sfr}$), we impose a density-dependent temperature floor that extends smoothly below the transition, motivated by \citet{schaye2008relation,schaye2023flamingo}:
\begin{equation}
\resizebox{0.88\linewidth}{!}{$
u_{\rm min} = u_{\rm ISM} (n_{\rm sfr})
\begin{cases}
       \left( \frac{n_{\rm Hp}}{n_{\rm sfr}} \right)^{1/3}\!\!\!\!\!\!\!\!\!, & n_{\rm sfr}\ge n_{\rm Hp}>10^{-4}\; {\rm cm^{-3}}\\
        1, & n_{\rm Hp} > n_{\rm sfr}
\end{cases}
$}
\end{equation}
where by construction $u_{\rm min} = u_{\rm ISM}$ at the density interface.\footnote{This condition is applied in addition to the global temperature floor imposed on all baryonic particles, which ensures that no gas cools below the temperature expected from adiabatic expansion since thermal decoupling, i.e. $T > T_{\rm cmb}/(1+z_{\rm decouple}) = 0.02~{\rm K}$, for $z_{\rm decouple}=129$.} Gas particles evaluate $u_{\rm ISM}(n_{\rm sfr})$ using the local metallicity and helium fraction (subject to the minimum ISM values $Z_{\rm ISM,\,min}$ and $Y_{\rm ISM,\,min}$), to maintain energy continuity even when enrichment occurs upon crossing the star formation threshold.

To complete the implementation, gas particles governed by the ISM equation of state are not only thermally regulated but also act as sites of star formation by converting cold cloud mass into stars at a rate $\dot{\rho}_{\rm sfr} \propto\rho_c/t_{\rm sfr}$. Accordingly, the star formation rate (SFR) $\dot{M}_{\rm sfr}$ for ISM particles of mass $m$ is given by 
\begin{equation} \label{eq:mstar}
  \dot{M}_{\rm sfr} = \alpha\, x_c\, m / t_{\rm sfr}, 
\end{equation} 
where $\alpha = (1 - \beta)$ in the original \citetalias{springel2003cosmological} formulation accounts for prompt stellar mass loss from core-collapse supernovae. In our implementation, stellar mass loss is modeled explicitly through the chemical enrichment module (Section~\ref{subsec:chem}), which tracks the evolving stellar population and its ejecta over time. Consequently, adopting an SFR scaling factor of $\alpha = 1$ is more consistent. All star-forming gas particles store the instantaneous local SFR as an additional attribute, which supports downstream analyses of galaxy star formation histories and sets the rate at which gas is converted into stars.

To produce stars at a rate $\dot{M}_{\rm sfr}$, \cref{eq:mstar} is integrated stochastically. Each star-forming gas particle is converted into a collisionless star particle of equal mass with probability
\begin{equation}\label{eqn:psfr}
p_{\rm \star} = 1-e^{-\frac{\dot{M}_{\rm sfr}}{m}\Delta t}
\end{equation}
evaluated at each timestep.\footnote{For star-forming gas particles, a timestep constraint $\Delta t < m/\dot{M}_{\rm sfr}$ is imposed in addition to the standard hydrodynamic CFL conditions described by Eqs.~(47)--(49) of \citetalias{frontiere2022simulating}.} 
This probabilistic formulation ensures that, on average, stars form at the expected rate while avoiding fractional particle spawning or continuous mass transfer. The integration procedure is extensible and can accommodate additional physics, such as galactic wind modeling, as described in the next section. 

Once formed, star particles evolve purely under gravity and no longer participate in hydrodynamic interactions. Each particle represents a single stellar population (SSP) and stores additional properties such as stellar age, birth SFR, and initial stellar mass. These quantities are retained for analysis and for use in subsequent subgrid models, including chemical enrichment as described in Section~\ref{subsec:chem}.

\subsection{Galactic Winds}\label{subsec:wind} 

Galactic winds are large-scale gas outflows driven by stellar feedback processes, including core-collapse supernovae, radiation pressure from young stars, and stellar winds from massive stars. Although the \citetalias{springel2003cosmological}-inspired model depicts the multiphase structure of the ISM and regulates star formation through an effective equation of state, it does not model bulk outflows from galaxies. As a result, galactic winds must be implemented separately to regulate star formation, mitigate runaway cooling, and enrich the circumgalactic medium (CGM). In our implementation, we adopt a subgrid wind model similar to that used in the \TNG\ simulations~\citep{vogelsberger2013model,pillepich2018}.

In this model, wind particles are stochastically launched from star-forming regions at an outflow rate $\dot{M}_{\rm w}$ proportional to the local star formation rate,
\begin{equation}\label{eqn:wind_rate}
\dot{M}_{\rm w} = \eta_{\rm w} \, \dot{M}_{\rm sfr},
\end{equation}
where $\eta_{\rm w}$ is the mass loading factor that controls the amount of gas ejected per unit stellar mass formed.

Wind particles are hydrodynamically decoupled from the surrounding gas upon launch, allowing them to escape the dense ISM and propagate into the CGM before recoupling based on local gas conditions. We have found that decoupled wind models facilitate robust convergence and simplify the calibration of multiredshift observables, such as the galaxy stellar mass function (see Figure~\ref{fig:gsmf}). However, fully coupled wind schemes (e.g., \citealt{dalla2008}) can yield markedly different internal galaxy morphologies. While \CRKHACC does not aim to resolve internal galactic structure in detail, both decoupled and coupled wind formulations are implemented to enable broader experimentation and model flexibility.

The wind launch velocity $v_{\rm w}$ is tied to the local dark matter velocity dispersion $\sigma_{\rm DM}$, following \citet{oppenheimer2006}, with an added redshift-dependent scaling from the \TNG model:
\begin{equation}\label{eqn:vw}
    v_{\rm w} = \kappa_{\rm w}\, \sigma_{\rm DM} \left( \frac{H_0}{H(z)} \right)^{1/3}\!\!\!\!\!\!\!\!,
\end{equation}
where $\kappa_{\rm w}$ is a dimensionless calibration parameter. The Hubble scaling compensates for the higher characteristic densities of virialized halos at early times, yielding wind velocities that are roughly redshift-independent (see Figure~6 of \citealt{pillepich2018}).

Unlike the \TNG\ model, we do not impose a minimum velocity floor or include a separate thermal energy component in the wind. Instead, the mass loading factor is computed directly from the available supernova energy per unit stellar mass:
\begin{equation}
    \eta_{\rm w} = \frac{2}{v_{\rm w}^2}\, E_{\rm w}, \quad \text{with} \quad E_{\rm w} = e_{\rm w} \cdot E_0.
\end{equation}
Here $e_{\rm w}$ is a dimensionless calibration parameter, and $E_0$ is the specific energy available from core-collapse supernovae per solar mass of stars formed. For consistency, we adopt the normalization of \citet{vogelsberger2013model}, $E_0 = 1.73 \times 10^{-2} \times 10^{51}\, \mathrm{erg}\,\mathrm{M}_\odot^{-1}$, corresponding to $10^{51}$~erg per event and $1.73 \times 10^{-2}$ supernovae per solar mass of star formation.\footnote{\citet{vogelsberger2013model} assumes a \citet{chabrier2003} IMF, whereas our subgrid models adopt a \citet{kroupa2001} distribution. The latter yields a core-collapse event rate of $1.05 \times 10^{-2}$ per solar mass, implying that our $e_{\rm w}$ values would increase by a factor of 1.65 to match the IMF normalization used in our simulations.}

\subsubsection{Numerical Implementation}\label{subsec:wind_algo}

The calibrated wind parameters at the resolution used in this study are $\kappa_{\rm w} = 3.0$ and $e_{\rm w} = 0.5$ (see Section~\ref{subsec:calib}). The dark matter velocity dispersion $\sigma_{\rm DM}$ required in \cref{eqn:vw} is computed by evaluating the velocity moments of nearby dark matter particles within a fixed proper radius of $30$~pkpc around each gas particle. This radius provides a stable, local estimate of the gravitational potential while minimizing sensitivity to small-scale velocity noise.

For each wind particle at position $\vb{x}_i$, we compute an SPH-weighted mean dark matter velocity using the standard \cite{Wendland1995} kernel with smoothing length selected to be $h_{\sigma} = 30~\mathrm{pkpc}$:
\begin{equation}
\langle \vb{v}_{\rm DM} \rangle_i = \frac{ \sum_{j\in \rm DM} \mathbf{v}_j \,  W_{ij}(h_{\sigma}) }{ \sum_{j\in \rm DM} W_{ij}(h_{\sigma}) },
\end{equation}
summing over all neighboring dark matter particles within the kernel support. The local velocity dispersion is then given by  
\begin{equation}
\sigma_{{\rm DM,}\,i}^2 = \frac{1}{3} \left( \langle \vb{v}_{\rm DM} \cdot \vb{v}_{\rm DM} \rangle_i - \langle \vb{v}_{\rm DM} \rangle_i \cdot \langle \vb{v}_{\rm DM} \rangle_i \right),
\end{equation}
where angle brackets denote SPH-weighted spatial averages throughout. 

The stochastic procedure for producing wind particles at the proper outflow rate $\dot{M}_{\rm w}$ follows \citet{vogelsberger2013model}. Since both stars and winds originate from the ISM reservoir, we calculate a total mass conversion rate
\begin{equation}\label{eqn:mtot}
    \dot{M}_{\rm tot} = \dot{M}_{\rm sfr}+\dot{M}_{\rm w} = (1+\eta_{\rm w})\,\dot{M}_{\rm sfr}.
\end{equation}
The probability that a star-forming gas particle is converted into either a star or a wind particle during a timestep $\Delta t$ is then
\begin{equation}\label{eqn:pevent}
    p_{\rm event} = 1-e^{-\frac{\dot{M}_{\rm tot}}{m}\Delta t}.
\end{equation}
By definition, this expression reduces to \cref{eqn:psfr} when the wind outflow rate is zero. If an event occurs, the conditional probabilities for the resulting particle type follow
\begin{align}
    p({\rm \star \, |\, event}) &= \frac{\dot{M}_{\rm sfr}}{\dot{M}_{\rm tot}},  \nonumber\\
    p({\rm w\, |\, event}) &= \frac{\dot{M}_{\rm w}}{\dot{M}_{\rm tot}} \equiv 1-p({\rm \star \, |\, event}).
\end{align}

Numerically, each star-forming gas particle draws two random numbers, $r_1$ and $r_2$, from a uniform distribution in the range $[0, 1]$. If $r_1 < p_{\rm event}$, a conversion occurs: a star particle is spawned if $r_2 < p({\rm \star \, |\, event})$; otherwise, a wind particle is launched. In both cases, the spawned particle inherits the parent gas mass, ensuring that, on average, mass is converted at the total rate $\dot{M}_{\rm tot}$, with stars and winds forming at their respective rates $\dot{M}_{\rm sfr}$ and $\dot{M}_{\rm w}$.

Following \citet{pillepich2018}, winds are launched isotropically with velocity $\vb{v}_{\rm w} = v_{\rm w} \vb{n}_{\rm w}$, where $\vb{n}_{\rm w}$ is a randomly oriented unit vector assigned for each launch event. As described in Section~\ref{subsec:stochastic}, we employ a counter-based random number generator (CBRNG) that uses the particle ID (unique integer label) and global timestep as input keys. This approach ensures that wind particles spawned in overloaded regions are consistent across MPI ranks without requiring communication. The same procedure is applied to assign directions for kinetic AGN feedback, as discussed in Section~\ref{subsec:bhfeedback}.

Once launched, wind particles record their age $t_{\rm wind}$, launch velocity $v_{\rm w}$, and mass loading factor $\eta_{\rm w}$ for diagnostics and analysis. While wind particles evolve without hydrodynamic force interactions, they are explicitly included in local gas density estimates:
\begin{equation}\label{eqn:rhow}
      \rho_{i} = \frac{\sum_{j\in g\pluscup\rm w} m_{j} V_{j,\, g\pluscup\rm w} \Wrij}{\sum_{j\in g\pluscup\rm w} V_{j, \,g\pluscup\rm w}^2 \Wrij},
\end{equation}
where $\Wrij$ is the \CRKHACC reproducing kernel function described by \cref{eq:RK} in Appendix~\ref{app:RKRelax}. 
This formulation is identical to the gas density equation used in our previous studies~(\citetalias{frontiere2017}~\&~\citetalias{frontiere2022simulating}), with the modification that the particle volume definition extends to include wind particle neighbors if present, as does the summation above (denoted by $g\pluscup\rm w$, the disjoint union of gas and wind particle sets), as discussed in Appendix~\ref{app:MS}.

\subsubsection{Wind recoupling}\label{subsec:windR}

Wind particles remain hydrodynamically decoupled until one of two recoupling criteria is met, following the \TNG model: (a) the local gas density drops below $5\%$ of the star formation threshold, $\rho_{\rm couple} = 0.05\,\rho_{\rm sfr}$; or (b) the particle has traveled for longer than $2.5\%$ of the current Hubble time, $t_{\rm wind} > 0.025\,H^{-1}_0$. Consistent with \citet{pillepich2018}, we find that the time criterion is rarely exceeded before the density threshold is satisfied. We support an option for decoupled wind particles to remain subject to radiative cooling and UV photoheating (see Section~\ref{subsec:radcool}), which reduces spurious transients during reintegration. Once a recoupling condition is met, the wind particle is converted back into a hydrodynamically active gas particle.

\subsubsection{Wind Metal Deposition}\label{subsec:windZ}

In addition to carrying mass and energy, galactic winds transport metals from the ISM into the circumgalactic medium. To model this process, we follow the approach of \citet{vogelsberger2013model} and introduce a metal loading factor, $\gamma_{\rm w}$, which controls the fraction of metals entrained by wind particles.

When a wind particle is launched, it inherits a fraction $\gamma_{\rm w}$ of the parent star-forming gas particle metal mass:
\begin{equation}
    m_{Z_{\rm w}} = \gamma_{\rm w} m_{Z_{\rm ISM}}.
\end{equation}
The remaining fraction, $(1 - \gamma_{\rm w})$, is deposited locally into the ISM. The deposition follows SPH weighting, where each gas neighbor \( j \) with volume $V_{j,g}$ (see \cref{eqn:Vg}) receives a metal mass 
\begin{equation}\label{eqn:windmetal}
    m_{Z_{{\rm deposit,\,} j}} = A\, V_{j,g}\, W_{ij}(h_{\rm w}) \, (1 - \gamma_{\rm w})\, m_{Z_{\rm ISM}},
\end{equation}
with normalization
\begin{equation}\label{eq:Anorm}
A = \left( \sum_j V_{j,g}\, W_{ij}(h_{\rm w}) \right)^{-1}\!\!\!\!\!.
\end{equation}
The kernel shape and wind smoothing length, $h_{\rm w}$, are identical to those used in the hydrodynamic force solver for regular gas. Metallicities are updated accordingly as $Z_{\rm w} = m_{Z_{\rm w}}/m_{\rm w}$ and $Z_{{\rm ISM},\,j} = (m_{Z_{\rm ISM},\,j} + m_{Z_{{\rm deposit,\,} j}})/m_{\rm ISM, \, j}$, leaving the total particle masses unchanged. Neighbors with metallicities exceeding $Z_{\max} = 20 \,Z_\odot$ are excluded from both enrichment and normalization, although such cases are rare.

This deposition model is motivated by both physical and numerical considerations. Physically, galactic winds may preferentially entrain newly synthesized metals from supernova ejecta, enhancing metal transport into the CGM. Numerically, at finite mass resolution, unregulated metal loading can excessively deplete galaxies of metals, particularly when the resolution is insufficient to capture recycling flows and mixing within the ISM.

In our implementation, we find that maintaining realistic galaxy metallicities at the mass resolution used in this paper (baryon mass $\sim 2\times 10^8\,\massh$) requires a small value of $\gamma_{\rm w} \in [0,0.05]$, which yields the most robust results. We therefore adopt a fiducial value of $\gamma_{\rm w} = 0$, meaning all metals are deposited locally in the ISM at wind launch.\footnote{Although we set $\gamma_{\rm w} = 0$, metals deposited into the ISM at wind launch can still be transported into the CGM through subsequent gas flows, AGN-driven outflows, and large-scale dynamical evolution.} The choice of $\gamma_{\rm w}$ is resolution dependent and partly degenerate with the seed metallicity floor imposed on the ISM ($Z_{\rm ISM,\,min}$; see Section~\ref{subsec:SFImplementation}) when evaluating the galaxy mass--metallicity relation (MZR). The resulting ISM and stellar MZRs at the calibrated resolution of this study are shown in Figure~\ref{fig:galmet} and discussed in Section~\ref{subsec:MZR}.

\subsection{Chemical Enrichment}\label{subsec:chem}

Each star particle in \CRKHACC represents a single stellar population (SSP) whose formation time is recorded at the moment it is converted from ISM gas. From that point onward, we integrate stellar mass-loss rates at every timestep to compute the total mass ($\Delta M^{\rm enrich}_T$), helium ($\Delta M^{\rm enrich}_Y$), and metal ($\Delta M^{\rm enrich}_Z$) returned to the ISM by each SSP. 

In the following, we consider stellar enrichment from two primary channels: discrete supernovae (Type~Ia and core-collapse SNe) and continuous stellar winds from OB and AGB stars. By default, \CRKHACC adopts enrichment rates and yields from the \FIIIRE\ model (\citealt{hopkins2023fire}; hereafter \citetalias{hopkins2023fire}), which provides time- and metallicity-dependent stellar evolution and nucleosynthetic prescriptions. 

Briefly summarizing the implementation, \FIIIRE employs the 2021 release of \smaller{STARBURST99} with rotating stellar tracks from \citet{leitherer2014}, assuming a three-part \citet{kroupa2001} IMF. Core-collapse SNe yields follow modifications to \citet{sukhbold2016}; Type~Ia SNe use the delay-time distribution of \citet{maoz2017} with ejecta compositions given by the average of the W7 and WDD2 models from \citet{leung2018}; and continuous OB- and AGB-star winds adopt the \smaller{STARBURST99} mass-loss rates combined with the yield models of \citet{cristallo2015} and \citet{limongi2018}. The earlier population-averaged \FIIRE fits from \citet{hopkins2018fire} are also available as an alternative configuration for the enrichment module in \CRKHACC.

\subsubsection{Supernovae Mass Loss}\label{subsec:SNloss}

Mass loss attributed to supernovae events over a given timestep $\Delta t$ is calculated as:
\begin{equation}
\Delta M_c^{\rm SN} = M_*\int_{t_0}^{t_0 + \Delta t} M^{\rm SN}(t) R^{\rm SN}(t) f^{\rm SN}_c(t)\ dt,
\label{eqn:snint}
\end{equation}
where $M_*$ is the initial star particle mass, $M^{\rm SN}(t)$ is the ejecta mass per event, $R^{\rm SN}(t)$ is the supernova rate per unit stellar mass, and $f^{\rm SN}_c(t)$ is the time-dependent mass fraction yield for components $c = \{T,Y,Z\}$.

For Type~Ia supernovae, the ejecta mass per event is fixed at $M^{\rm Ia} = 1.4\,M_\odot$, and the expelled material is assumed to be hydrogen- and helium-free ($f^{\rm Ia}_T = f^{\rm Ia}_Z = 1$, $f^{\rm Ia}_Y = 0$). Thus, $\Delta M^{\rm Ia}_T = \Delta M^{\rm Ia}_Z$ and $\Delta M^{\rm Ia}_Y = 0$. The SNIa rate, $R^{\rm Ia}(t)$, is modeled as a power law (Eq.~(3) in \citetalias{hopkins2023fire}) and is non-zero for stellar ages $t > 44$ Myr.

For core-collapse supernovae, the event rate $R^{\rm CC}(t)$ is a piecewise power law (Eq.~(2) in \citetalias{hopkins2023fire}), active over the stellar-age interval $3.7~\text{Myr} \leq t \leq 44~\text{Myr}$. CC events adopt a time-dependent ejecta mass $M^{\rm CC}(t)$ and yields $f^{\rm CC}_c(t)$ with coefficients given in Eq.~(7) and Table~1 of \citetalias{hopkins2023fire}.\footnote{\citetalias{hopkins2023fire} tabulates yields for He and nine individual metal species. In \CRKHACC, we use the total metal return $\Delta M^{\rm CC}_Z$ and fit the summed fraction $f_Z(t) = \sum_{i=1}^9 f_{Z_i}(t)$ to the same piecewise form.} In both cases, \cref{eqn:snint} is integrated analytically, with closed-form solutions provided in Appendix~\ref{app:SNloss}.

\subsubsection{Stellar Outflows}\label{subsec:stellaroutflow}
Continuous stellar mass loss over the SSP lifetime, primarily from OB and AGB winds, is computed as
\begin{equation}\label{eqn:swind}
\Delta M_c^{\rm w} = M_* \int_{t_0}^{t_0 + \Delta t} f_{\rm w}(t) f^{\rm w}_c(t)\, dt,
\end{equation}
where $f_{\rm w}(t)$ is the total wind mass-loss rate per unit stellar mass, and $f^{\rm w}_c(t)$ is the yield fraction for each component. 

The wind rate combines separate OB and AGB contributions, $f_{\rm w}(t) = f_{\rm w}^{\rm OB}(t) + f_{\rm w}^{\rm AGB}(t)$, both given by Eq.~(4) in \citetalias{hopkins2023fire}. OB winds are active for $t < \unit{20}{\Myr}$, while AGB winds become significant at later times ($t \approx \unit{800}{\Myr}$). OB winds depend on progenitor metallicity via fitted coefficients in a piecewise power law model, whereas AGB winds are metallicity-independent but have a more complex time dependence. For most elements, stellar wind yields match the progenitor abundances, but special channels for He, C, N, and O are tracked separately (Eq.~(8) in \citetalias{hopkins2023fire}).  

\Cref{eqn:swind} cannot be integrated analytically for all inputs, as several fitted segments from \citetalias{hopkins2023fire} lack exact closed-form solutions. We therefore numerically integrate and fit the cumulative mass loss for the total, helium, and metal components (combining all individual channels). Table~\ref{tab:integrals} in Appendix~\ref{app:stellaroutflow} lists the fitted functions used in practice.

\subsubsection{Numerical Implementation}

\begin{figure}
    \centering
    \includegraphics[width=\linewidth]{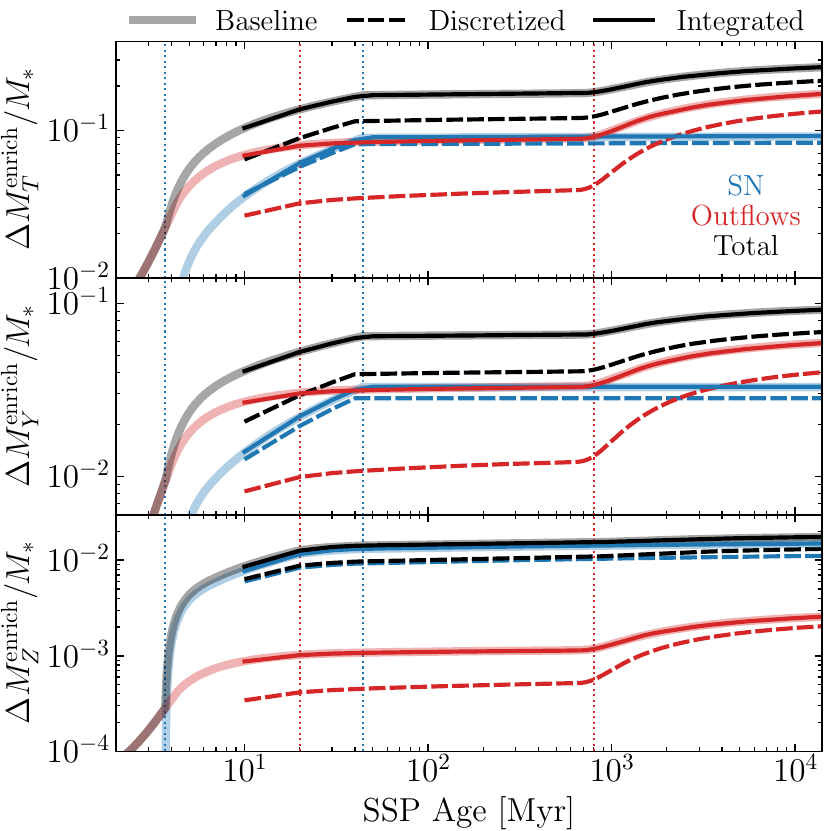}
    \caption{Cumulative mass lost for the total (top), helium (middle), and metal (bottom) mass components of an SSP with solar metallicity, a primordial helium fraction, and initial mass $M_*$. Blue and red curves separate the mass-loss contributions from supernovae and stellar winds, respectively, and the black curve shows their sum. Solid lines show results from the integrated method using a timestep of $\unit{10}{\Myr}$, representative of the largest hydrodynamic steps in \CRKHACC. Shaded lines indicate the baseline solution obtained by summing instantaneous FIRE-3 rates over very fine timesteps, while dashed lines show the discretized case, where instantaneous rates are sampled only at $\unit{10}{\Myr}$ intervals. Vertical dotted blue lines mark 3.7 and $\unit{44}{\Myr}$, the characteristic timescales for core-collapse and Type Ia supernovae, and vertical dotted red lines mark 20 and $\unit{800}{\Myr}$, corresponding to OB and AGB stellar winds. The timestep-independent integration method matches the baseline to within $2\%$ (set by the accuracy of the implemented fits), whereas coarse discretization of the instantaneous rates suppresses mass loss, primarily due to unresolved early stellar evolution.}
    
    \label{fig:mloss}
\end{figure}

At each timestep, the total mass for each component returned by all enrichment sources is
\begin{equation}
\Delta M^{\rm enrich}_c = \Delta M^{\rm Ia}_c + \Delta M^{\rm CC}_c + \Delta M^{\rm w}_c.
\label{eq:dmcet}
\end{equation}
The terms in \cref{eq:dmcet} correspond to the supernova and stellar wind contributions described above, with analytic and fitted formulations summarized in Appendix~\ref{app:Enrich}. Although the resulting functional form is somewhat complex, the GPU implementation is straightforward: at each timestep, the cumulative solutions yield $\Delta M^{\rm enrich}_T$, $\Delta M^{\rm enrich}_Y$, and $\Delta M^{\rm enrich}_Z$, which are then redistributed to gas neighbors.

\Cref{fig:mloss} shows the cumulative mass loss for the total, helium, and metal components, separating contributions from supernovae and stellar outflows and illustrating the underlying stellar evolution. The integration scheme provides an accurate, timestep-independent cumulative mass loss, analogous to the exact integration cooling treatment described in Section~\ref{subsec:exact}. The integrated method agrees with the baseline solution, obtained by summing the instantaneous FIRE-3 rates over very fine timesteps, with a small residual arising from AGB segments that cannot be integrated analytically. In contrast, evaluating the instantaneous rates only at coarse, discretized hydrodynamic intervals underestimates the total mass loss by tens of percent due to incomplete sampling of the rapidly varying early stages of stellar evolution. Thus, this integration approach enables large-scale cosmological simulations to capture the cumulative effects of detailed stellar evolution without requiring prohibitively fine temporal integration.

The ejected mass is transferred to neighboring gas particles using SPH kernel weighting, while the star particle mass is reduced accordingly. Specifically, the mass component for gas neighbor $j$ increases as
\begin{equation}\label{eqn:starmetal}
    \Delta m_{c,j} = A\, V_{j,g}\, W_{ij}(h_{\star})\, \Delta M^{\rm enrich}_c,
\end{equation}
where $h_{\star}$ is the stellar smoothing length and $A$ is the normalization constant spanning all gas neighbors of the star, analogous to \cref{eq:Anorm} for wind ISM enrichment. This local coupling conserves mass and species, thereby ensuring consistent ISM enrichment. 

To limit mass growth, we enforce that any gas particle with a mass exceeding four times the initial baryon mass is not enriched and is excluded from the normalization. As ISM mass increases only through incremental stellar enrichment, this case is rarely encountered. Consistent with the wind metal loading prescription (Section~\ref{subsec:windZ}), outlier gas particles with metallicities exceeding $Z_{\rm max} = 20\, Z_\odot$ are likewise excluded from stellar enrichment.

In addition to mass transfer, linear momentum is also explicitly conserved. Star particles are treated as collisionless, so we do not directly apply momentum changes to them during enrichment. Instead, the lost momentum is transferred to neighboring gas particles according to
\begin{equation}\label{eqn:stellarmom}
\vb{v}_{\rm g}^{\,\rm new} = \frac{m_{\rm g} \vb{v}_{\rm g} + \sum_{k^{\star}} \Delta m_{T,\,k^{\star}} \vb{v}_{j^{\star}}}{m_{\rm g} + \sum_{k^{\star}} \Delta m_{T,\,k^{\star}}},
\end{equation}
where $k^{\star}$ indexes all stellar neighbors that contribute mass to a given gas particle. Thus, star particle dynamics are minimally impacted by enrichment, and momentum is transferred to the gas, as is done in other subgrid operators (e.g., wind and AGN feedback).

\subsection{Black Holes and AGN Feedback}
\label{sec:bhmodel}

Feedback from accreting supermassive black holes, commonly referred to as active galactic nuclei (AGN), plays a central role in regulating galaxy formation across cosmic time. Observational and theoretical studies indicate that AGN are required to quench star formation in massive halos, maintain hot gaseous atmospheres in galaxy clusters, and prevent runaway cooling in the cores of massive galaxies (e.g., \citealt{fabian2012,heckman2014}). In large-scale cosmological simulations, these effects are typically modeled using subgrid prescriptions for black hole (BH) seeding, accretion, growth, and feedback energy injection, following approaches originating from, e.g., \citet{springel2005AGN}, \citet{sijacki2007}, and \citet{booth2009}.

In our simulations, BHs are represented as collisionless sink particles that serve as localized sources of energy and momentum injection into the surrounding medium. They are seeded during the early stages of galaxy formation and grow through gas accretion and mergers with other BHs. Although they do not exert hydrodynamic forces directly and interact with gas only through subgrid prescriptions, they contribute to the gravitational potential and evolve dynamically under gravity like any other collisionless particle (e.g., dark matter or stars). The following subsections summarize the implementation of each major component of AGN evolution.

\subsubsection{Seeding}
\label{subsec:bhseed}

We initialize BHs directly within resolved galaxies. At regular intervals, galaxies are identified on the fly using a GPU-accelerated density-based spatial clustering (DBSCAN; \citealt{ester1996}) algorithm applied to the star-particle distribution within each rank, together with a friends-of-friends (FOF; \citealt{davis1985}) halo finder executed on all dark matter particles (see Section~\ref{subsec:cosmotools} for details). Owing to the efficiency of our GPU-based in situ analysis framework, AGN seeding is performed by default at every PM step. In practice, the results are largely insensitive to the specific interval, provided early galaxy formation is adequately captured.

For each identified galaxy, we define its center as the most bound star particle and construct a $50$~pkpc aperture around it to extract all constituent particles, including baryons and nearby dark matter. Galaxies are considered for seeding only if they do not already host a BH particle. To ensure that the galaxy resides within a virialized structure, we locate the dark matter particle closest to the galactic center and verify its FOF group membership. If multiple galaxies exist within the same halo, only the \emph{central} system --- defined as the galaxy with the largest stellar mass --- is eligible for seeding.

These criteria help suppress numerical artifacts, such as the reseeding of black holes displaced during interactions, and provide greater flexibility across multiple subgrid configurations. In most cases, a single galaxy in a low-mass halo is seeded only once; the additional logic primarily guards against special cases such as flybys or interacting substructures.

This galaxy-based strategy contrasts with the more common approach of seeding black holes at the centers of FOF halos once a fixed mass threshold is exceeded \citep[e.g.,][]{sijacki2007, dimatteo2008}. By instead tying BH formation to the resolved stellar mass, we avoid populating halos prematurely and ensure that seeding occurs only once a galaxy is well formed. Consequently, the seeding prescription remains consistent across different simulation mass resolutions, adapting naturally as higher particle sampling resolves smaller and earlier galaxies that are then seeded accordingly. The FOF halo finder is used solely to verify that the galaxy resides within a virialized halo and to prevent duplicate seeding, not to impose an explicit mass threshold.

A BH is seeded by converting the gas particle closest to the galactic center into a BH particle and repositioning it exactly at that location. The new BH inherits the mass of its parent gas particle. However, this particle mass, $m_{\rm BH}$, is typically much larger than a realistic astrophysical seed mass. Following the standard practice introduced by \citet{springel2005AGN}, we evolve a separate internal mass, $M_{\rm BH}$, representing the physical black hole mass used in accretion and feedback calculations. This internal mass is initialized to a calibration parameter $M_{\rm seed}$, typically in the range $10^5$--$10^6\,\massh$, depending on the simulation resolution. $M_{\rm BH}$ then grows continuously through AGN accretion, while the particle mass remains fixed until synchronization events such as mergers or explicit mass updates, as detailed below.

\subsubsection{Repositioning and Mergers}\label{subsec:bhrepos}

To keep each BH centered within its host galaxy, we apply an explicit repositioning step at every PM update. Candidate locations are selected from star particles within a fixed interaction radius of $10\,{\rm pkpc}$, which also sets the maximum allowed BH displacement. The local gravitational potential is computed from these neighboring stars using a softening length of $1\,{\rm pkpc}$, and the BH is relocated to the position of the particle with the minimum potential. The BH velocity is then updated to the mass-weighted mean of nearby stars.

The adopted distance scales reflect typical galaxy sizes at intermediate redshift and are chosen to suppress noise in the local potential while preventing the BH from spuriously jumping between galaxies. Although these parameters are tunable, any choice consistent with characteristic galaxy extents yields stable and reliable repositioning. Only stars contribute to the potential calculation and candidate selection, as they trace the collisionless stellar core and are less affected by hydrodynamic forces or transient perturbations. This treatment helps keep the BH anchored to the true galactic center, avoiding misidentification with gas clumps or diffuse dark matter substructure.

BHs are merged when, after repositioning, two or more black hole particles occupy the same spatial location. This situation naturally occurs during galaxy mergers, as the repositioning scheme drives BHs toward the local stellar potential minimum, ultimately placing them at the same central star particle. This method has proven robust in avoiding premature coalescence or artificial displacements and removes the need for additional merger criteria based on relative velocity or separation, such as those employed by \citet{springel2005AGN}, \citet{booth2009}, and \citet{bahe2022}. These alternatives rely on assumptions about black hole kinematics that are not well resolved in our simulations. Likewise, we do not adopt more complex prescriptions such as dynamical friction models \citep[e.g.,][]{tremmel2015}, as they offer no clear advantage in our context.

Merging is executed in parallel on the GPU using a lock-based protocol that guarantees unique, non-overlapping pairings during each iteration. If a black hole is eligible to merge with multiple partners, atomic locks ensure that only a single merge is carried out per pass, and the routine is repeated until no eligible pairs remain. To maintain consistency across domain boundaries, mergers are allowed only when both black holes are either owned by the same MPI rank or by none (i.e., both are overloaded particles). This restriction may delay a merger by at most one PM step but ensures consistent behavior across overlapping rank domains.

After merging, the resulting BH inherits the center-of-mass position (which, in our implementation, coincides with the original input locations) and the center-of-momentum velocity of the pair. Both the internal and particle masses are summed, and the less massive BH is deactivated (flagged as ``dead'') but retained in the simulation for post-processing analysis.

\subsubsection{Accretion}
\label{subsec:bhaccretion}

We model black hole accretion using the canonical Bondi--Hoyle--Lyttleton formalism \citep{hoyle1939,bondi1944}, as is standard in many cosmological simulations. The accretion rate is
\begin{equation}
\dot{M}_{\rm Bondi} = \alpha \frac{4\pi G^2 M_{\rm BH}^2 \rho}{(c_s^2 + v_{\rm rel}^2)^{3/2}},
\end{equation}
where $M_{\rm BH}$ is the internal black hole mass, $\rho$ is the ambient gas density and $c_s$ is the local sound speed (both measured within an SPH kernel of smoothing length $h_{\rm BH}$), and $v_{\rm rel}$ is the relative velocity between the black hole and surrounding gas. The dimensionless factor $\alpha$ serves as a calibration parameter to account for unresolved small-scale gas structure.
Following \citet{weinberger2016simulating}, we fix $\alpha = 1$, as its impact is largely degenerate with the seed mass $M_{\rm seed}$.

Rather than estimating $v_{\rm rel}$ directly, we adopt a fixed sub-resolution value of $v_{\rm rel} = 8\,\mathrm{km\,s^{-1}}$, motivated by the typical velocity dispersion of ISM gas \citep{dib2006}. This choice avoids artificial divergences in the Bondi denominator and reflects the fact that BH--gas relative velocities are not reliably resolved at our simulation scale. Although $v_{\rm rel}$ is formally tunable, its impact is modest across physically reasonable values (a few to tens $\mathrm{km\,s^{-1}}$) and does not significantly affect global black hole growth or the star formation history.

To prevent unphysically large accretion rates, we impose an upper limit set by the Eddington rate,
\begin{equation}
\dot{M}_{\rm Edd} = \frac{4\pi G M_{\rm BH} m_p}{\epsilon_{\rm r} \sigma_T c},
\end{equation}
where $m_p$ is the proton mass, $\sigma_T$ is the Thomson cross section, $c$ is the speed of light, and $\epsilon_{\rm r}$ is the radiative efficiency.\footnote{Super-Eddington accretion is observed and simulated in some contexts \citep[e.g.,][]{jiang2019,du2018}, but lies beyond the scope of the simplified subgrid model adopted here, where the Eddington limit serves as a numerical safeguard against unresolved or extreme accretion rates.} Typical values of $\epsilon_{\rm r}$ range from 0.1 to 0.2; we adopt $\epsilon_{\rm r} = 0.2$, consistent with \citet{weinberger2016simulating}. The resulting accretion rate is then
\begin{equation}
\dot{M}_{\rm BH} = \min(\dot{M}_{\rm Bondi}, \dot{M}_{\rm Edd}),
\end{equation}
which is computed and stored for each black hole after every AGN operator call, including repositioning and merging steps at the PM level, as well as post-feedback updates executed at the end of each gravity subcycle.

The internal black hole mass is updated continuously, following the procedure of \citet{bahe2022}. At each drift timestep (synchronized across all particles; see Appendix~\ref{app:integrator}), the internal mass increases by
\begin{equation}
\Delta M_{\rm BH} = (1 - \epsilon_{\rm r}) \, \dot{M}_{\rm BH} \, \Delta t.
\end{equation}
When the internal mass exceeds the particle mass ($M_{\rm BH} > m_{\rm BH}$), the excess is drawn from neighboring gas particles using SPH kernel weights during each gravity subcycle. The mass contribution from gas neighbor $j$ is
\begin{equation}\label{eqn:agnmass}
    \Delta m_j = A\, V_{j,g}\, W_{ij}(h_{\rm BH})\, (M_{\rm BH} - m_{\rm BH}),
\end{equation}
where $A$ is a normalization factor defined analogously to \cref{eq:Anorm}. No gas particle is allowed to fall below half its initialized baryon mass; in such rare cases, the residual mass is redistributed to other eligible neighbors in the following subcycle.

Mass transfer also carries momentum. To conserve linear momentum, we preserve gas particle velocities and update the black hole velocity according to
\begin{equation}
\vb{v}_{\rm BH}^{\,\rm new} = \frac{m_{\rm BH} \vb{v}_{\rm BH} + \sum_j \Delta m_j \vb{v}_j}{m_{\rm BH} + \sum_j \Delta m_j}.
\end{equation}
This procedure mirrors \cref{eqn:stellarmom} for stellar enrichment, where the mass recipient also absorbs the momentum change. Given our fiducial implementation of manual AGN repositioning, this correction has minimal impact on AGN evolution, but is more consequential when employing alternative repositioning or dynamical prescriptions implemented in \CRKHACC.

\subsubsection{Feedback}
\label{subsec:bhfeedback}

We implement a two-mode AGN feedback model using a modified \TNG thermal and kinetic scheme \citep{weinberger2016simulating}.
The feedback mode is determined by the dimensionless accretion state parameter
\begin{equation}
\chi = \frac{\dot{M}_{\rm BH}}{\dot{M}_{\rm Edd}}.
\end{equation}
We adopt the same mode-switching threshold,
\begin{equation}
\chi_{\rm thresh} = \min\left[\chi_0\left(\frac{M_{\rm BH}}{10^8 \,M_\odot}\right)^\beta, \;\;\chi_{\rm max}\right],
\end{equation}
using identical parameter values to those in \TNG: $\chi_0 = 0.002$, $\beta = 2.0$, and $\chi_{\rm max} = 0.1$, which regulate the thermal-to-kinetic transition.

At each gravity subcycle, the energy released by AGN feedback is computed as
\begin{equation}
\Delta E_{\rm AGN} = \epsilon \, \dot{M}_{\rm BH} \, c^2 \, \Delta t,
\end{equation}
where the effective coupling efficiency $\epsilon$ depends on the current accretion mode:
\begin{equation}
\epsilon = 
\begin{cases}
    \epsilon_{\rm high} \, \epsilon_{\rm r}, & \chi \ge \chi_{\rm thresh} \\
    \epsilon_{\rm kin}, & \chi < \chi_{\rm thresh}
\end{cases}
\end{equation}
with $\epsilon_{\rm high} = 0.1$ as in \TNG, and $\epsilon_{\rm kin}$ a tunable calibration parameter. Evaluating feedback at the subcycle interval, rather than on the local hydrodynamic timestep, ensures that all gas particles are synchronized at each injection event and that subsequent CFL constraints are immediately enforced. 

If the feedback mode is thermal ($\chi \ge \chi_{\rm thresh}$), the AGN energy $\Delta E_{\rm AGN}$ is injected immediately.
In the kinetic mode ($\chi < \chi_{\rm thresh}$), the energy instead accumulates until it exceeds a minimum injection threshold defined as
\begin{equation}
E_{\rm thresh} = \frac{1}{2} M_{\rm enclosed} \, v_{\rm jet}^2,
\end{equation}
where $M_{\rm enclosed} = N_{\rm BH} \times m_0$ is the idealized enclosed gas mass within the AGN kernel, with $N_{\rm BH}=48$ neighbors (see Appendix~\ref{app:MS}) and $m_0$ the initialized baryon particle mass. The parameter $v_{\rm jet}$ serves as a second tunable calibration constant. We found this constant-threshold formulation numerically preferable to the potentially noisier dark matter velocity dispersion parameterization used in \TNG (see Eq.~13 in \citealt{weinberger2016simulating}).

When feedback energy is injected --- either immediately (thermal mode) or after exceeding the threshold (kinetic mode) --- it is distributed to neighboring gas particles using SPH kernel weighting:
\begin{equation}\label{eqn:AGNE}
\Delta E_j = A\, V_j\, W_{ij}(h_{\rm BH})\, \Delta E_{\rm AGN},
\end{equation}
where the normalization constant $A$ is defined similarly to that used in the accretion and wind enrichment calculations (\cref{eqn:agnmass,eqn:windmetal}).

If the feedback mode is thermal, the injected energy is deposited as internal energy:
\begin{equation}
\Delta u_j = \Delta E_j / m_j.
\end{equation}
If the mode is kinetic, the energy is instead converted into a velocity impulse directed along a single, randomly chosen unit vector $\vb{n}_{\rm jet}$:
\begin{equation}
\Delta \vb{v}_j = \sqrt{2\,\Delta E_j/m_j}\;\vb{n}_{\rm jet}.
\end{equation}

To ensure consistent directionality across MPI ranks for AGNs located in overloaded zones, we employ a counter-based random number generator (CBRNG) seeded by a composite key derived from the unique particle ID and the global accumulated subcycle timestep (see Section~\ref{subsec:stochastic}). This guarantees synchronized and reproducible injection without inter-node communication, a practical advantage of injecting feedback at the regular subcycle interval.  

If the AGN switches from kinetic to thermal mode before the accumulated energy exceeds the threshold, all stored energy is immediately converted to thermal feedback and injected accordingly. Each black hole particle continuously tracks the accumulated feedback energy as an additional attribute, along with other evolved quantities such as the internal mass $M_{\rm BH}$, the accretion rate $\dot{M}_{\rm BH}$, and the particle age (time since seeding). These quantities are updated throughout the simulation and together define the instantaneous state of the AGN model.

\subsection{Stochastic Integration and Parallel Reproducibility}\label{subsec:stochastic}

Many of the subgrid models introduced above involve stochastic integration or randomized sampling --- for example, probabilistically converting star-forming gas into stars, launching galactic winds, or selecting AGN feedback orientations. As the \CRKHACC\ solver advances particles independently on each MPI rank during a PM step (using particle overloading), it is important to evaluate stochastic processes consistently across overlapping domain boundaries to maintain parallel reproducibility.

This poses a challenge for subgrid operators that act at the hydrodynamic timestep level, which may vary between ranks depending on local conditions. Unlike the PM and gravity subcycle intervals --- which are fully synchronized --- local hydrodynamic timesteps may desynchronize stochastic updates if not carefully controlled.

To resolve this, we employ a CBRNG with a key defined by the combination of each particle unique ID and a global timestep counter.\footnote{Since multiple subgrid operators may require random draws during the same interval, an additional subgrid index key is included to allow for multiple independent random sequences using the same particle ID and timestep counter.}
The global timestep represents an artificially small integration interval, specified at the start of the simulation, that defines the finest level of the timestep hierarchy. Although particles do not explicitly evolve on this interval, all stochastic integrators do. For example, rather than evaluating the probability of a star-forming gas particle spawning a wind or star over its entire local timestep, we integrate that probability repeatedly at the global timestep scale. This guarantees that stochastic events remain synchronized across ranks, even when local integration intervals differ due to domain decomposition.

Although this approach can require many more random number draws, it remains computationally tractable because the chosen CBRNG (``squares'' RNG; \citealt{widynski2020}) relies solely on simple multiplication and bitwise operations that are highly efficient on GPUs. Moreover, since stochastic events are integrated independently for each particle, these operations add negligible cost compared to neighbor traversal and force interaction kernels. As such, this strategy ensures consistent stochastic behavior and fully reproducible results across arbitrary parallel decompositions, without requiring additional MPI communication or incurring significant computational cost.

\subsection{In Situ Galaxy Finding}\label{subsec:cosmotools}

The new physics modules implemented in \CRKHACC, while coarsening complex physical processes, are computationally expensive. Increasing the fidelity and diversity of modeled feedback sources drives smaller timesteps and more frequent updates, placing heavy demands on the numerical infrastructure.

To maintain performance at exascale, \CRKHACC leverages an optimized GPU architecture featuring a tree solver that minimizes rebuild overhead, fully GPU-resident short-range gravity and hydrodynamic kernels that eliminate host--device transfers, and a ``warp-splitting'' strategy that improves thread occupancy and memory efficiency across interacting particle groups. Together, these optimizations keep more than $90\%$ of the total runtime on device and enable state-of-the-art simulations to complete within a week on modern supercomputers --- computations that would otherwise require nearly a year on equivalently scaled CPU systems \citep{frontiere2025GB}.

A crucial component of this architecture is the in situ analysis module, implemented through the \COSMOTOOLS framework (see Section~3.8 in \citetalias{frontiere2022simulating}) and extended to meet the demands of subgrid physics. In addition to its native friends-of-friends (FOF; \citealt{davis1985,klypin1983}) and spherical-overdensity (SO; \citealt{lacey1994}) halo finders, \COSMOTOOLS now supports galaxy identification and measurement.

Galaxies are identified using a GPU-accelerated implementation of the density-based spatial clustering algorithm (DBSCAN; \citealt{ester1996}), applied to all stellar particles on each rank. DBSCAN generalizes FOF by introducing a minimum neighbor count parameter, $\texttt{minPts}$, where FOF corresponds to the special case $\texttt{minPts} = 1$. We identify galaxies as DBSCAN clusters using a fixed proper linking length of $50$~pkpc and $\texttt{minPts} = 10$. This compact search criterion provides a robust measure of the stellar galactic extent, enabling reliable identification of the stellar potential minimum.

Galaxy properties are then measured using an aperture cutout centered on this position, including all particle species within a fixed radius of $r_{\mathrm{cutout}} = 50$~pkpc. This choice is motivated by \citet{deGraaf2022}, who demonstrated using the \EAGLE simulation that a $50$~pkpc spherical aperture yields stable integrated galaxy measurements across resolution and feedback variations. The same aperture size has since been adopted by more recent simulations such as \FLAMINGO \citep{schaye2023flamingo} and \COLIBRE \citep{schaye2025}.

All clustering operations are executed entirely on GPUs using the open-source ArborX library~\citep{arborx2020,prokopenko2024}, enabling structure identification at every PM step with negligible computational overhead. The GPU-accelerated galaxy finder, in particular, supports rapid AGN seeding and efficient galaxy--halo membership tracking, as discussed in Section~\ref{subsec:bhseed}.

With the full suite of subgrid and analysis components in place, the remaining task is to determine the numerical values of the calibration parameters --- specifically, the galactic wind variables $\kappa_{\rm w}$ and $e_{\rm w}$ from Section~\ref{subsec:wind}, and the AGN model quantities $M_{\rm seed}$, $\epsilon_{\rm kin}$, and $v_{\rm jet}$ described in Section~\ref{sec:bhmodel}. The calibration of these parameters to observational targets is presented in the following section.

\section{Model Calibration}\label{subsec:calib}

In gravity-only simulations, the primary model parameters are determined entirely by the chosen cosmology. Hydrodynamic simulations, however, introduce additional complexity through subgrid models that account for a wide range of unresolved baryonic processes. These subgrid models include their own parameters, which must be calibrated against observations to produce realistic predictions.

Model calibration can be performed by running suites of simulations in which subgrid parameters are systematically varied, followed by emulation of the resulting observables to identify the regions of parameter space that best match specific observational targets \citep[e.g.,][]{bower2010, jo2023, kugel2023, chaikin2025b}. This approach enables both the constraint of model parameters and the quantification of degeneracies and uncertainties inherent in subgrid assumptions.

While, in principle, all subgrid parameters could be treated as free, many are physically motivated or theoretically constrained. For example, quantities such as the star formation timescale or metal yields may be based on empirical or theoretical priors and held fixed. In contrast, parameters governing feedback processes --- particularly those controlling energy or momentum injection --- often encode complex, unresolved physics and require empirical calibration to reproduce key observables such as the galaxy stellar mass function, halo gas fractions, or cluster properties.

A further complication for large-volume simulations is that coarsened subgrid models may exhibit explicit dependence on mass resolution, as is the case in our \CRKHACC implementation. As such, the calibration procedure must be repeated as a function of resolution to maintain consistency across simulation campaigns, similar to the efforts described in \cite{kugel2023}.

The full details of our calibration campaign are presented in \RAMACHANDRA. Here, we briefly summarize the approach and identify the primary observational targets used to constrain our fiducial model. 

\subsection{Calibration Parameters}

In Section~\ref{sec:solvers}, we described the primary subgrid models implemented in \CRKHACC. For calibration, we vary five physically motivated parameters that govern the strength and character of galactic winds and AGN feedback: the wind velocity scaling $\kappa_{\rm w}$, the supernova energy fraction available for wind injection $e_{\rm w}$, the black hole seed mass $M_{\rm seed}$, the AGN kinetic feedback efficiency $\epsilon_{\rm kin}$, and the jet velocity $v_{\rm jet}$.

As discussed in Section~\ref{subsec:bhaccretion}, the black hole seed mass is largely degenerate with the accretion boost factor $\alpha$ at our examined mass resolution. Rather than varying both simultaneously, we fix $\alpha = 1$ and calibrate $M_{\rm seed}$ to regulate AGN growth and thermal feedback efficiency at our target resolution.

The wind parameters $e_{\rm w}$ and $\kappa_{\rm w}$ directly set the mass loading and velocity of galactic outflows, thereby modulating stellar mass assembly and the star formation rate. Likewise, $\epsilon_{\rm kin}$ and $v_{\rm jet}$ determine the onset and effectiveness of AGN-driven kinetic feedback, with important consequences for gas content in massive halos.

Calibration parameters were sampled using a symmetric Latin hypercube design of 64 simulations in an $L_\mathrm{box} = 128~h^{-1}\mathrm{Mpc}$ box, complemented by a nested suite of 16 larger-volume runs ($L_\mathrm{box} = 256~h^{-1}\mathrm{Mpc}$) to constrain cluster-scale gas profiles. The full parameter ranges and additional calibration results are presented in \RAMACHANDRA, and the resulting calibrated values adopted in our fiducial model are listed in Table~\ref{tab:calib_params}.

\begin{table}[h]
\caption{Calibrated subgrid model parameters.}
\label{tab:calib_params}
\noindent
\begin{minipage}{\linewidth}
\footnotesize
\begin{tabularx}{\linewidth}{@{}lXc@{}}
\toprule
\textbf{Parameter} & \textbf{Description} & \textbf{Value} \\
\midrule
$\kappa_{\rm w}$     & Wind velocity scaling factor                         & 3.0 \\
$e_{\rm w}$          & Wind energy factor                                   & 0.5 \\
$\epsilon_{\rm kin}$ & AGN kinetic efficiency                       & 1.3 \\
$v_{\rm jet}$        & AGN jet velocity [km/s]                              & 5.1 \\
$M_{\rm seed}$       & AGN seed mass [$M_\odot~h^{-1}$]             & $8 \times 10^5$ \\
\bottomrule
\end{tabularx}
\end{minipage}
\end{table}

\subsection{Calibrated Observables}

\begin{figure*}[htp]
    \centering
    \includegraphics[width=\linewidth]{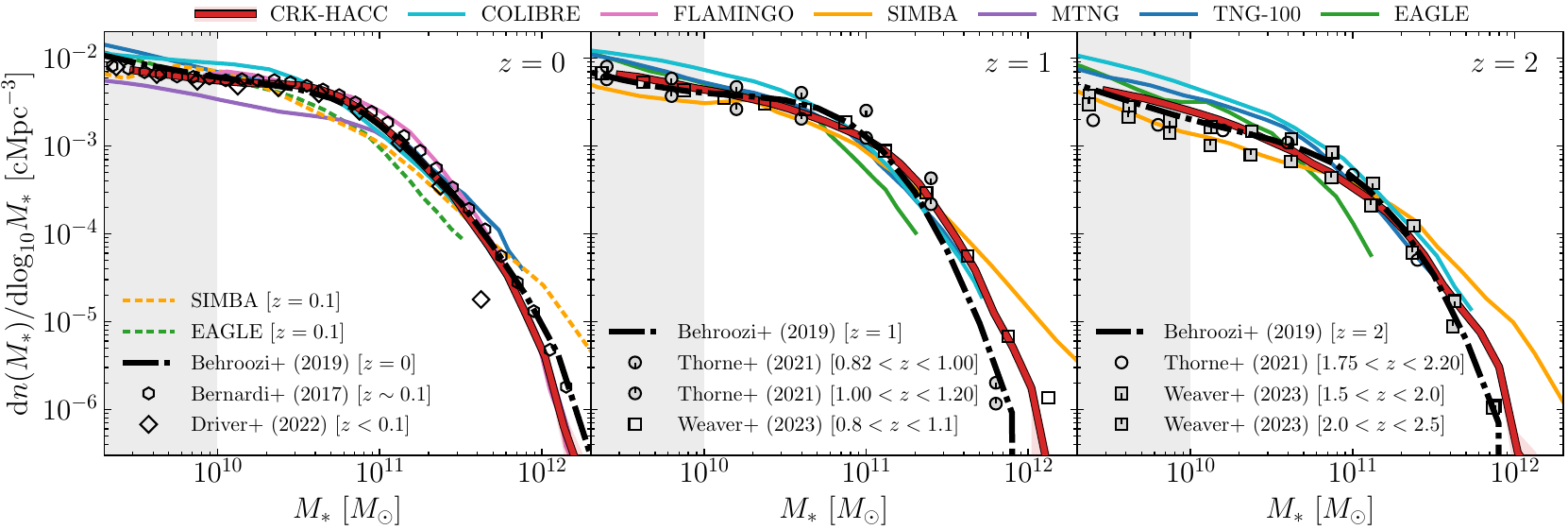}
    \caption{Galaxy stellar mass function (GSMF) measurements from the calibrated \CRKHACC\ simulation at redshifts $z = 0$, $1$, and $2$, with red shaded regions indicating Poisson uncertainties. Gray shaded regions mark the resolution limit of the \CRKHACC\ results. Observational datasets \citep{bernardi2017,driver2022,thorne2021,weaver2023} and empirical predictions from \UM\ (\citealt{behroozi2019}; the calibration target) are included for reference. Results from selected cosmological hydrodynamic simulations --- \EAGLE \citep{furlong2015}, \SIMBA\citep{dave2019}, \COLIBRE\citep{chaikin2025}, \TNG\citep{pillepich2018b}, \FLAMINGO\citep{schaye2023flamingo}, and \MTNG \citep{pakmor2023} --- are also shown for qualitative comparison. The \CRKHACC\ GSMFs closely match both the observational constraints and the \UM\ predictions across all presented redshifts, including at high redshift where the spread among different simulation predictions becomes more pronounced. \\}
    \label{fig:gsmf}
\end{figure*}

To constrain our fiducial subgrid model, we calibrate against two categories of observational measurements: galaxy stellar mass functions (GSMFs) and intracluster gas density profiles. The GSMF provides a sensitive probe of the efficiency of star formation and stellar feedback across cosmic time. We use measurements of the GSMF over $0 < z < 2$ to calibrate the galactic wind parameters $\kappa_{\rm w}$ and $e_{\rm w}$, as well as thermal AGN feedback through the seed mass $M_{\rm seed}$. The results presented here are based on an $L_\mathrm{box} = 256~h^{-1}\,\mathrm{Mpc}$ calibration run using the fiducial parameter set summarized in Table~\ref{tab:calib_params}. 

The simulation is initialized at \mbox{$z=200$} with $1024^3$ dark matter particles and an equal number of baryonic gas particles. These correspond to particle masses of $m_\mathrm{dm} = \unit{1.13\times10^9}{\massh}$ and $m_\mathrm{b} = \unit{2.12\times10^8}{\massh}$, matching the mass resolution calibrated in \RAMACHANDRA. The gravitational softening length is set to the minimum of $10$\,\kpch\,(comoving) and $6$\,\kpch\,(proper), such that the softening follows the comoving scale at high redshift and transitions to a fixed proper value at late times. The adopted cosmology follows \citet{Planck2018}, with parameters ($\Omega_c$, $\Omega_b$, $\sigma_8$, $n_s$, $h$) = ($0.26067$, $0.04897$, $0.8102$, $0.9665$, $0.6766$). This setup represents a down-scaled realization of the \FRONTIERE simulation \citep{frontiere2025GB}, employing an identical parameter configuration.

\Cref{fig:gsmf} shows the GSMF measured from the calibrated simulation at redshifts $z = 0$, $1$, and $2$, using stellar masses within a 50~pkpc aperture. Gray shaded regions denote stellar masses below $M_* \lesssim 10^{10}\,M_\odot$, corresponding to $\sim30$ star particles, where the mass function becomes resolution limited. At low redshift, we compare with measurements from the SDSS and GAMA surveys, reported by \citet{bernardi2017} and \citet{driver2022}, respectively. At intermediate and high redshift, we include GSMFs from the DEVILS survey \citep{thorne2021} and the COSMOS2020 catalog \citep{weaver2023}. The observational datasets correspond to survey redshift bins that bracket each simulation snapshot, ensuring consistent coverage around the targeted epochs. Across all redshifts, we also show the empirical predictions from the \UM\ model \citep{behroozi2019}, which self-consistently reproduces the observed stellar mass functions, star formation rates, and quenched fractions over $0 < z < 10$. These predictions provide continuous redshift coverage and serve as a reference calibration target for the evolving GSMF.

At all three redshifts, the \CRKHACC results track closely with the \UM\ calibration target and, by construction, are consistent with the presented observational datasets. For qualitative comparison, we also include GSMF measurements from several recent cosmological hydrodynamic simulations, including \EAGLE\ \citep{furlong2015}, \SIMBA\ \citep{dave2019}, \COLIBRE\ \citep{chaikin2025}, \TNG\ \citep{pillepich2018b}, \FLAMINGO\ \citep{schaye2023flamingo}, and \MTNG\ \citep{pakmor2023}. These provide representative examples of current large-scale galaxy formation models and help contextualize the calibration of our fiducial subgrid parameters. As expected, the scatter among simulations is modest at $z = 0$, where most models are calibrated to local observables, and increases toward higher redshift as differences in feedback implementation and star formation efficiency become more pronounced.

\begin{figure}
    \centering
    \includegraphics[width=\linewidth]{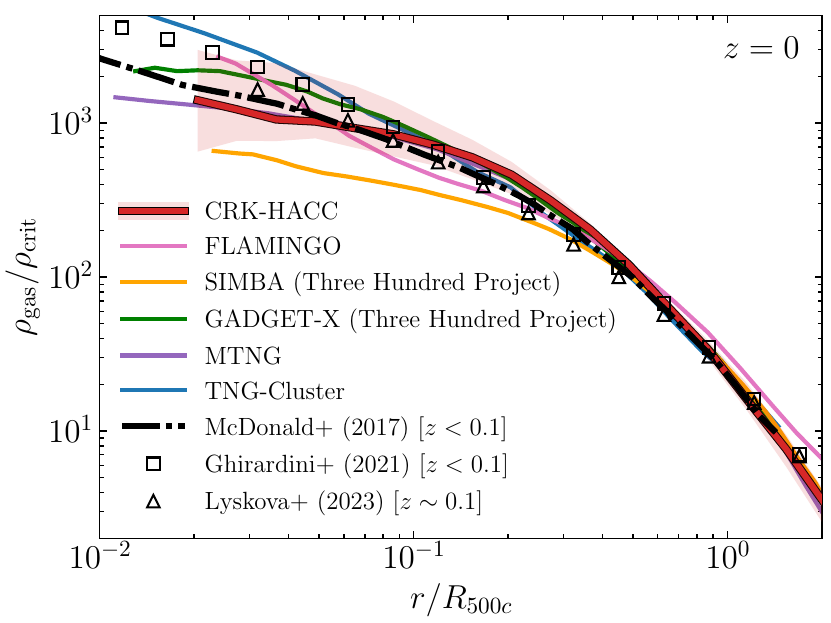}
    \caption{Median cluster gas density profiles from the fiducial \CRKHACC\ simulation at $z\!= \!0$ for halos with $M_{500c} > 3\times10^{14}\,M_\odot$. The shaded region shows the 16th-84th percentile range. Observational measurements are shown from \citet{mcdonald2017} (calibration target), \citet{ghirardini2021}, and \citet{lyskova2023}. Results from recent cosmological simulations include \TNGCLUSTER\ \citep{lehle2024}, \MTNG\ \citep{pakmor2023}, \FLAMINGO\ \citep{braspenning2024}, and the \SIMBA\ and \GX\ clusters from \emph{The Three Hundred} project \citep{li2023}. The calibrated \CRKHACC profile lies near the midpoint of both the observational and simulation ranges, indicating consistency with current constraints on intracluster gas structure.}
    \label{fig:rhogasprofile}
\end{figure}

To constrain the AGN kinetic feedback parameters $\epsilon_{\rm kin}$ and $v_{\rm jet}$, we calibrate against low-redshift X-ray measurements of intracluster gas density profiles. Our primary calibration target is the stacked profile of \citet{mcdonald2017}, derived from deep \textit{Chandra} observations of 27 massive, X-ray--selected clusters at $z \lesssim 0.1$ (originally from the \citealt{vikhlinin2009} sample). As shown in \Cref{fig:rhogasprofile}, the stacked \CRKHACC profiles are consistent with the target measurement after applying a mass cut of $M_{500c} > 3 \times 10^{14}\,M_\odot$, which yields 37 clusters, comparable to the 27-cluster sample analyzed by \citet{mcdonald2017}.

We also include the nearby X-COP reference from \citet{ghirardini2019,ghirardini2021}, drawn from a 12-cluster Planck-selected sample, and the \textit{eROSITA} all-sky stack of 38 Planck-selected clusters from \citet{lyskova2023}, based on the CHEX-MATE sample \citep{chexmate2021}. The \citet{mcdonald2017} and \citet{lyskova2023} profiles show consistent normalization with \CRKHACC, while the X-COP profiles are systematically higher, collectively spanning the plausible range of gas densities in low-redshift clusters.

For qualitative context, we again include results from several recent cosmological simulations, noting that the exact mass selections and measurement methodologies differ between studies. These comprise \TNGCLUSTER \citep{lehle2024}, \MTNG \citep{pakmor2023}, \FLAMINGO \citep{braspenning2024}, and the \SIMBA and \GX clusters from \emph{The Three Hundred Project} \citep{li2023}. The \CRKHACC profiles lie near the midpoint of the simulation spread and are nearly coincident with the selected \MTNG curve, which corresponds to clusters in the $10^{14.6} < M_{500c}/M_\odot < 10^{14.8}$ range. 

As an additional validation, we simulate the nIFTy high-mass cluster from \citet{sembolini2016nifty,sembolini2016niftyradiative} in Appendix~\ref{sec:nifty}, benchmarking the calibrated \CRKHACC model against published radiative cluster simulations. The comparison likewise shows consistency across modern hydrodynamic solvers, with \CRKHACC agreeing with other codes on the cluster profile measurements considered.

Finally, while we calibrate to high-mass cluster gas-density profiles, several large-volume simulations instead target gas fractions in group-scale and low-mass clusters \citep[e.g.,][]{mccarthy2016,schaye2023flamingo,kugel2023}. In \RAMACHANDRA, we tested both observables but found they could not be simultaneously matched within a single subgrid parameter set. Gas-fraction calibration required significantly stronger feedback, whereas the gas-density approach adopted here achieves a better balance with observed low-redshift, high-mass cluster structure.

This calibration choice is particularly well-suited to next-generation, exascale-class simulations, where reproducing realistic massive-cluster profiles is a key objective given the large statistical samples available for study. The \FRONTIERE\ flagship simulation, for example, resolves more than 500{,}000 clusters with $M_{500c} > 10^{14}\massh$, enabling robust statistical population analyses \citep{frontiere2025GB}. The alternative \CRKHACC\ parameter set calibrated to gas fractions is described in \RAMACHANDRA\ and will be explored in future targeted studies.

In summary, the GSMF and halo profile calibrations jointly constrain our feedback model from galactic to cluster scales, ensuring consistency with the selected observational benchmarks and yielding physically reasonable galaxies and high-mass halos within the calibrated regimes. In the following section, we examine how the fiducial model performs for a broader suite of non-calibrated observables, spanning galaxy scaling relations, baryon fractions, and metallicity trends, to evaluate its predictive fidelity and any residual tensions beyond the direct calibration targets.

\section{Cosmological Simulation Measurements}\label{sec:results}

\begin{figure*}[ht!]
    \centering
    \includegraphics[width=\textwidth]{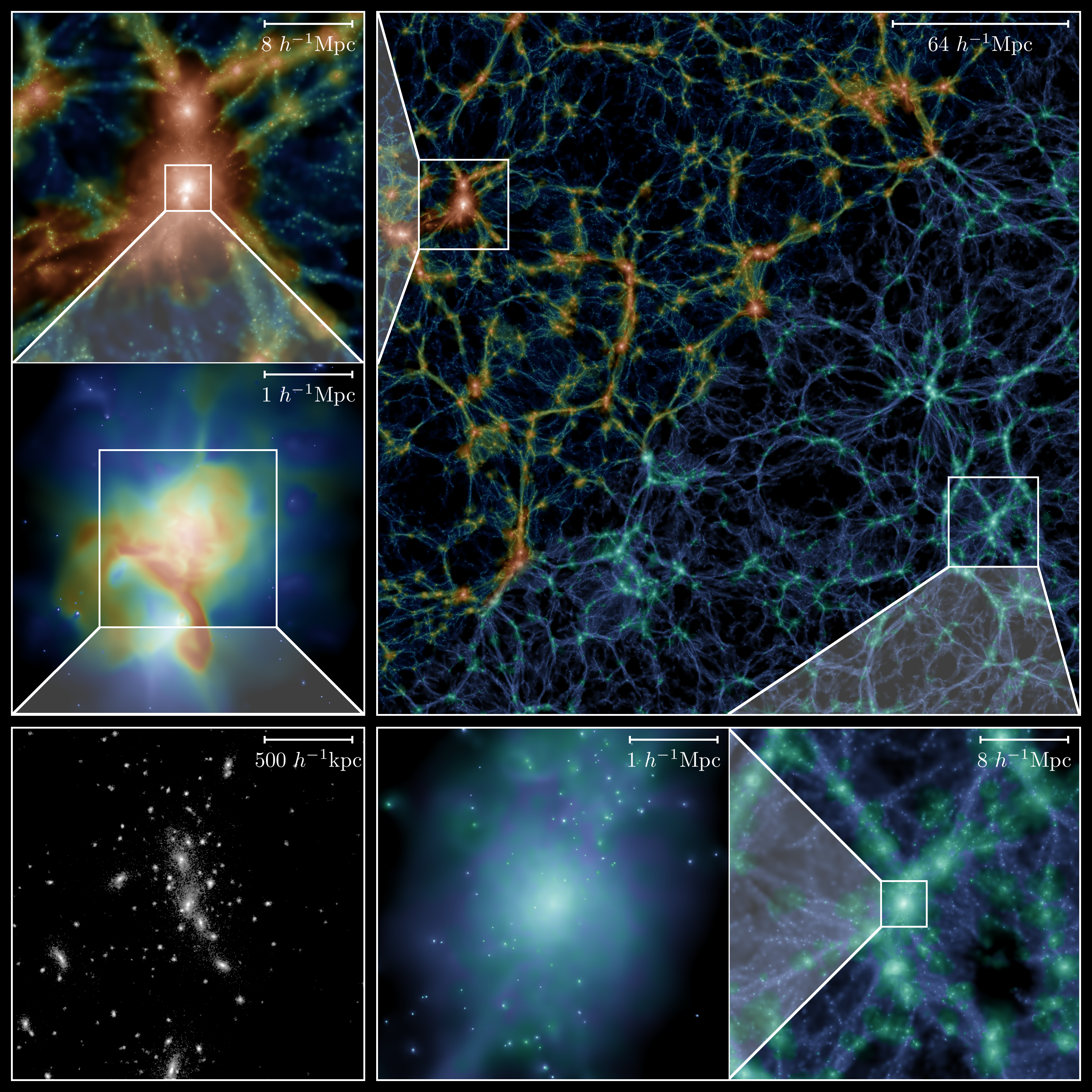}
    \caption{
    The large central panel shows a slice of the gas density field across the full simulation domain ($256\,h^{-1}\mathrm{Mpc}$ on a side) with a slice depth of $4\,h^{-1}\mathrm{Mpc}$. The left half of this panel is color-mapped by temperature (cold in blue, hot in red), while the right half is color-mapped by metallicity (metal-poor in blue, metal-rich in green). The panels along the left column present successive zooms into the most massive halo in the volume ($M_{200c} = 1.2\times 10^{15}\,h^{-1}M_\odot$), and the panels along the bottom row show progressive zooms into a second massive cluster ($M_{200c} = 1.4\times 10^{14}\,h^{-1}M_\odot$). The lower-left panel shows the stellar density field in the central region of the largest halo. \\
    }
    \label{fig:sim}
\end{figure*}

We now evaluate the predictive performance of the fiducial \CRKHACC\ model against a broad set of observational benchmarks using the calibrated $L_\mathrm{box}=256\,h^{-1}\,\mathrm{Mpc}$ simulation introduced in Section~\ref{subsec:calib}. With its modest volume, this run enables a converged study of galaxy- and halo-scale properties and provides sufficient statistical power to assess how the calibrated feedback and enrichment prescriptions shape galaxy and halo populations across mass and environment. Synthetic sky-projected observables, light-cone measurements, and massive-cluster analyses are deferred to forthcoming studies based on our leadership-class simulations (e.g., \FRONTIERE; \citealt{frontiere2025GB}), which possess the necessary volume to robustly render these observables.

We begin by examining global stellar evolution and mass buildup across cosmic time (Sections~\ref{subsec:CSFRD}--\ref{subsec:CSMD}), then turn to galaxy-scale properties --- sizes, star-formation activity and quenching, and scaling relations in stellar mass, halo mass, and metallicity (Sections~\ref{subsec:Gsize}--\ref{subsec:MZR}). We next assess black-hole growth and its co-evolution with host galaxy mass (Section~\ref{subsec:bhsm}), and conclude with baryon fraction measurements in massive halos (Section~\ref{subsec:gasfrac}) to evaluate model performance across environments.

A visual overview of the simulation domain is shown in Figure~\ref{fig:sim}, illustrating the large-scale distributions of gas temperature, metallicity, and density, together with the stellar component produced by the calibrated model. The accompanying analyses draw on a suite of derived physical quantities and structure catalogs generated in situ and subsequently post-processed using our parallel, optimized Python framework \HAVOCC (the \HACC Analysis and Validation to Observational Constraints Code),\footnote{\url{https://git.cels.anl.gov/hacc/HAvoCC}} which performs distributed measurements of halo, galaxy, and environmental properties directly from particle-data outputs.

Throughout this study, we compare results from \CRKHACC with both large-volume simulations, such as \MTNG (\MTNGNUM;  \citealt{pakmor2023}) and \FLAMINGO (L2p8\_m9; \citealt{schaye2023flamingo}), and higher-resolution reference suites, including \EAGLE (Ref-L100N1504; \citealt{schaye2015eagle}), \TNG (\TNGONE and \TNGTHREE; \citealt{nelson2019}), \SIMBA (m100n1024; \citealt{dave2019}), and \COLIBRE (L200m6; \citealt{schaye2025}).  
Key numerical parameters for these simulations --- box size, particle counts, mass resolution, and gravitational softening --- are summarized in Table~\ref{tab:sims}.

\setlength{\tabcolsep}{8.5pt}
\begin{deluxetable*}{@{\extracolsep{\fill}}lcccccc}

    \tablecaption{Summary of comparison simulations referenced in this study. Listed are the comoving box length $L_\mathrm{box}$, number of gas and dark matter particles ($N_\mathrm{gas,DM}$), baryon and dark matter particle masses ($m_\mathrm{b}$, $m_\mathrm{dm}$), gravitational softening length $\epsilon_{\rm soft}$, and the primary reference.\label{tab:sims}}
    
    \tablehead{
        \textbf{Name} & 
        \colhead{$L_\mathrm{box}$ [Mpc]} & 
        \colhead{$N_\mathrm{gas,CDM}$} & 
        \colhead{$m_\mathrm{b}$ [$M_\odot$]} & 
        \colhead{$m_\mathrm{dm}$ [$M_\odot$]} & 
        \colhead{$\epsilon_{\rm soft}$ [pkpc]} & 
        \colhead{Reference}
     }
    \startdata
        \CRKHACC\tablenotemark{a} & 378.4 & $1024^3$ & $3.1\times10^8$ & $1.7\times10^9$ & 8.9 & this work \\
        \FLAMINGO\ (L2p8\_m9) & 2800 & $5040^3$ & $1.1\times10^9$ & $5.7\times10^9$ & 5.7 & \citet{schaye2023flamingo} \\
        \MTNG\  & 740 & $4320^3$ & $3.1\times10^7$ & $1.7\times10^8$ & 3.7 & \citet{pakmor2023} \\
        \TNGTHREE\ & 302.6 & $2500^3$ & $1.1\times10^7$ & $5.9\times10^7$ & 1.5 & \citet{nelson2019} \\
        \SIMBA\ (m100n1024) & 147.1 & $1024^3$ & $1.8\times10^7$ & $9.6\times10^7$ & 0.7 & \citet{dave2019} \\
        \EAGLE\ (Ref-L100N1504) & 100 & $1504^3$ & $1.8\times10^6$ & $9.7\times10^6$ & 0.7 & \citet{schaye2015eagle} \\
        \TNGONE\  & 110.7 & $1820^3$ & $1.4\times10^6$ & $7.5\times10^6$ & 0.7 & \citet{nelson2019} \\
        \COLIBRE\ (L200m6) & 200 & $3008^3\,(4\times3008^3)$\tablenotemark{b} & $1.8\times10^6$ & $2.4\times10^6$ & 0.7 & \citet{schaye2025} \\
    \enddata
    {\tiny
    \tablenotetext{a}{Down-scaled calibration run of the 4655.6 Mpc \FRONTIERE\ simulation \citep{frontiere2025GB}.}
    \vspace{-4pt} 
    \tablenotetext{b}{\COLIBRE\ simulates four times more dark matter than gas particles to maintain nearly uniform particle masses.}
    }
\end{deluxetable*}

When comparing to observational datasets, all stellar quantities have been converted to a \cite{chabrier2003} initial mass function (IMF) for consistency. For all presented \CRKHACC\ results involving galaxy stellar mass, gray shaded regions denote $M_* \lesssim 10^{10}\,M_\odot$ (approximately 30 star particles), indicating the mass resolution limit as in the calibration figures.

These comparisons are used to identify where the coarsened model predictions are consistent with established simulation results despite differences in resolution and calibration, and where reduced sampling and force resolution begin to limit physical fidelity. In doing so, we aim both to demonstrate the physical realism of stellar and baryonic buildup within the survey-scale regime targeted by \CRKHACC and to delineate where resolution limitations impact predictions --- for example, in high-redshift star formation or the internal structure of individual galaxies. Naturally, the reference simulations themselves exhibit non-negligible differences across several regimes, as well as inconsistencies with observations, so agreement or tension among models should be interpreted as an indicator of relative rather than absolute physical fidelity.

\subsection{Cosmic Star Formation Rate Density}\label{subsec:CSFRD}

We begin by examining the cosmic star formation rate density (CSFRD), a well-studied global diagnostic of galaxy formation models that traces the buildup of stellar mass across cosmic time. Observationally, the CSFRD rises rapidly from the early universe, peaks near $z \sim 2$, and declines by nearly an order of magnitude to the present day. In \CRKHACC, we measure the CSFRD from the instantaneous star formation rates of gas particles at each PM timestep.\footnote{Integrating the stellar mass formed over each step, using the initial masses of newly created star particles, yields a consistent CSFRD measurement.}

\begin{figure}
    \centering
    \includegraphics[width=\linewidth]{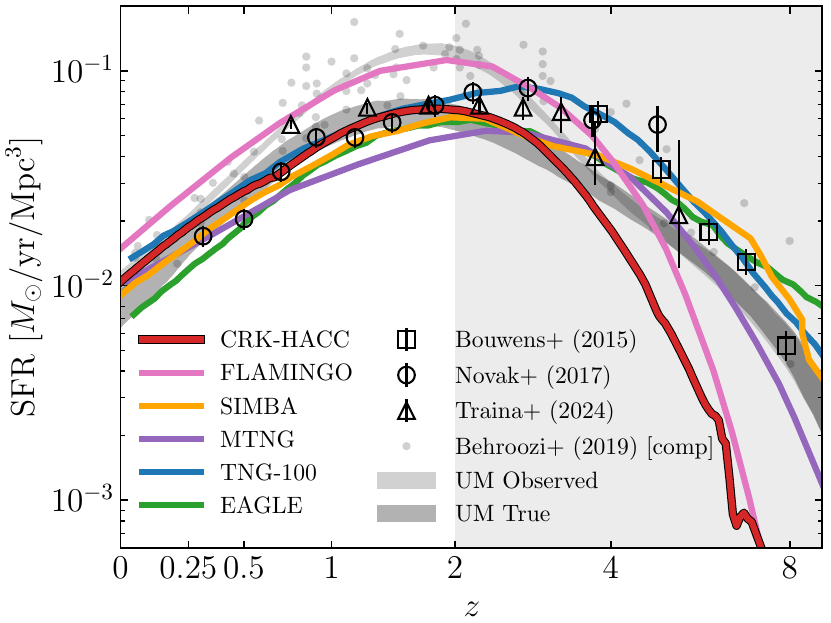}
    \caption{Cosmic star formation rate density (CSFRD) as a function of redshift. Results from the fiducial \CRKHACC simulation are compared with observational estimates from \citet{bouwens2015} (UV), \citet{novak2017} (radio), and \citet{traina2024} (submillimeter), and with predictions from hydrodynamical simulations (\EAGLE; \citealt{furlong2015}, \TNG and \MTNG; \citealt{pakmor2023}, \SIMBA; \citealt{dave2019}, and \FLAMINGO; \citealt{schaye2023flamingo}). Also shown are the results of \citet{behroozi2019}, including their compiled pre-2016 multi-wavelength dataset and empirical \UM model curves, which distinguish ``observed'' (bias-affected) and ``true'' (bias-corrected) CSFRD histories. Over the resolved redshift range $0 \!\leq \!z\! \lesssim \!3$, \CRKHACC is consistent with other simulations and closely follows the \UM ``true'' prediction as well as recent observations. At higher redshifts, both \CRKHACC and \FLAMINGO underpredict the observational estimates and higher-resolution simulations, as expected from the reduced contribution of low-mass halos in coarser-resolution runs. For reference, the region at $z > 2$ is shaded to indicate epochs beyond the redshift range calibrated by the GSMF (see Section~\ref{subsec:calib}).} 
    \label{fig:csfr}
\end{figure}

Figure~\ref{fig:csfr} compares the CSFRD from the fiducial \CRKHACC simulation with recent observational constraints \citep{bouwens2015, novak2017, traina2024} and with predictions from representative hydrodynamical simulations (\EAGLE; \citealt{furlong2015}, \TNG and \MTNG; both from \citealt{pakmor2023}, \SIMBA; \citealt{dave2019}, and \FLAMINGO; \citealt{schaye2023flamingo}). We also include the data compilation of \citet{behroozi2019}, showing both their aggregated pre-2016 multiwavelength CSFRD measurements (listed in their Table~C3) and the associated empirical \UM model curves that self-consistently link galaxy star formation to halo growth. The observational datasets span complementary tracers, including rest-frame UV at high redshift \citep{bouwens2015}, dust-unbiased radio estimates \citep{novak2017}, and infrared/submillimeter measurements \citep{traina2024}, while the \UM framework provides a synthesis of earlier surveys and a bias-corrected baseline for comparison.

At high redshift ($z \gtrsim 3$), \CRKHACC predicts a lower CSFRD than both the UV-based estimates of \citet{bouwens2015} and the higher-resolution hydrodynamical simulations. This difference reflects the relatively coarse mass resolution of the \CRKHACC results --- set to match the configuration of survey-scale production runs --- compared to simulations such as \TNG, \MTNG, \EAGLE, and \SIMBA, which better resolve the population of faint, low-mass halos that dominate early star formation (see Table~\ref{tab:sims}). By contrast, \FLAMINGO employs a comparable resolution and likewise shows a suppression at high redshift, though with a somewhat steeper recovery toward intermediate epochs.

Between $z = 3$ and $z = 0$, \CRKHACC converges with the other hydrodynamical simulations, generally tracking the middle of their spread. Relative to the observational data, the \CRKHACC CSFRD agrees well with the measurements of \citet{novak2017} and \citet{traina2024} across most of the resolved redshift interval. Divergences appear primarily at the highest measured redshifts, where observational uncertainties increase and the \CRKHACC estimate falls below the data due to limited resolution. In contrast, the simulation remains below the \citet{behroozi2019} compilation and the ``observed'' CSFRD predicted by \UM, which reproduces the higher normalization of those datasets (light-gray shaded curve and points in Figure~\ref{fig:csfr}).

The discrepancy between star-formation rate indicators and the growth of the stellar mass density is a long-recognized problem (e.g., \citealt{wilkins2008, leja2015, yu2016}), and it provides the motivation for the bias framework adopted in \UM. In this model, redshift-dependent corrections are applied, with the strongest adjustments at $z \sim 2$, the epoch of peak star-formation activity, where the tension is most severe, and modest reductions at higher redshift. This yields a lower, bias-corrected ``true'' CSFRD estimate that falls below the observational compilations near the peak epoch (dark-gray shaded curve).

Interestingly, \CRKHACC\ follows the ``true'' \UM\ prediction very closely, matching both shape and normalization from $z \sim 4$ to $z = 0$. This consistency is notable, given that the CSFRD was not part of the calibration (Section~\ref{subsec:calib}), which relied solely on the galaxy stellar mass function (spanning $0 < z < 2$) and low-redshift cluster density profiles. While the agreement may be partly coincidental, it suggests that the calibrated model reproduces the global efficiency of star formation in a manner consistent with empirically bias-corrected expectations.

\subsection{Cosmic Stellar Mass Density}\label{subsec:CSMD}

\begin{figure}
    \centering
    \includegraphics[width=\linewidth]{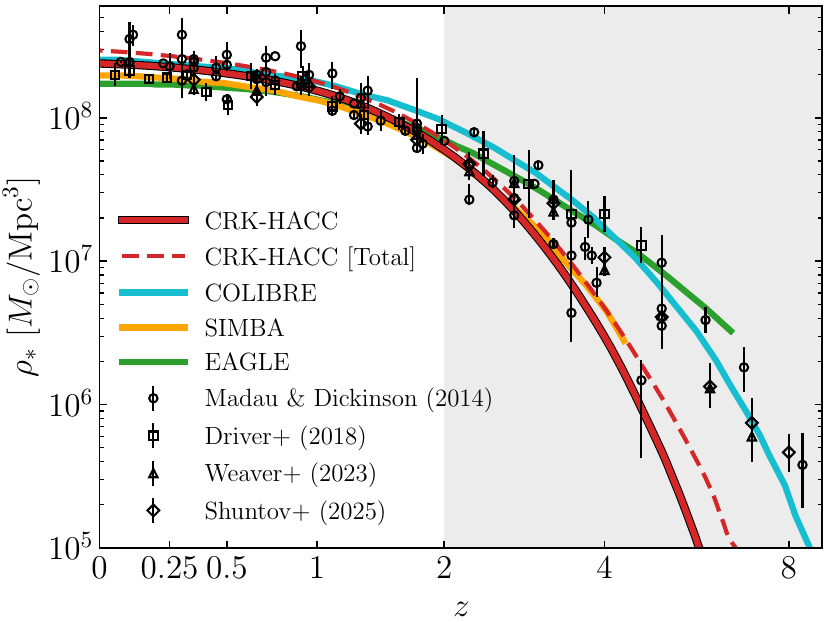}
    \caption{Cosmic stellar mass density (CSMD) as a function of redshift. Results from the fiducial \CRKHACC simulation are compared with observational estimates from \citet{madau2014}, \citet{driver2018}, \citet{weaver2023}, and \citet{shuntov2025}, and with predictions from the \COLIBRE \citep{chaikin2025}, \EAGLE \citep{furlong2015}, and \SIMBA \citep{thomas2019} simulations. The solid \CRKHACC curve shows the stellar mass density from stars residing in galaxies, while the dashed curve labeled [Total] includes all stars in the simulation, capturing both galaxy and intra-halo light contributions. Comparisons among simulations are based on galaxy-only measurements, adopting fixed apertures of $50$~pkpc for \CRKHACC and \COLIBRE, $30$~pkpc for \EAGLE, and all bound stars for \SIMBA. The \CRKHACC results closely match the observed stellar mass density over the resolved redshift range ($0 \!\le \!z \!\lesssim \!3$), consistent with the level of agreement seen for the CSFRD in Figure~\ref{fig:csfr}, beyond which resolution effects begin to suppress early stellar mass formation. For reference, the region at $z > 2$ is shaded to indicate epochs beyond the redshift range calibrated by the GSMF (see Section~\ref{subsec:calib}).}
    \label{fig:sden}
\end{figure}
 
A complementary observable to the cosmic star formation rate density is the buildup of the cosmic stellar mass density (CSMD), which provides a cumulative record of how efficiently baryons are converted into long-lived stars. Whereas the CSFRD measures the instantaneous rate of star formation, the CSMD integrates this history over cosmic time (accounting for stellar mass loss) and tests whether a simulation reproduces the observed stellar mass assembly.

Figure~\ref{fig:sden} compares the evolution of the CSMD from the fiducial \CRKHACC run (computed for stars in galaxies and for the total simulated volume) against recent observational estimates and predictions from three other hydrodynamical simulations. The observational datasets span a broad redshift range and include the compilation of \citet{madau2014}, the homogeneous multi-survey analysis of \citet{driver2018}, the COSMOS2020 stellar mass functions of \citet{weaver2023}, and the recent JWST-based measurements of \citet{shuntov2025}. We also include predictions from the \COLIBRE\ \citep{chaikin2025}, \EAGLE\ \citep{furlong2015}, and \SIMBA\ \citep{thomas2019} simulations, which serve as representative benchmarks for galaxy formation models calibrated to different observational constraints.

In all cases, the stellar mass density reported by the simulations (solid lines) is measured from stars residing in galaxies. This approach mirrors that of observational estimates, which derive stellar mass densities from galaxy stellar mass functions and therefore exclude intra-halo light (IHL) contributions. Following our analysis pipeline (Section~\ref{subsec:cosmotools}), \CRKHACC measures galaxy properties within fixed $50$~pkpc apertures, as does \COLIBRE. The \EAGLE\ results adopt $30$~pkpc apertures, while \SIMBA\ reports stellar masses including all bound star particles. For reference, the dashed \CRKHACC curve in Figure~\ref{fig:sden} includes all stars in the simulation (both galaxy and diffuse IHL components) and serves as an upper bound on the stellar mass locked in galaxies.

The \CRKHACC results demonstrate the expected buildup of stellar mass across cosmic time. At high redshift ($z \gtrsim 3$), the predicted stellar mass density gradually falls below the observational estimates, consistent with the suppressed CSFRD at early times shown in Figure~\ref{fig:csfr}. From $z \sim 2$--$3$ to the present, \CRKHACC\ follows the overall spread of observational datasets, remaining within their combined scatter across this interval. As emphasized by \citet{behroozi2019}, systematic uncertainties in stellar mass functions are notably smaller than those affecting SFR indicators and diminish further toward low redshift. The close agreement between \CRKHACC\ and the observed stellar mass density over the resolved redshift range is therefore reassuring.

Relative to other hydrodynamical simulations, \CRKHACC\ closely follows the shape and normalization of \COLIBRE\ at low redshift ($z < 1$), remaining systematically above the \EAGLE\ and \SIMBA\ predictions. Toward intermediate epochs ($1 \lesssim z \lesssim 2$), the \CRKHACC\ stellar mass density falls slightly below \COLIBRE, aligning more closely with \SIMBA. At higher redshift, it declines more steeply, crossing below \EAGLE\ and later \SIMBA, reflecting the increasing sensitivity of early stellar mass assembly to numerical resolution and feedback efficiency.

\subsection{Galaxy Sizes}\label{subsec:Gsize}

With the global buildup of stellar mass established, we next examine the internal structure of galaxies beginning with morphology. A central structural property of galaxies is their size, commonly quantified by the radius enclosing half the stellar mass, $R_{1/2}$. In simulations, galaxy sizes are particularly sensitive to numerical resolution and the adopted feedback prescriptions, which together shape the morphology and spatial distribution of stars. 

\begin{figure}
    \centering
    \includegraphics[width=\linewidth]{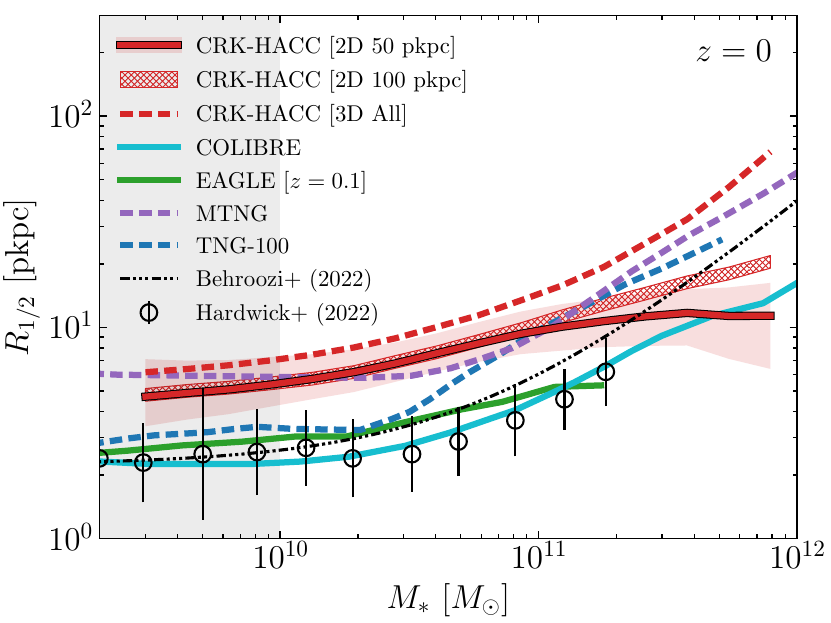}
    \caption{Stellar half-mass radius as a function of stellar mass at $z\!=\!0$. The \CRKHACC\ results are shown for three size definitions: projected radii within $50$~pkpc (solid), projected radii within $100$~pkpc (hatched), and three-dimensional radii from all DBSCAN-assigned stars (dashed). Lines indicate median relations, with the red shaded region denoting the 16th--84th percentile range for the default measurement. Gray shaded regions mark the regime where the \CRKHACC\ results are resolution limited. Observational constraints include the 3D empirical fit to SDSS ($z\! < \!0.2$) data from \citet{behroozi2022} and the 2D projected measurements from \citet{hardwick2022} for nearby xGASS galaxies ($z\! < \!0.05$). Comparison simulation results for 2D measurements (solid lines) are drawn from \EAGLE\ (\citealt{schaye2015eagle}; at $z\!=\!0.1$) and \COLIBRE\ \citep{schaye2025}, while 3D results (dashed) are from \TNG\ and \MTNG\ \citep{pakmor2023}. 2D observations and predictions are systematically lower than 3D owing to projection and aperture effects. High-resolution simulations (\EAGLE, \COLIBRE, \TNG) reproduce the observed size--mass relation, while lower-resolution runs (\MTNG, \CRKHACC) yield larger radii at low masses but converge toward the others at high masses.}
    \label{fig:galsize}
\end{figure}

Consequently, large-volume simulations with gravitational force resolutions approaching galactic scales ($\sim$ kpc), such as \CRKHACC, \MTNG, and \FLAMINGO\ (see Table~\ref{tab:sims}), are not expected to reproduce precise size--mass relations but can still provide meaningful qualitative benchmarks. Reported sizes are further influenced by measurement details such as projection effects and aperture cutoffs. In Figure~\ref{fig:galsize}, we present a set of \CRKHACC\ size measurements for comparison, spanning both 2D projected and 3D configurations.

The \CRKHACC curves illustrate three representative size definitions that together bracket the range of predicted galaxy radii. Our default measurement (solid line) adopts a 2D half-mass radius within a fixed $50$~pkpc aperture, where stellar mass is summed over the projected stellar distribution. This aperture choice causes the measured radii to saturate at high stellar masses. Expanding the projected region to $100$~pkpc leaves the low-mass sizes unchanged but increases radii at the massive end, where extended stellar envelopes become significant. Finally, we include a 3D half-mass radius computed from all DBSCAN-assigned stars (see Section~\ref{subsec:cosmotools}). This definition does not saturate at high mass and, as expected, yields systematically larger radii than the 2D measurements because projection compresses stars along the line of sight (see, e.g., Appendix C of \citealt{Furlong2016}).

To contextualize these measurements, we compare them to recent observational datasets. \citet{behroozi2022} provide an empirical fit (their Eq.~13) to SDSS measurements of galaxy sizes, converted to three-dimensional half-mass radii, which we show as a continuous curve. \citet{hardwick2022} present direct measurements of projected stellar half-mass radii, derived from stellar mass surface density profiles in deep imaging of xGASS galaxies, which we include as discrete points. Together, these datasets provide both a 3D empirical relation and its direct 2D observational counterpart, spanning a broad range of stellar masses at $z=0$. As shown in Figure~\ref{fig:galsize}, the datasets are mutually consistent, with the \citet{hardwick2022} points lying systematically lower, as expected from projection effects.

For theoretical context, we also compare with predictions from several hydrodynamical simulations, which span different mass resolutions. The \EAGLE\ \citep{schaye2015eagle} and \COLIBRE\ \citep{schaye2025} results are reported as projected half-mass radii within fixed apertures and were directly calibrated against the observed size--mass relation. We also include results from \TNG\ and \MTNG, taken from \citet{pakmor2023}, which present intrinsic three-dimensional half-mass radii measured from all stars bound to each subhalo. Owing to these differing measurement conventions, the comparison to \CRKHACC\ should be regarded as qualitative rather than quantitative.

As shown in Figure~\ref{fig:galsize}, the high-resolution simulations reproduce the observed size--mass relation well. The \EAGLE\ and \COLIBRE\ curves closely follow the \citet{hardwick2022} data points, as expected given that these simulations report projected half-mass radii and were calibrated to match the observed relation. The three-dimensional \TNG measurements are broadly consistent with the \citet{behroozi2022} fit, though they lie slightly above it at the highest stellar masses.

The lower-resolution \MTNG\ curve lies systematically above both the observational constraints and the higher-resolution simulations. This offset reflects the impact of finite force softening and mass resolution, which broaden stellar distributions and inflate measured radii. At high stellar masses, however, \MTNG\ converges toward the \TNG\ relation once galaxies become sufficiently resolved. A similar behavior is seen in other large-volume runs, such as \FLAMINGO, where both star-forming and quenched galaxies exhibit systematically larger radii at lower masses, as shown in Figure~14 of \citet{schaye2023flamingo}.

Our \CRKHACC results follow the same overall trend. At low stellar masses, the predicted radii are consistent with \MTNG, as expected given our coarser resolution. The three-dimensional half-mass definition provides the closest analog to the \MTNG measurement, though it may be slightly inflated because DBSCAN-defined galaxies can include additional intra-halo light. At high masses, the 2D \CRKHACC measurements begin to approach the smaller radii reported by \COLIBRE, while the 3D definition remains consistent with the \TNG\ and \MTNG\ curves. Together, these comparisons show how measurement choices bracket the predicted range of galaxy sizes and emphasize that \CRKHACC\ results are resolution-limited but aligned with expectations from other large-volume simulations.

\subsection{Galaxy Star Formation Activity and Quenching}\label{subsec:sfrq}

\begin{figure*}[htp]
    \centering
    \includegraphics[width=\linewidth]{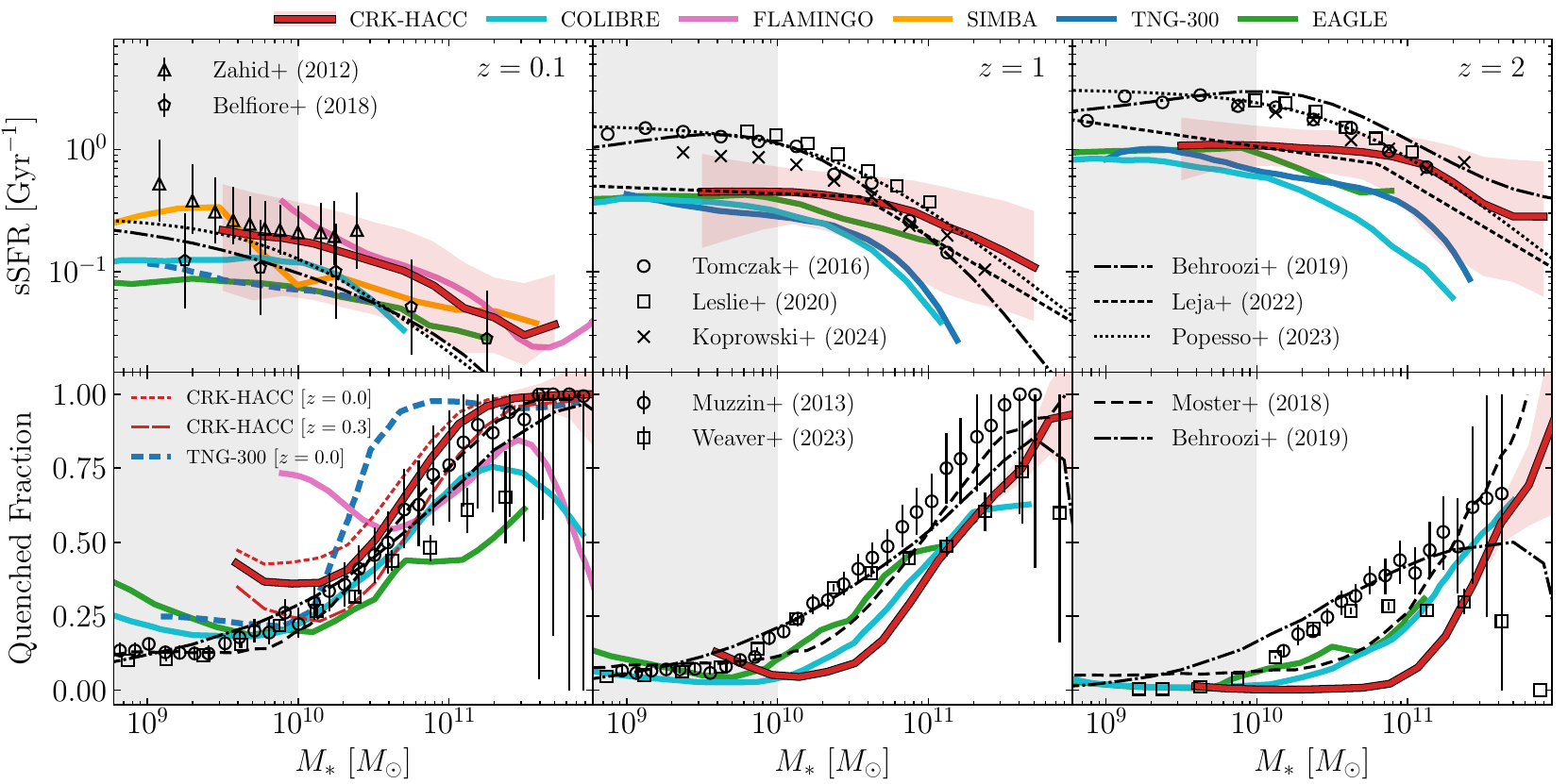} 
    \caption{\textbf{Top:} Median sSFR as a function of stellar mass for \CRKHACC galaxies at $z\!=\!0.1$, $z\!=\!1$, and $z\!=\!2$ (red shaded 16th--84th percentile range), compared with simulation results from \TNG (sSFR: \citealt{donnari2019}; quenched: \citealt{donnari2021b}), \EAGLE \citep{furlong2015}, \COLIBRE \citep{chaikin2025}, \SIMBA \citep{dave2019}, and \FLAMINGO \citep{schaye2023flamingo}. Observational datasets include optical spectroscopy \citep{zahid2012,belfiore2018}, NIR photometry \citep{tomczak2016}, radio stacking \citep{leslie2020}, and FIR stacking \citep{koprowski2024}, as well as empirical (\citealt{behroozi2019,popesso2023}) and SED-based (\citealt{leja2022}) fits. Data points correspond to \citet{zahid2012} ($z\! <\! 0.1$), \citet{belfiore2018} ($z\! \sim\! 0.028$), \citet{tomczak2016} ($0.75 \!< \!z \!< \!1$; $1.5\! <\! z \!< \!2$), \citet{leslie2020} ($0.8\! <\! z\! <\! 1.1$; $1.5\! <\! z \!< \!2.0$), and \citet{koprowski2024} ($0.75\! \leq\! z\! < \!1.0$;  $1.6\! \leq\! z\! < \!2.2$). \textbf{Bottom:} Quenched fractions for the full galaxy population in \CRKHACC compared to the same set of simulations, with observational constraints from \citet{muzzin2013} ($0.2\! \leq\! z\! <\! 0.5$; $0.5\! \leq\! z\! < \!1$; $1.5\! \leq \! z \!< \!2$) and \citet{weaver2023} ($0.2 \!<\! z\! \leq\! 0.5$; $0.8 \!< \!z\! \leq\! 1.1$; $1.5\! <\! z\! \leq \!2$) and semi-empirical predictions from \EMERGE \citep{moster2018} ($0\! < \!z \!\leq\! 0.5$; $0.5\! < \!z\! \leq \!1$; $1.5\! < \!z\! \leq\! 2$) and \UM \citep{behroozi2019}. Additional $z\!=\!0.0$ and $z\!=\!0.3$ \CRKHACC curves are shown in the $z\!=\!0.1$ panel to illustrate rapid low-$z$ evolution. In all panels, gray shaded regions mark the regime below the \CRKHACC\ mass resolution limit. Overall, \CRKHACC follows the trends of other simulations, showing low-$z$ sSFR agreement with observations but systematic high-$z$ offsets, except for the lower SFRs inferred by \citet{leja2022}. The resolved high-mass \CRKHACC\ prediction shows good agreement with data. Quenched fractions are similar, with residual differences primarily set by the resolution of each simulation --- limited SFR sampling drives the low-mass upturn, while unresolved faint systems suppress early growth and delay quenching at high redshift (see text). \\}
    \label{fig:ssfr}
\end{figure*}

Accurately reproducing galactic stellar evolution poses a challenging constraint on astrophysical subgrid models. Although the GSMF was calibrated to match the observed stellar mass distribution over $0<z<2$, this constraint alone does not determine the star-formation activity or quenching efficiency of the galaxy population. To assess these processes in \CRKHACC, we examine the specific star formation rate (sSFR) and the passive fraction of galaxies as a function of redshift. 

Throughout this analysis, we distinguish star-forming (main-sequence) galaxies from passive systems using the commonly adopted threshold ${\rm sSFR} > 0.2\,t_{\rm H}(z)^{-1}$, where $t_{\rm H}(z)$ is the Hubble time at redshift $z$ \citep[e.g.,][]{gallazzi2014,pacifici2016}. We compare our results with corresponding measurements from \COLIBRE \citep{chaikin2025}, \TNG \citep{donnari2019,donnari2021b}, \EAGLE \citep{furlong2015}, \SIMBA \citep{dave2019}, and \FLAMINGO \citep{schaye2023flamingo}. Measurement and aperture definitions, as well as the adopted main-sequence thresholds, differ among simulations; the comparisons shown here are therefore qualitative.

\subsubsection{Specific Star Formation Rate}\label{subsec:ssfr}
The specific star formation rate is defined as ${\rm sSFR} \equiv {\rm SFR}/M_*$, quantifying the instantaneous growth of stellar mass relative to the existing stellar population. The top panel of Figure~\ref{fig:ssfr} presents the median sSFR--$M_*$ relation for \CRKHACC\ galaxies at $z = 0.1$, $1$, and $2$, with red shaded regions showing the 16--84th percentile range. 

Alongside the comparison simulation curves, we include observational datasets spanning optical spectroscopy \citep{zahid2012,belfiore2018}, near-infrared photometry \citep{tomczak2016}, radio stacking \citep{leslie2020}, and far-infrared stacking \citep{koprowski2024}. As broader references across all redshifts, we overlay the empirical relation from \citet{behroozi2019} and the harmonized fit of modern SFR observations from \citet{popesso2023}. Lastly, we show the analysis of \citet{leja2022}, which infers systematically lower SFRs (by $\sim$0.2--0.5 dex) from panchromatic SED modeling that accounts for dust heating by older stellar populations.

At $z = 0.1$, the selected simulation results cluster together and are in broad agreement with the mutually consistent reference relations from \citet{behroozi2019} and \citet{popesso2023}. The \CRKHACC\ measurement lies within this grouping, toward the upper envelope alongside \FLAMINGO. All curves fall between the optical spectroscopic constraints of \citet{zahid2012} and \citet{belfiore2018}, which bracket the local normalization of the star-forming main sequence.

At higher redshifts ($z = 1$ and $z = 2$), the simulations remain closely grouped but diverge systematically from the bulk of observational measurements. While surveys and empirical fits indicate higher sSFR normalizations, the simulations are uniformly lower, reflecting a long-standing tension between models and observations \citep[e.g.,][]{daddi2007,somerville2015physical}. This offset is often attributed to differences in how SFRs are defined, being measured directly in simulations but inferred observationally through modeling assumptions.

An important exception is the analysis of \citet{leja2022}, which, as noted above, infers lower SFRs and agrees more closely with the simulation trends. In this regime, \CRKHACC\ again follows the upper edge of the simulation grouping, matching particularly well with the \citet{leja2022} relation. Notably, this agreement extends to the high-mass end, where all observational datasets converge. A similar pattern appears in the \EAGLE\ results at $z = 1$ and persists in the \CRKHACC\ measurements through $z = 2$. We find a corresponding agreement in the quenched fraction at high masses, as discussed below.

\subsubsection{Quenched Fraction}\label{subsec:qf}

A complementary measurement to the sSFR is the evolution of the passive fraction of galaxies over cosmic time, tracing the buildup of the quenched population and the efficiency with which galaxies transition out of the star-forming sequence. The lower panels of Figure~\ref{fig:ssfr} show the \CRKHACC quenched fraction as a function of stellar mass at $z=0.1$, $z=1$, and $z=2$. We include the simulation results from \TNG, \FLAMINGO, \EAGLE, and \COLIBRE where available.

For observational context, we include quenched fractions derived from near-infrared--selected samples \citep{muzzin2013} and from wide-area multiwavelength photometric surveys \citep{weaver2023}. We also show semi-empirical predictions from \EMERGE\ \citep{moster2018} and \UM\ \citep{behroozi2019}, which connect galaxy growth to dark matter halo assembly. We again note that sSFR thresholds, SFR timescales, and apertures differ among simulations, and that many observational constraints are both redshift-binned and inference-dependent (e.g., SED modeling and dust corrections). Consequently, cross-survey and cross-simulation comparisons should be interpreted qualitatively. 

At $z = 0.1$, the quenched fractions from the selected simulations form a relatively tight grouping, with \CRKHACC\ lying near the upper edge of the band. The supplemental $z = 0$ and $z = 0.3$ curves illustrate the rapid recent growth of the passive population in \CRKHACC, bracketing the observational estimates from \citet{muzzin2013} and \citet{moster2018}, and lying at a slightly higher normalization than \citet{behroozi2019} and \citet{weaver2023}.

A notable feature of the \CRKHACC\ relation is the upturn in the quenched fraction below the mass resolution limit at $M_* \lesssim 10^{10}\,M_\odot$. Similar behavior is seen across all simulations at mass scales set by their respective resolutions, where galaxies with sparsely sampled star-forming gas can artificially appear quenched. The upturn occurs at a higher mass for the lower-resolution \FLAMINGO\ run and at smaller masses for the higher-resolution simulations \TNG, \EAGLE, and \COLIBRE, as expected.

At earlier redshifts, the observed quenched fractions are systematically higher than those predicted by the simulations, except for the semi-empirical results from \citet{moster2018}, which fall closer to the data. Within the simulation set, \CRKHACC\ consistently lies at the low end, reflecting the elevated sSFRs seen in the top panels. This combination of enhanced star-formation activity and reduced quenching aligns with the deficit in the CSFRD at high redshift (see Figure~\ref{fig:csfr}), where unresolved low-mass galaxies suppress early growth and shift stellar assembly to later times. As a result, \CRKHACC\ maintains a larger star-forming population at $z = 1$--2, delaying quenching relative to observations and higher-resolution simulations. At the well-sampled high-mass end, however, the \CRKHACC\ relation converges toward the observational estimates, consistent with the agreement seen in the sSFR relations.

\begin{figure}
    \centering
    \includegraphics[width=\linewidth]{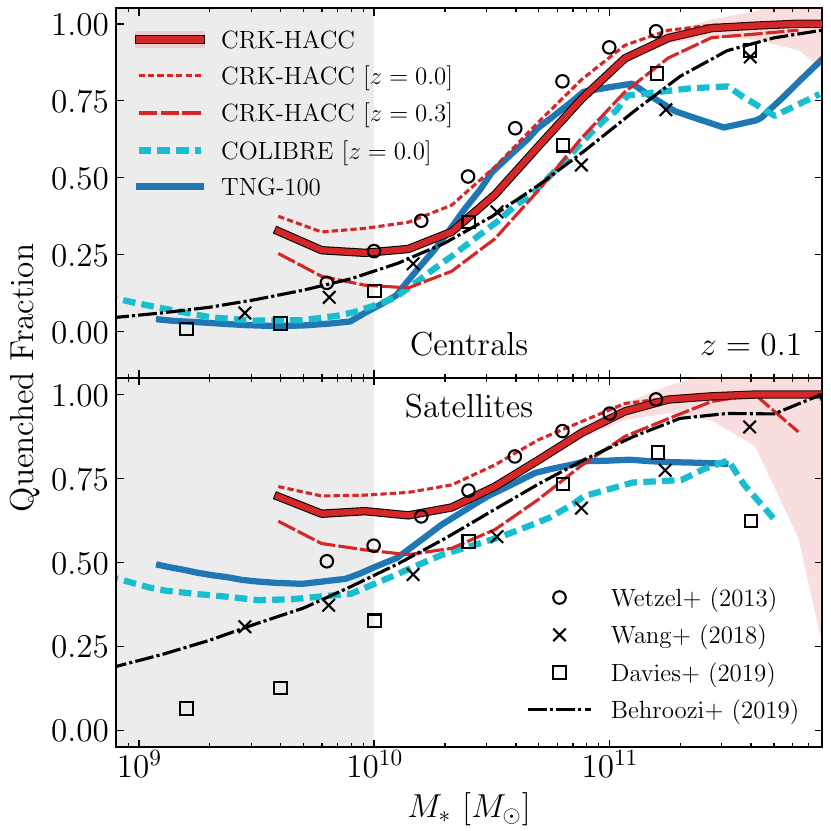}
    \caption{Quenched fractions for central (top) and satellite (bottom) galaxies at $z\!=\!0.1$ in \CRKHACC, with additional $z\!=\!0.0$ and $z\!=\!0.3$ curves shown to illustrate rapid late-time evolution. Red shaded regions indicating Poisson uncertainties are shown for the $z\!=\!0.1$ curve. Gray shaded regions mark the regime below the \CRKHACC\ mass resolution limit. Results are compared with \TNG\ \citep{donnari2021b} and \COLIBRE (\citealt{chaikin2025}; measured at $z\!=\!0$), as well as local ($z\!<\!0.2$) observational estimates from SDSS \citep{wetzel2013,wang2018} and GAMA \citep{davies2019}. Semi-empirical predictions from \UM\ at $z\!=\!0.1$ are also shown. Observational surveys primarily probe group and cluster environments, whereas the simulations represent the full satellite population. \CRKHACC\ lies between the other simulations for centrals and near the upper edge for satellites, in reasonable agreement with the spread of observational estimates. Each simulation exhibits a resolution-dependent low-mass upturn in the quenched fraction, consistent with the trends shown in Figure~\ref{fig:ssfr}.}
    \label{fig:quenched_cs}
\end{figure}

We next examine the quenched fraction separately for centrals and satellites, two populations expected to follow distinct quenching pathways. Satellites are typically more quenched than centrals at fixed stellar mass because infall halts fresh gas accretion and exposes galaxies to environmental processes (such as ram-pressure and tidal stripping) that accelerate gas depletion. Figure~\ref{fig:quenched_cs} shows the \CRKHACC\ passive fractions for both populations at $z = 0.1$, compared with \COLIBRE\ (measured at $z = 0$) and \TNG, as well as observational estimates from SDSS \citep{wetzel2013,wang2018} and GAMA \citep{davies2019}. We also include empirical predictions from \UM\ \citep{behroozi2019}.

For centrals, the predictions from \TNG, \COLIBRE, and \CRKHACC\ all fall within the envelope of the observational datasets, with the \CRKHACC\ curve lying near the middle for intermediate and high stellar masses. Specifically, \CRKHACC\ lies between the estimates of \citet{wetzel2013} and \citet{davies2019}, and above the lower normalizations reported by \citet{wang2018} and \UM. At lower masses, we again see the upturn noted in Figure~\ref{fig:ssfr}, which reflects numerical resolution limits. The supplemental $z = 0.0$ and $z = 0.3$ curves further highlight the rapid recent evolution of the central passive population in \CRKHACC, consistent with the trends already evident in the full galaxy sample.

For satellites, shown in the lower panel of Figure~\ref{fig:quenched_cs}, the overall interpretation is similar, though in this case the \UM\ predictions track the \CRKHACC\ results more closely. The observational constraints show a wider spread, partly because the relevant surveys target satellites in specific group or cluster environments rather than the full satellite population represented in simulations. Even so, \CRKHACC\ lies near the upper edge of the simulation set, predicting a more quenched satellite population at low redshift.

\subsection{Stellar Mass--Halo Mass Relation}\label{subsec:smhm}

\begin{figure}
    \centering
    \includegraphics[width=\linewidth]{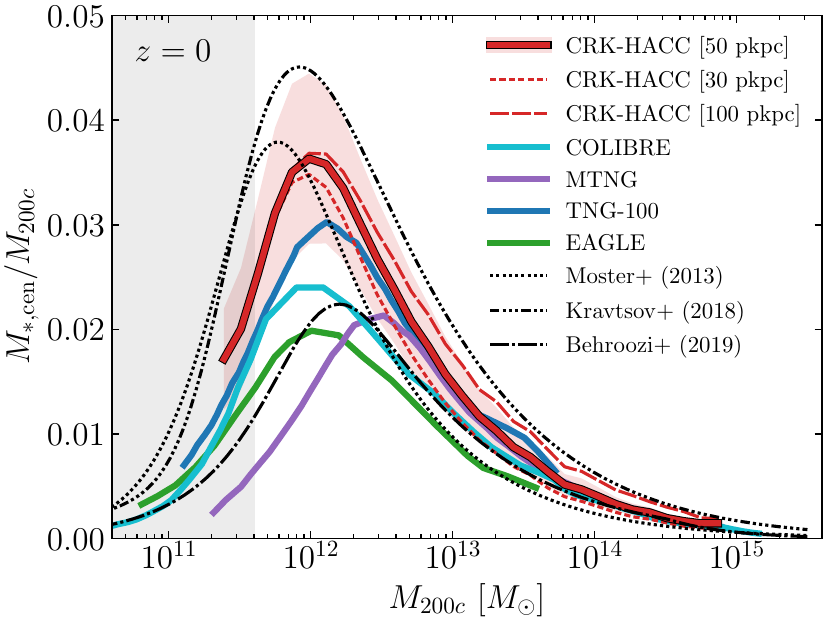}
    \caption{Stellar mass--halo mass (SMHM) relation for central galaxies at $z = 0$. The median \CRKHACC\ prediction is shown for fixed physical apertures of $30$, $50$ (default), and $100$~pkpc, with the red shaded region indicating the $16$th--$84$th percentile range for the default choice. Observational constraints from \citet{moster2013}, \citet{behroozi2019}, and \citet{kravtsov2018} are shown, spanning a mutual spread of $\sim0.3$~dex across halo masses. Results from the \EAGLE\ \citep{schaye2015eagle}, \TNG\ \citep{pakmor2023}, \MTNG\ \citep{pakmor2023}, and \COLIBRE\ \citep{schaye2025} simulations are included for comparison, each shown over its reported mass range. Gray shaded regions mark the regime below the \CRKHACC\ mass resolution limit ($M_{*,\,{\rm cen}} \lesssim 10^{10}\,M_\odot$). The fiducial $50$~pkpc \CRKHACC measurement lies within the observational envelope with a normalization similar to \citet{moster2013}, and is consistent with other simulations and observations at higher masses. The $30$~pkpc and $100$~pkpc curves illustrate the aperture dependence of stellar mass, with the latter converging toward the \citet{kravtsov2018} prediction, consistent with capturing intra-halo light contributions at group and cluster scales.
    }
    \label{fig:smhm}
\end{figure}

The stellar mass--halo mass (SMHM) relation measures the fractional stellar content of central galaxies as a function of host halo mass. Empirically, it peaks near Milky Way--mass halos ($M_{200c}\sim10^{12}\,M_\odot$), where galaxy formation is most efficient, and declines toward both lower and higher halo masses. In simulations, the SMHM relation traces how star formation and feedback prescriptions regulate galactic growth within the context of hierarchical halo assembly.

Figure \ref{fig:smhm} compares the \CRKHACC SMHM relation for central galaxies with three observational baselines. The \citet{moster2013} curve represents a classical abundance‐matching result that links the observed stellar and halo mass functions. The median \citet{behroozi2019} (\UM) relation employs forward modeling to jointly reproduce multiple galaxy statistics and provides a posterior distribution for the intrinsic scatter. At high halo masses, the \citet{kravtsov2018} relation accounts for extended stellar envelopes and intracluster light (ICL), increasing the inferred stellar fractions by approximately $0.2$--$0.3$~dex at $M_{200c}\gtrsim10^{13.5}\,M_\odot$. Together, these three curves span the roughly factor‐of‐two range among current observational estimates, with ICL treatment remaining the dominant systematic uncertainty at the high‐mass end.

The fiducial \CRKHACC SMHM measurement in Figure~\ref{fig:smhm} adopts a $50$~pkpc aperture and includes only central galaxies, defined as the most stellar-massive system within each halo. We report the relation for halos with $M_{200c}\gtrsim10^{11}\,M_\odot$, corresponding to the smallest galaxies reliably resolved in the simulation. For comparison, we also show results from the \EAGLE\ \citep{schaye2015eagle}, \TNG\ \citep{pakmor2023}, \MTNG\ \citep{pakmor2023}, and \COLIBRE\ \citep{schaye2025} simulations, which extend to lower halo masses owing to their higher mass resolution.

The \CRKHACC relation lies well within the observational envelope and most closely follows the shape and normalization of the \citet{moster2013} relation, though it is slightly shifted toward higher halo masses. Relative to other simulations, the halo mass of peak efficiency ($M_{200c}\sim10^{12}\,M_\odot$) lies near the center of the published range, while the corresponding stellar fraction is somewhat higher in \CRKHACC at this mass scale. At larger halo masses, all simulations and most observational inferences converge, with the \citet{kravtsov2018} relation remaining elevated due to its inclusion of ICL at group and cluster scales.

To illustrate the impact of aperture choice, Figure~\ref{fig:smhm} also shows \CRKHACC\ measurements using $30$~pkpc and $100$~pkpc apertures. The smaller aperture is consistent with all other simulations shown here except \COLIBRE\ (which uses our default $50$~pkpc aperture) and yields a lower inferred stellar fraction by excluding extended stellar light. The larger aperture includes more diffuse stellar components and shifts the relation upward, converging toward the \citet{kravtsov2018} curve at the highest halo masses where ICL contributions become significant. For large-volume simulations, this underscores the importance of matching the level of intra-halo light included in specific observational analyses when comparing stellar mass fractions in groups and clusters.

\subsection{Stellar Mass--Metallicity Relations}\label{subsec:MZR}

\begin{figure}
    \centering
    \includegraphics[width=\linewidth]{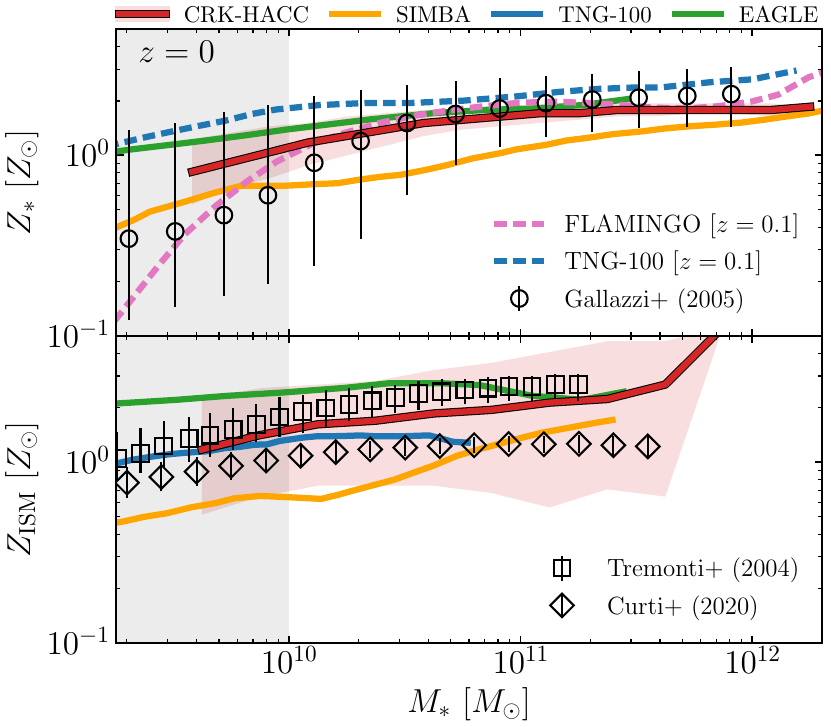}
    \caption{Galaxy mass--metallicity relations (MZR) from the fiducial \CRKHACC simulation (red shaded 16th--84th percentile range) compared to observational measurements and predictions from the \EAGLE\ \citep{schaye2015eagle}, \SIMBA\ \citep{dave2019}, \TNG\ (stellar: \citealt{nelson2018}; ISM: \citealt{torrey2019}), and \FLAMINGO\ \citep{schaye2023flamingo} simulations. All metallicities are expressed in solar units with $Z_\odot = 0.0129$. Gray shaded regions mark the regime below the \CRKHACC\ mass-resolution limit. \textbf{Top:} Stellar metallicity relation, with local ($z < 0.22$) SDSS-based estimates from \citet{gallazzi2005}. \textbf{Bottom:} ISM metallicity relation, compared to local ($z \sim 0.1$) SDSS strong-line results from \citet{tremonti2004} and $T_e$-based calibrations from \citet{curti2020}. The \CRKHACC\ results reproduce observed stellar metallicities across the resolved mass range and predict ISM abundances bracketed by observational systematics.}
    \label{fig:galmet}
\end{figure}

The galaxy mass--metallicity relation (MZR) provides an important constraint on galaxy evolution models, linking stellar mass growth to the efficiency of metal production and retention. Two complementary forms of the MZR are commonly considered: the stellar metallicity relation, which reflects the integrated enrichment history imprinted on long-lived stars, and the ISM metallicity relation, which captures the present-day abundance of heavy elements in star-forming gas. Together, these observables probe both the cumulative and instantaneous regulation of baryons and metals in galaxies.

The top panel of Figure~\ref{fig:galmet} shows the stellar mass--metallicity relation from the fiducial \CRKHACC\ simulation compared with SDSS-based measurements from \citet{gallazzi2005} and predictions from other hydrodynamical simulations. The observational relation indicates that galaxies with $M_* \gtrsim 10^{9}\,M_\odot$ are already substantially enriched, with only a weak residual dependence on stellar mass.

\CRKHACC\ reproduces the observed stellar metallicities well across the resolved mass range. This agreement supports the parameter choices in our metallicity model, where star-forming gas is initialized with a minimum metallicity floor of $Z_{\mathrm{ISM,\,min}} = 0.25\,Z_\odot$ (Section~\ref{subsec:SFImplementation}), and galactic winds are assumed to be metal-poor ($\gamma_w = 0$; Section~\ref{subsec:windZ}) at the coarse resolution. Together, these settings seed unresolved early galaxies with the enrichment they would have produced and ensure that those metals remain in the ISM rather than being expelled in outflows. \footnote{At higher resolution, this choice becomes less restrictive, as metal loading factors near the conventional value of $\gamma_w = 0.4$ (as used in \TNG) can be adopted in our model without oversuppressing enrichment, and the imposed $Z_{\mathrm{ISM,\,min}}$ correspondingly decreases as earlier enrichment is directly resolved.}

Relative to other simulations, \CRKHACC\ most closely follows the predictions of \EAGLE\ \citep{schaye2015eagle}, despite operating at a much lower mass resolution. The \TNG\ relation \citep{nelson2018} lies slightly above both \CRKHACC\ and the observational mean but remains consistent within the scatter, whereas \SIMBA\ \citep{dave2019} predicts a shallower relation with systematically lower stellar metallicities. \FLAMINGO\ \citep{schaye2023flamingo} shows a spurious ramp at low masses caused by resolution effects, but converges toward the observational relation for $M_* \gtrsim 10^{10}\,M_\odot$. 

The comparison with \FLAMINGO\ is particularly informative, as both simulations operate at similar resolution. \FLAMINGO\ models metal production self-consistently, without imposing an artificial metallicity floor, which leads to a delayed enrichment ramp-up until galaxies are sufficiently resolved. In contrast, \CRKHACC\ enforces early enrichment through a resolution-dependent initialization, allowing the stellar metallicity relation to match observations across the reported mass range. 

We now turn to the metallicity relation of the star-forming ISM, shown in the bottom panel of Figure~\ref{fig:galmet}. The canonical SDSS analysis of \citet{tremonti2004} inferred gas-phase abundances from strong emission-line ratios using photoionization models, yielding relatively high metallicities at fixed stellar mass. In contrast, calibrations tied to direct $T_e$-based abundance measurements \citep[e.g.,][]{curti2020} produce a lower normalization by about $0.2$--$0.3$~dex. The $T_e$ method is generally regarded as the more reliable standard, as it avoids the model dependencies inherent to strong-line techniques, though systematic uncertainties remain. Together, these two determinations are representative of the range spanned by current observations of the ISM MZR.

\CRKHACC\ predicts ISM metallicities that fall between the two observational measurements, with substantial scatter such that both the \citet{tremonti2004} and \citet{curti2020} relations lie within its spread. The median trend most closely resembles that of \TNG\ (ISM data from \citealt{torrey2019}), whose predictions align well with the $T_e$-based calibration of \citet{curti2020}. In contrast, \EAGLE\ yields systematically higher metallicities consistent with the \citet{tremonti2004} relation, while \SIMBA\ produces lower values across the full mass range. Overall, \CRKHACC\ predicts an ISM MZR intermediate between the two observational calibrations and bracketed by their systematic uncertainties.

We note that all results in Figure~\ref{fig:galmet} have been renormalized to our adopted solar abundance of $Z_\odot = 0.0129$ (see Section~\ref{subsec:mlc}) for consistency. In addition, the simulations differ in how metallicities are defined and averaged (e.g., mass-weighted, as used here, versus SFR- or light-weighted), so comparisons should be regarded as qualitative.

\subsection{Black Hole Mass--Stellar Mass Relation}\label{subsec:bhsm}

\begin{figure}
    \centering
    \includegraphics[width=\linewidth]{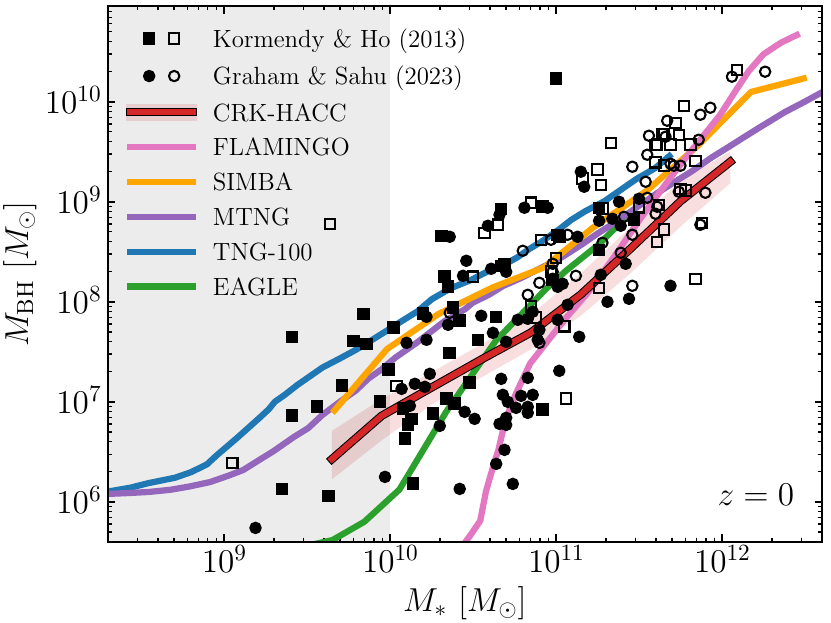}
    \caption{Black hole--stellar mass relation at $z=0$ from the fiducial \CRKHACC\ simulation, with the red shaded region denoting the 16th--84th percentile range. Gray shaded regions mark the regime below the \CRKHACC\ mass resolution limit. Results are compared to hydrodynamical models including \EAGLE\ \citep{schaye2015eagle}, \TNG\ \citep{pakmor2023}, \MTNG\ \citep{pakmor2023}, \FLAMINGO\ \citep{schaye2023flamingo}, and \SIMBA\ \citep{thomas2019}, as well as local ($z \lesssim 0.05$) observational compilations \citep{kormendy2013, graham2023}. In the observational data, filled and open markers denote disk and elliptical galaxies, respectively. The \citet{kormendy2013} points are reported in terms of bulge masses ($M_{\rm bulge}$). The \CRKHACC\ relation exhibits a slope consistent with \TNG, \MTNG, and \SIMBA, and lies within the broad observational scatter.}
    \label{fig:bhsm}
\end{figure}

The black hole--stellar mass (BHSM) relation characterizes the coevolution of galaxies and their central supermassive black holes. Observationally, black hole masses correlate with bulge stellar mass and velocity dispersion, though both the slope and scatter depend strongly on galaxy morphology. In cosmological simulations, black holes cannot be resolved directly and are instead modeled through subgrid prescriptions that capture their growth and feedback. As a result, relations such as the BHSM are not predicted from first principles but serve as informative benchmarks for assessing whether these coarse-grained models reproduce the integrated buildup of black hole mass and provide a physically consistent feedback energy budget within their host galaxies.

Figure~\ref{fig:bhsm} compares the BHSM relation from the fiducial \CRKHACC\ simulation with predictions from other hydrodynamical models, including \EAGLE\ \citep{schaye2015eagle}, \TNG\ \citep{pakmor2023}, \MTNG\ \citep{pakmor2023}, \FLAMINGO\ \citep{schaye2023flamingo}, and \SIMBA\ \citep{thomas2019}. Local observational compilations from \citet{kormendy2013} and \citet{graham2023} are also shown. The \citet{kormendy2013} dataset reports bulge stellar masses ($M_{\rm bulge}$), while the \citet{graham2023} sample provides total galaxy stellar masses ($M_{\ast,\,\mathrm{gal}}$). Filled and open symbols denote disk and elliptical galaxies, respectively.

The implementation of AGN feedback in \CRKHACC\ closely follows that of \TNG\ \citep{weinberger2016simulating}, adopting identical values for the radiative accretion and high-accretion mode feedback efficiencies, $\epsilon_r = 0.2$ and $\epsilon_{\rm high} = 0.1$ (see Section~\ref{sec:bhmodel}). Furthermore, in \CRKHACC\ the calibrated seed mass independently converged to $M_{\rm seed} = 8\times10^5\,\massh$, the same value adopted in \TNG.\footnote{The primary feedback distinction relative to \TNG\ is the kinetic implementation, where \CRKHACC\ employs a constant low-accretion efficiency of $\epsilon_{\rm kin} = 1.3$ and a jet velocity of $v_{\rm jet} = 5.1\,{\rm km\,s^{-1}}$, calibrated to reproduce the observed cluster gas density profiles at the fiducial resolution (see Section~\ref{subsec:calib}). The kinetic mode contributes only weakly to shaping the BHSM relation compared to the other feedback parameters.}

Given this alignment, it is not surprising that Figure~\ref{fig:bhsm} shows similar BHSM slopes for \CRKHACC, \TNG, and \MTNG. The horizontal offsets among these models arise in part from resolution: in \CRKHACC, black holes are seeded only once host galaxies exceed $M_*\sim10^9\,M_\odot$, the effective resolution limit, whereas \TNG\ and \MTNG\ reach lower masses owing to their finer mass resolution. That said, given differences in the black hole evolution model, as well as the broader \CRKHACC\ subgrid framework and calibration targets, the apparent agreement in normalization may be partly coincidental.

Relative to observations, the scatter in black hole mass at fixed stellar mass is large, reflecting both intrinsic variation and morphological dependence. The predictions from all simulations fall within this broad observational envelope. \EAGLE\ and \FLAMINGO\ seed black holes at low masses and exhibit a more rapid subsequent growth, while \TNG, \MTNG, \SIMBA, and \CRKHACC\ follow higher, more elliptical-like relations. The \CRKHACC\ trend lies on the lower edge of the observed scatter, consistent with the adopted efficiency parameters and the coarse simulated mass resolution. Overall, the adopted black hole growth and feedback prescriptions appear to plausibly capture the black hole--galaxy coevolution.

\subsection{Halo Gas Fraction}\label{subsec:gasfrac}

\begin{figure}
    \centering
    \includegraphics[width=\linewidth]{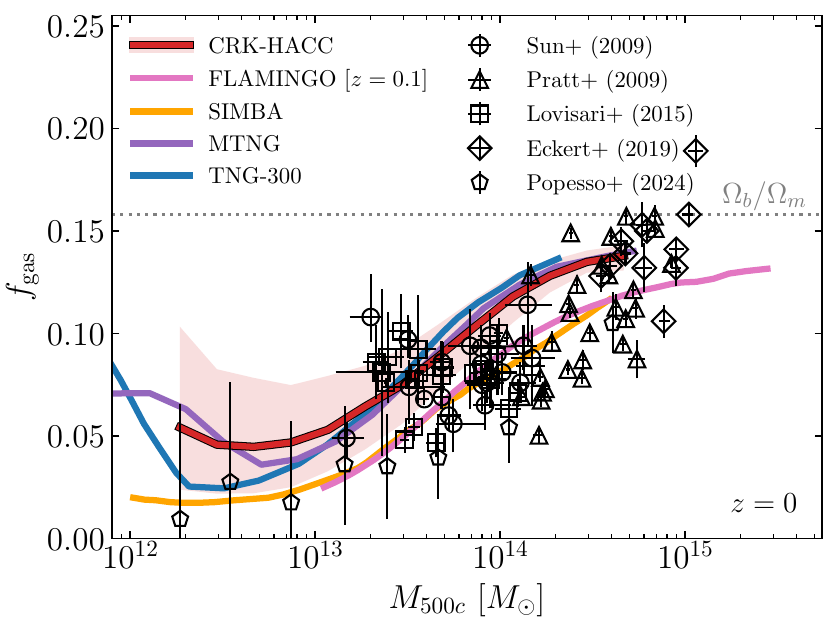}
    \caption{Halo gas fraction as a function of mass at $z\!=\!0$ for the calibrated \CRKHACC\ simulation, with the shaded region indicating the 16th--84th percentile range. Results are compared to local ($z \!<\! 0.2$) X-ray measurements from \citet{sun2009}, \citet{pratt2009}, \citet{lovisari2015}, \citet{eckert2019}, and \citet{popesso2024}, where all but \citet{eckert2019} assume hydrostatic equilibrium without explicit bias correction. For reference, we also include predictions from the \TNG\ \citep{pakmor2023}, \MTNG\ \citep{pakmor2023}, \SIMBA\ \citep{dave2019}, and \FLAMINGO\ \citep{schaye2023flamingo} simulations. The \CRKHACC\ calibration closely matches the \TNG\ and \MTNG\ relations while lying above \SIMBA\ and \FLAMINGO, which employ more aggressive AGN feedback. This modestly higher normalization reflects the choice to calibrate on observed cluster gas density profiles, prioritizing accurate modeling of the intracluster gas distribution over achieving lower group-scale gas fractions.}
    \label{fig:fgas}
\end{figure}

We conclude with a measurement of the halo gas mass fraction, $f_{\mathrm{gas}} = M_{\mathrm{gas}}(r<R_{500c})/M_{500c}$, for halos spanning group to cluster scales, comparing our results with a representative suite of X-ray observations and large-volume hydrodynamic simulations.  The gas fraction provides a sensitive probe of feedback strength, as more energetic models drive stronger outflows that deplete halos and lower their total baryonic content.

The predicted \CRKHACC\ $f_{\mathrm{gas}}$--$M_{500c}$ relation is shown in Figure~\ref{fig:fgas}. Our observational comparison sample includes the \textit{Chandra} study of nearby galaxy groups by \citet{sun2009}, the \textit{XMM-Newton} REXCESS cluster sample analyzed by \citet{pratt2009}, the \textit{ROSAT}-selected \textit{XMM-Newton} group sample of \citet{lovisari2015}, and the joint \textit{XMM-Newton}+\textit{Planck} SZ analysis of massive clusters by \citet{eckert2019}. We also include a recent \textit{eROSITA} measurement from \citet{popesso2024} based on optically selected groups stacked in X-ray maps, which reports systematically lower gas fractions at fixed mass than earlier X-ray--bright samples. Taken together, these measurements span two orders of magnitude in halo mass but exhibit substantial intrinsic scatter, underscoring the challenge of constraining $f_{\mathrm{gas}}$ observationally.

The \CRKHACC\ results in Figure~\ref{fig:fgas} are in close agreement with the predicted $f_{\mathrm{gas}}$--$M_{500c}$ relations from the \TNG\ and \MTNG\ simulations, reproducing both the normalization and the characteristic rise of the gas fraction toward the cluster mass scale. By contrast, the predicted gas fractions lie systematically above those of \SIMBA\ and \FLAMINGO. This behavior is expected, as \SIMBA\ employs more aggressive AGN jet feedback that efficiently removes gas from group-scale halos, while \FLAMINGO\ explicitly tuned its parameters to reproduce observed baryon fractions, resulting in lower $f_{\mathrm{gas}}$ by construction.

One of the most important systematics affecting observational measurements is the assumption of hydrostatic equilibrium (HSE) when deriving total masses from X-ray data. Cosmological simulations predict that non-thermal pressure support leads to a $\sim10$--$30\%$ underestimate of $M_{500c}$ in typical systems \citep[see][]{pratt2019}, with smaller biases in dynamically relaxed clusters \citep[e.g.,][]{eckert2019}. As no correction for this bias is applied in Figure~\ref{fig:fgas}, the observed $f_{\mathrm{gas}}$ values are systematically overestimated, which in turn increases the apparent tension with models that predict higher gas retention.

The close agreement with \TNG\ and \MTNG\ is noteworthy given that the \CRKHACC\ AGN feedback calibration was performed independently, using cluster gas density profiles as the primary target observable. As discussed in Section~\ref{subsec:calib} and in \RAMACHANDRA, this calibration strategy was motivated by the goal of producing realistic cluster populations for survey-scale predictions.  Attempts to calibrate directly to $f_{\mathrm{gas}}$ --- as done in e.g., \FLAMINGO --- were found to strongly alter cluster density profiles under our chosen subgrid implementation. The adopted calibration therefore preserves the observed intracluster gas distribution at the cost of slightly elevated $f_{\mathrm{gas}}$ at group scales. Modified AGN feedback models, including redshift-dependent prescriptions, are currently being explored to help reconcile this remaining tension.

Finally, we emphasize that the intrinsic scatter in $f_{\mathrm{gas}}$ measurements remains considerably larger in the data than in simulations. Future work employing synthetic X-ray pipelines to estimate $M_{\mathrm{gas}}$ and $M_{500c}$ in an observationally consistent manner will be necessary for determining whether these residual offsets primarily reflect measurement biases or limitations of the feedback model itself. For applications where reproducing the mean observed $f_{\mathrm{gas}}$ is the primary objective, \RAMACHANDRA\ also provide an alternative calibration with stronger AGN feedback that yields lower gas fractions, suitable for targeted investigations.

\section{Conclusions}\label{sec:conclusion}
\label{sec:conclude}

Modern precision cosmology efforts have delivered percent-level constraints on the growth of structure through increasingly sophisticated multi-probe surveys, setting new requirements for theoretical accuracy. At this level of precision, baryonic processes significantly affect the matter distribution probed by these surveys, and must therefore be modeled alongside dark matter to ensure reliable predictions of observed signals. Moreover, simulations must encompass full survey-scale volumes in order to capture the statistical reach of current and upcoming experiments. Meeting this challenge remains difficult. The inclusion of baryonic physics substantially increases computational cost, while the underlying astrophysical processes span wide dynamic ranges and require sub-resolution models whose fidelity and consistency are continually being refined. 

Fortunately, the necessary components for a new generation of survey-scale modeling efforts are now in place. Exascale computing platforms provide the throughput to include complex baryonic physics at cosmological volumes, while decades of progress in hydrodynamic simulation have yielded physically motivated formulations that can be systematically refined and integrated into modern frameworks. In recent years, we have already witnessed a dramatic increase in the scope and volume of hydrodynamic cosmological simulations --- a trend that must continue to meet the demands of upcoming survey analyses.

This study extends the \CRKHACC\ framework with a physically motivated suite of subgrid models for galaxy formation and feedback, enabling self-consistent treatment of baryonic physics within cosmological volumes comparable to gravity-only simulations. The implementation incorporates radiative cooling, a subgrid multiphase interstellar medium with stochastic star formation, kinetic galactic winds, and two-mode AGN feedback, all coupled to the hydrodynamics and gravity solvers through an operator-split source-term formalism. Designed for GPU acceleration, the framework achieves high performance and scaling on leadership-class supercomputers, providing the computational throughput required for full-physics simulations at exascale volumes and resolutions relevant for survey-scale synthetic-sky generation.

To calibrate the models, we adopt a subgrid parameter set tuned to reproduce the observed galaxy stellar mass function over a survey-relevant redshift range ($0 < z < 2$) as well as the gas-density profiles of massive clusters --- an important target for large-volume simulations aimed at studying statistically representative cluster populations. We then carried out an $L_\mathrm{box}=256\,h^{-1}\,\mathrm{Mpc}$ cosmological simulation using this fiducial model to evaluate the fidelity of the resulting galaxy and halo populations produced by the hydrodynamic framework.

The simulation predictions yield realistic global stellar evolution, reproducing the observed buildup of stellar mass and the decline of star formation across cosmic time. On galactic scales, it produces plausible levels of star-formation activity and quenching, along with chemical enrichment, black-hole growth, and halo gas fractions comparable to those found in other modern simulations. Across these comparisons, there are clear indications of where limited resolution constrains predictive accuracy, particularly in reproducing detailed galactic structures, while the model nevertheless demonstrates the accuracy required to capture the global trends and baryonic couplings that underpin survey-scale observables. In general, balancing model complexity and resolution with simulation scale will remain an ongoing challenge, motivating continued efforts to refine subgrid physics and numerical treatments as computational capabilities improve.

Looking ahead, we will continue to expand the \CRKHACC\ subgrid framework through increasingly ambitious large-volume hydrodynamic simulations and targeted studies. The framework has already been deployed in an exascale production run, the \FRONTIERE simulation \citep{frontiere2025GB}, demonstrating that hydrodynamic cosmological simulations at trillion-particle scales are now computationally achievable. Building on this foundation, upcoming campaigns will extend both the physical realism and statistical reach of our models, combining flagship large-volume runs with ensembles of hundred-megaparsec simulations that systematically vary cosmological parameters and subgrid prescriptions. Collectively, these efforts will strengthen the predictive power of \CRKHACC\ and advance its role in supporting next-generation survey science.

\begin{acknowledgments}

We recognize the efforts of past and current \HACC team members David Daniel, Patricia Fasel, Hal Finkel, Patricia Larsen, Vitali Morozov, Adrian Pope, Esteban Rangel, and Tom Uram. We thank Damien~Lebrun-Grandi\'{e}, Andrey~Prokopenko, and the ArborX team for their critical contributions to the analysis framework. We thank Andy Marszewski for providing \FIRE simulation data. We acknowledge the continued collaboration with Mike Owen and Cody Raskin on \CRKSPH solver discussions and development. NF, SH, and KH are also grateful for the insightful discussions and guidance provided during the 2023 summer program ``Groups and Clusters of Galaxies at the Crossroad between Astrophysics and Cosmology'' at the Aspen Center for Physics.

This research was supported by the Exascale Computing Project (17-SC-20-SC), a collaborative effort of the U.S. DOE Office of Science and NNSA and by the U.S. Department of Energy, Office of Science, Office of Advanced Scientific Computing Research and Office of High Energy Physics, Scientific Discovery through Advanced Computing (SciDAC) program. Work at Argonne National Laboratory was supported under the U.S. Department of Energy contract DE-AC02-06CH11357. This research used resources of the National Energy Research Scientific Computing Center, a DOE Office of Science User Facility supported by the Office of Science of the U.S. Department of Energy under Contract~DE-AC02-05CH11231. This research also used resources of the Argonne Leadership Computing Facility, which is a DOE Office of Science User Facility supported under Contract~DE-AC02-06CH11357. Additionally, this work used resources of the Oak Ridge Leadership Computing Facility, which is a DOE Office of Science User Facility supported under Contract~DE-AC05-00OR22725. We gratefully acknowledge use of the Improv cluster in the Laboratory Computing Resource Center at Argonne National Laboratory. CAFG was supported by NSF through grants AST-2108230 and AST-2307327; by NASA through grants 21-ATP21-0036 and 23-ATP23-0008; and by STScI through grant JWST-AR-03252.001-A. Lastly, NF thanks his mother for helping translate his thoughts into something resembling English, for which the other authors wish to express their undying gratitude.
\end{acknowledgments}

\appendix

\section{CRK Solver Modifications}\label{app:CRKUpdate}

The inclusion of subgrid models necessitates several key modifications to the original reproducing kernel (RK) solver described in \citetalias{frontiere2022simulating}. In particular, coupling complex astrophysical processes introduces deeper timestepping hierarchies, species conversions, and irregular particle topologies resulting from rapid cooling and strong feedback. These effects pose new challenges for numerical stability and solver efficiency. In the following, we briefly summarize the principal solver updates and highlight the most relevant implementation details. 

\subsection{Reproducing Kernel Relaxation}\label{app:RKRelax}

High-order reproducing kernels are a defining feature of \CRKSPH, providing improved accuracy over traditional SPH by exactly reproducing linear fields. The corrected kernel is given by  
\begin{equation}
    \Wrij = A_i (1 + \vb{B}_i\cdot\vb{x}_{ij})W_{ij}(h_i), \label{eq:RK}
\end{equation}
where $A_i$ and $\vb{B}_i$ are correction coefficients derived from local geometric moments (see Appendix~A in \citetalias{frontiere2017}). These coefficients depend solely on the spatial distribution of neighboring particles, making the corrections sensitive to unphysical, high-frequency irregularities in the local point topology. This tradeoff is typical for higher-order methods, where Gibbs-like ringing effects can appear when the local distribution is poorly resolved and is usually handled by artificial viscosity damping. However, when unresolved subgrid sources drive abrupt local changes in momentum, energy, or mass, they can still generate irregular topologies that degrade interpolation accuracy and produce spurious over-expansion or collapse.

To mitigate unphysical forces arising from irregular kernel shapes, we introduce a simple reproducing kernel regularization procedure based on the velocity divergence of each particle. Recall that the RK formulation provides an accurate measurement of the velocity gradient,
\begin{equation}\label{eq:gradv}
\nabla \vb{v}_i = -\sum_j V_j (\vb{v}_i - \vb{v}_j)\otimes\nabla\Wrij,
\end{equation}
which is critical for constructing an artificial viscosity limiter that suppresses excessive dissipation in smooth flows (see Section~2.2 in \citetalias{frontiere2022simulating}).

In the context of regularization, the RK correction is relaxed whenever the particle volume is predicted to expand or contract by more than a factor of eight (equivalently, by a factor of two radially). The expected volume change over a timestep $\Delta t$ can be approximated using the volumetric continuity equation,
\begin{equation}\label{eqn:contV}
    \frac{dV_i}{dt} = V_i\,\, \nabla \!\cdot \!\vb{v}_i  \Longrightarrow  V(\Delta t) = V_i\, \exp\!\Big[\nabla \!\!\cdot \!\!\vb{v}_i \,\Delta t\Big],
\end{equation}
where the divergence is computed from the trace of the velocity gradient, $\nabla \!\cdot\! \vb{v}_i \equiv \text{Tr}\,\nabla \vb{v}_i$. The relaxation condition is enforced by reverting to the standard SPH kernel if
\begin{equation}
    \exp\!\Big[|\nabla \!\!\cdot \!\!\vb{v}_i|\, \Delta t\Big] > 8 \quad \Longrightarrow \quad A_i = 1,\; \vb{B}_i = 0,
\end{equation}
which prevents unphysical over-expansion or collapse. This condition does not imply that the timestep is too large; rather, it guards against unphysical forces that arise from poorly conditioned RK corrections even when the timestep is properly limited.

While more sophisticated metrics were tested --- such as directly examining higher moments of the local point distribution --- the simple velocity divergence condition proved both robust and computationally efficient in practice. When triggered, the relaxation is applied symmetrically: any pairwise force interaction involving a regularized particle is computed without RK corrections for that pair, ensuring consistent momentum exchange without requiring all neighbors to be regularized.

Overall, regularization events are rare because the RK scheme is generally robust for typical gas evolution. Only exceptionally strong subgrid events lead to unrecoverable topologies, and this safeguard prevents pathological configurations from forming. Notably, the stable evolution observed in our subgrid simulations allows us to retain several efficiency strategies from non-radiative runs, such as reusing previously computed correction coefficients for particles that are passive (i.e., not updating forces) during a given timestep. This avoids unnecessary recomputation and yields significant performance gains, especially given the deep timestepping hierarchies required for astrophysical modeling. One caveat is that subgrid species conversions can alter the spatial configuration of active neighbors, requiring any tree leaf that includes transformed particles to recompute its RK correction coefficients to maintain force accuracy. Such leaves are marked for update, minimizing cost by recalculating only where necessary.

\subsection{Multi-Species Interaction Domains}\label{app:MS}

The inclusion of astrophysical subgrid processes requires \CRKHACC\ to consistently handle multiple particle species, each with uniquely defined interaction domains for exchanging mass, energy, and momentum. In what follows, we describe the state properties of each species and how these domains are constructed in practice. Here, we refer to ``gas’’ as the combined set of normal and star-forming gas particles, since both contribute to the hydrodynamic force solver in the usual way (summarized in \citetalias{frontiere2022simulating}). Star-forming gas is distinct only in its ability to spawn wind and star particles and in being subject to thermal evolution governed by an equation of state (Section~\ref{subsec:SF}). We further refer to subgrid particles as the joint set of wind, star, and black hole particles.

We begin with the modified volume estimate for gas particles. As processes such as stellar enrichment and AGN accretion cause individual particle masses to deviate from the initialized baryon mass, we generalize the gas particle volume definition to
\begin{equation}\label{eqn:Vg}
    V_{i,g}^{-1} = m_{i,g}^{-1} \sum_{j\in g} m_{j,g}\, W(|\vb{x}_{ij}|, h_i),
\end{equation}
which reduces to the number density formulation used in \citetalias{frontiere2022simulating} when all $m_i$ are equal. This expression is equivalent to the standard SPH relation $V_i = m_i/\rho^{\rm SPH}_i$, where $\rho^{\rm SPH}_i$ is the usual SPH density estimate.

The summation in \cref{eqn:Vg} spans only gas particle neighbors, and the resulting volume is used for most hydrodynamic force operators.
The only exception is the gas density calculation, which instead uses the modified form given by \cref{eqn:rhow} and requires a volume that encompasses both gas and wind particles (labeled $V_{i,\,g\pluscup\rm w}$ in that expression). This alternate volume is obtained by expanding the sum in \cref{eqn:Vg} to include wind neighbors and is stored separately for use when updating wind and gas densities.

Subgrid particle volumes follow a similar definition: each subgrid species acts as an SPH-like element in the solver but defines its own interaction kernel based exclusively on the surrounding gas. For example, stars redistribute mass to gas through stellar enrichment, black holes inject feedback energy, and wind particles deposit metals into the ISM when launched. All such operations require that the subgrid neighbor sets be defined purely with respect to the physical gas they affect.

Unlike the gas volume estimate, which is mass-weighted, subgrid neighbor definitions are designed to act on a well-defined number of surrounding gas neighbors, emphasizing consistent local coupling rather than mass density alone. Accordingly, it is natural to define the effective volume of each subgrid particle using the SPH number density of local gas neighbors,
\begin{equation}\label{eqn:Vsg}
    V_{i, s}^{-1} =  \sum_{j\in g\pluscup i} W(|\vb{x}_{ij}|, \,h_{j}),
\end{equation}
where the index $s \in \{\star,\,{\rm w},{\rm BH}\}$ denotes each subgrid species. The summation explicitly includes the self-interaction, such that $g\pluscup i$ denotes the disjoint union of the (non-subgrid) gas neighbors with the subgrid particle itself. This formulation is similar to the original number density formalism in \citetalias{frontiere2022simulating}, except here the SPH sum uses the ``scatter'' definition \citep{Hernquist1989}, in which the smoothing length of each gas neighbor, $h_j$, determines the kernel contribution rather than the subgrid particle extent. This choice ensures that the interaction region of each subgrid element adapts self-consistently to the local gas distribution and allows its domain of influence to be determined entirely by the gas it affects.

The subgrid volume also determines the particle smoothing length, $h_s\propto V^{1/3}_{i, s}$, where all enrichment and feedback interactions are subsequently weighted by the neighboring gas volumes $V_{j,g}$ and kernels $W(|\vb{x}_{ij}|, \,h_s)$ using the ``gather'' definition (reflected in \cref{eqn:windmetal,eqn:starmetal,eqn:AGNE}). To support efficient multi-species neighbor interactions, we sort the particle members within each tree leaf by species type, ensuring that kernels (e.g., for hydrodynamic forces) operate only on the relevant contiguous subsets and skip unrelated particle types. 

All particles use the same \cite{Wendland1995} $C^4$ kernel of radius $h$ as in \citetalias{frontiere2022simulating}, ensuring consistent force estimates and neighbor sampling across species.\footnote{A common source of confusion in SPH implementations is that kernels are sometimes quoted with a radius of $2h$.} Each particle type defines its smoothing length based on a target neighbor count appropriate to its physical role. Following previous work, gas particles use $n_{h_g} \!= 4$ particles per kernel radius, yielding a neighbor count of $N_g = \frac{4\pi}{3}\,n^3_{h_g} \approx 268$. Wind particles adopt the same resolution, $n_{h_{\rm w}} \!= n_{h_g}$, to contribute consistently to gas density estimates and allow smooth recoupling.

In contrast, stars and black holes use more compact interaction domains with $n_{h_\star} \!= n_{h_{\rm BH}}\! = 2.25$, corresponding to approximately $N_{\star} = N_{\rm BH} \approx 48$ neighbors. This choice maintains localized enrichment and feedback near galaxies while avoiding excessive diffusion into low-density regions. 
For simulations run at different mass resolutions, these parameters can be adjusted to preserve consistent physical scales and overall accuracy.

The evolution of the smoothing length for all particles is computed using a modified approach based on \cite{thacker2000}, as described in Appendix~\ref{app:HInter}. For subgrid particle transformations, each $h_s$ is initialized from the volume of the progenitor particle according to the relation $h_s = n_{h_s}\,V^{1/3}$. This prescription applies when stars or wind particles are spawned from star-forming gas (with volume $V_g$) and when AGN particles are seeded or repositioned to stellar particle locations (with volume $V_\star$).

When considering the gravitational contribution of each particle species, we adopt a single uniform softening length, $\epsilon_{\rm soft}$, to maintain consistent force resolution between dark matter and baryons. Gas smoothing lengths are constrained to remain larger than $\epsilon_{\rm soft}$ to prevent spurious fragmentation near the local resolution limit. In contrast, stars and black holes, which do not participate in hydrodynamic forces, can employ much smaller smoothing lengths, down to $h_s = 0.01\,\epsilon_{\rm soft}$, enabling highly localized interaction kernels for enrichment and feedback. This approach preserves stable gravitational interactions even as individual smoothing lengths evolve dynamically.

\subsection{Smoothing Length Integration}\label{app:HInter}
 
In the original \CRKHACC implementation (Section~3.6 of \citetalias{frontiere2022simulating}), the smoothing length update followed the method of \citet{thacker2000}, which stabilizes the neighbor count without requiring an iterative solver,
\begin{equation}\label{eq:thacker}
    h^{\rm T}_i = h_i (1-\lambda+\lambda \Delta_N),
\end{equation}
where $h^{\rm T}_i$ is the Thacker estimate, $\lambda$ is a weighting coefficient, and $\Delta_N^3 \equiv N_{\rm res}/N_i = V_i\,n_{h_i}^3/h_i^3$ represents the ratio of target to measured neighbor counts defined for each species in the previous section.
During drift operations, the smoothing length was adjusted accordingly, with passive particles additionally updating their volume using the continuity equation estimate from \cref{eqn:contV}.

However, when coupled with deeper timestepping hierarchies and abrupt subgrid transformations, this scheme can produce oscillatory smoothing length estimates as particles transition between active and passive states, undermining the original motivation for adopting the Thacker method.

To address this, we now compute a total smoothing length derivative, $\dot{h}_i$, that combines the Thacker neighbor constraint with the local volumetric continuity condition. This approach ensures that both active and passive particles evolve $h_i$ smoothly and continuously, preventing discontinuities in the neighbor count.
 
For each active particle, $\dot{h}_i$ is computed as
\begin{equation}
    \dot{h}_i = \dot{h}^{\rm T}_i + \dot{h}^{\rm cont}_i,
\end{equation}
where
\begin{equation}
    \dot{h}^{\rm T}_i = \frac{h^{\rm T}_i - h_i}{\Delta t_{\rm a}},
    \quad
    \dot{h}^{\rm cont}_i = \frac{1}{3} h_i\, \nabla \!\cdot\! \vb{v}_i,
\end{equation}
and $\Delta t_{\rm a}$ is the current active timestep.  
This combined formulation is analogous to the smoothing length update in \cite{springel2001}. At each drift operation (integrated with timestep $\Delta t \leq \Delta t_{\rm a}$), all particles update their smoothing length using the stored $\dot{h}_i$, 
\begin{equation}
    h^{\rm new}_i = h_i\, \exp\!\Big[\,\frac{\dot{h}_i}{h_i}\,\Delta t\,\Big],
\end{equation}
yielding a smooth and continuous estimate even in the presence of deep integration hierarchies and rapid subgrid-driven changes.

Likewise, passive state variables such as volume and density are updated as
\begin{equation}
    V^{\rm new}_i = \Big(\frac{h^{\rm new}_i}{n_h}\Big)^3,
    \:\:
    \rho^{\rm new}_i = \rho_i\,\exp\!\Big[-\!\nabla \!\cdot\! \vb{v}_i\,\Delta t\,\Big].
\end{equation}
This modification retains the advantages of the original Thacker method --- avoiding a fully implicit solver --- while ensuring that the smoothing length evolves in a physically stable and well-behaved manner, even under the abrupt transitions imposed by the subgrid operators.

\subsection{Adaptive Kick Corrections for Timestep Refinement}\label{app:integrator}
  
In \CRKHACC, hydrodynamic evolution is advanced using a hierarchical kick--drift--kick (KDK) integrator applied to particle leaves (subdomains) grouped into power-of-two timestep bins (see Section 3.1 in \citetalias{frontiere2022simulating}). In typical configurations, each leaf contains a few hundred particles. The following describes the timestep refinement procedures implemented to handle abrupt changes in momentum and energy introduced by subgrid source terms.

All particles within a leaf evolve over an assigned active timestep interval $\Delta t_{\rm a}$, determined by the local CFL condition, while the smallest timestep among all leaves on a given rank is denoted $\Delta t_{\rm min}$. Over an active interval, the standard KDK update for each leaf is
\begin{alignat}{4}
\text{K:}\quad& \vb{v}^{1/2} &{}={}& \vb{v}^0 + \tfrac{1}{2}\vb{a}^0\,\Delta t_{\rm a} \nonumber\\
\text{D:}\quad& \vb{x}^{1}   &{}={}& \vb{x}^0 + \vb{v}^{1/2}\,\Delta t_{\rm a} \nonumber\\
\text{K:}\quad& \vb{v}^{1}   &{}={}& \vb{v}^{1/2}+\tfrac{1}{2}\vb{a}^{1}\,\Delta t_{\rm a}\nonumber \\
\quad& &{}={}& \vb{v}^0+\tfrac{1}{2}(\vb{a}^{0}+\vb{a}^{1})\,\Delta t_{\rm a}
\end{alignat}
where superscripts $0$ and $1$ denote quantities evaluated at the beginning and end of the timestep, respectively.

Particle leaves on coarser timestep levels become passive after completing their first kick. During this passive period, the drift operator is executed incrementally by accumulating position updates over smaller substeps. The total displacement across the full active interval is given by
\begin{equation}\label{eqn:driftSum}
\vb{x}^1 = \vb{x}^0 + \sum_j \vb{v}^{1/2} \, \Delta t_{\rm min}^{(j)}, 
\end{equation}
where each $\Delta t_{\rm min}^{(j)}$ denotes the smallest local-rank timestep applied during subinterval $j$, and $\sum_j \Delta t_{\rm min}^{(j)} = \Delta t_{\rm a}$. Once the accumulated drift spans the entire active interval, the second kick is applied to complete the KDK cycle.

Following \citet{saitoh2009}, we apply a timestep limiter that dynamically refines the interval assigned to each leaf when local conditions --- such as abrupt subgrid injections or rapid force variations --- demand higher temporal resolution. When the Saitoh limiter is triggered, certain particle leaves may transition to a shorter timestep partway through their KDK sequence. To maintain consistency in the velocity state, the partially completed half-kick is adjusted to match the refined interval.

Specifically, if the timestep assigned to a leaf is refined from an initial $\Delta t_{\rm a}$ to a shorter $\Delta t_{\rm b} < \Delta t_{\rm a}$ midway through a KDK cycle, the first half-kick is corrected as
\begin{equation}
    \vb{v}^{1/2} = \vb{v}^0 + \tfrac{1}{2} \vb{a}^0\, \Delta t_{\rm a}
    \Longrightarrow
    \vb{v}^{1/2} = \vb{v}^0 + \tfrac{1}{2} \vb{a}^0\, \Delta t_{\rm b}.
\end{equation}
This adjustment re-synchronizes the stored velocity with the smaller refined timestep, where superscripts $0$ and $1$ denote the start and end of the refined interval. After the drift step completes over the shorter interval $\Delta t_{\rm b}$, the second half-kick is applied using the updated local acceleration $\vb{a}^1$,
\begin{equation}
    \vb{v}^1 = \vb{v}^{1/2} + \tfrac{1}{2} \vb{a}^1\, \Delta t_{\rm b} = \vb{v}^0 +\tfrac{1}{2}(\vb{a}^0+\vb{a}^1)\Delta t_{\rm b}.
\end{equation}

If the updated acceleration necessitates an even shorter, CFL-constrained timestep $\Delta t_{\rm c}$ --- for example following a strong stochastic subgrid event --- the final kick is further refined using a piecewise-constant approximation:
\begin{equation}
    \vb{v}^1 = \vb{v}^0 + \vb{a}^0\,(\Delta t_{\rm b} - \Delta t_{\rm c})
               + \tfrac{1}{2}(\vb{a}^0 + \vb{a}^1)\,\Delta t_{\rm c}.
\end{equation}
In practice, this correction advances particles under constant acceleration $\vb{a}^0$ until the remaining portion of the step is completed with a standard KDK update using $\vb{a}^0$ and $\vb{a}^1$ over the refined interval $\Delta t_{\rm c}$. This ensures that the final velocity remains consistent with the resolved timestep hierarchy, preventing spurious excursions during completion of $\Delta t_{\rm b}$ and allowing the leaf to reduce its timestep level to $\Delta t_{\rm c}$ in the next KDK cycle. The approach mirrors the timestep adjustments described by \citet{durier2011} and maintains robustness under abrupt subgrid refinements.

A final modification concerns evaluation of the drift operation described in \cref{eqn:driftSum} for passive particles. Following \citetalias{frontiere2022simulating}, we adopt a piecewise-parabolic trajectory consistent with the constant-acceleration approximation used for passive motion in the KDK hierarchy. Instead of drifting with a fixed half-step velocity $\vb{v}^{1/2}$, a local half-step velocity $\vb{v}^{1/2}_{j}$ is computed for each drift subinterval, ensuring that the particle follows the parabolic path implied by constant acceleration rather than a linear trajectory across the full active step:
\begin{equation}
    \vb{v}^{1/2}_{j} = \vb{v}^{1/2} + \vb{a}^0\,\Big[\,\Delta t_{\rm elapse} + \tfrac{1}{2}(\Delta t_{\rm min}^{(j)} - \Delta t_{\rm a})\Big],
\end{equation}
where $\Delta t_{\rm elapse} \equiv \sum_{k=1}^{j-1}\Delta t_{\rm min}^{(k)}$ is the elapsed time since the start of the active interval up to (but not including) substep $j$. Applying this correction, the total drift displacement in \cref{eqn:driftSum} becomes
\begin{equation}
    \vb{x}^1 = \vb{x}^0 + \sum_j \vb{v}^{1/2}_{j} \, \Delta t_{\rm min}^{(j)},
\end{equation}
ensuring that passive drifts follow a trajectory consistent with piecewise-constant acceleration.

This drift modification does not alter the stored $\vb{v}^{1/2}$ state but is important for maintaining accurate passive positions during limiter-triggered mid-step refinements and for ensuring consistent neighbor interactions between active and passive particles at smaller timesteps. If no refinements occur, the final integrated position at the end of the active interval is equivalent to that obtained from a drift computed with a constant $\vb{v}^{1/2}$.

Since all particles drift on the smallest timestep interval --- and the integration hierarchy can extend to much deeper levels under subgrid physics --- single-precision arithmetic may become insufficient to accumulate small displacements accurately over many drift steps. To maintain long-term numerical stability, position updates are accumulated in double precision on the GPU, while force kernels and host-side particle states remain in single precision. This strategy preserves performance for the dominant force calculations and ensures accurate trajectories with only a modest increase in GPU memory footprint. Since \CRKHACC already evolves sub-volumes entirely on the GPU with numerous derived state arrays allocated, this additional storage cost is negligible in practice.

\section{Self-Similarity with Cooling}\label{sec:selfsimilarcool}

\begin{figure*}[htp]
    \centering
    \includegraphics[width=0.32\textwidth]{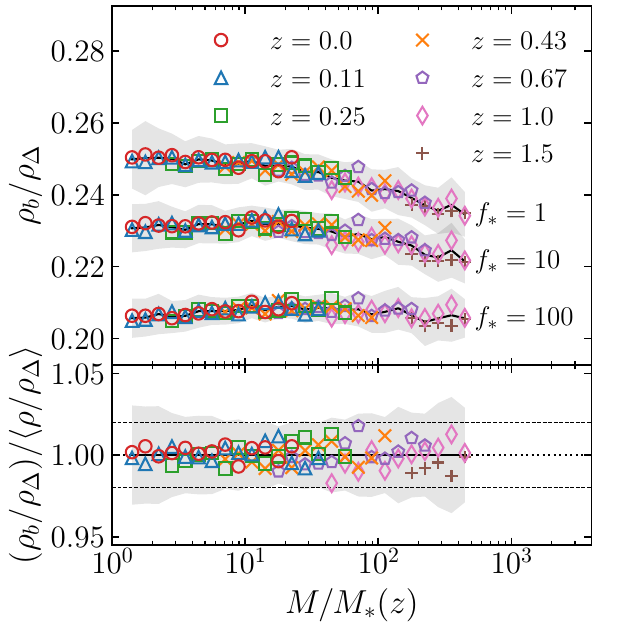}
    \includegraphics[width=0.32\textwidth]{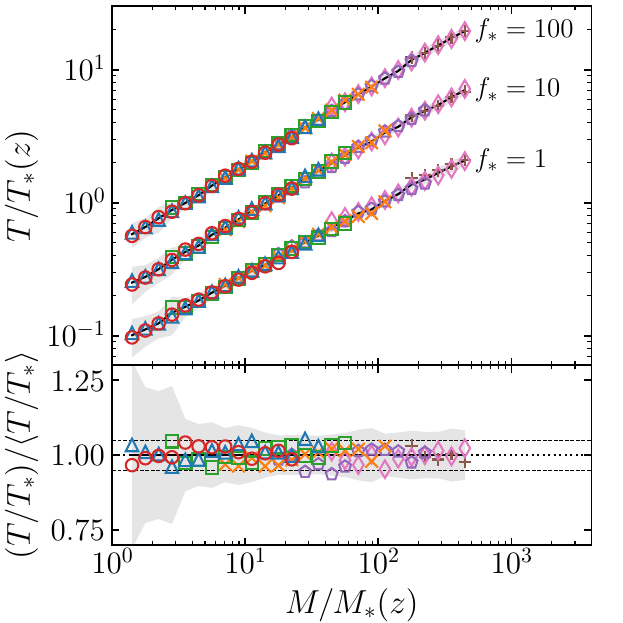}
    \includegraphics[width=0.32\textwidth]{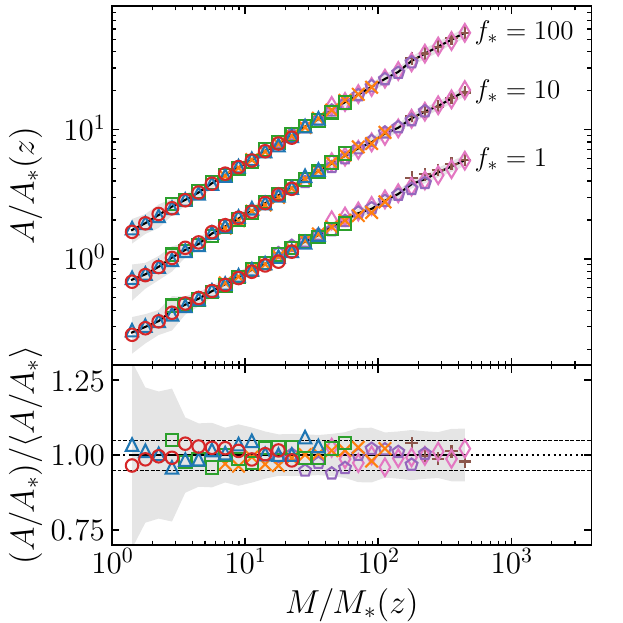}
    \caption{Baryon density (left), temperature (center), and entropy (right) as functions of scaled halo mass for three scale-free radiative simulations parameterized by the cooling amplitude $f_*$. Density is normalized to the spherical overdensity threshold $\rho_\Delta = 200 \bar{\rho}$, while mass, temperature, and entropy are normalized to the redshift-dependent nonlinear scales defined in \cref{eq:ss-scalingrelations}. We plot the entropic function $A$, related to the entropy by $S \propto \ln{A}$.  Each marker denotes the mean value for halos in a given redshift and mass bin containing at least ten halos; only bins with contributions from more than one redshift sample are shown. The solid black line traces the overall mean across all redshifts, and the shaded gray band indicates the corresponding standard deviation. The bottom panels show the deviation of all measurements from the mean curve. Horizontal dashed lines mark $\pm2\%$, $\pm5\%$, and $\pm5\%$ spreads for the left, center, and right panels, respectively. To reduce visual overlap in the top panel of the density plot, the $f_* = 10$ and $f_* = 1$ curves are shifted upward by 0.01 and 0.02, respectively. \\}
    \label{fig:selfsimilar}
\end{figure*}

The validation of cosmological simulations is complicated by the nonlinear nature of gravitational collapse, which precludes straightforward comparison to analytic predictions. An effective strategy to address this challenge is to perform simulations in which the only physical scale is the amplitude of density fluctuations. In this way, structure formation evolves in a temporally self-similar manner, enabling validation of the simulation solver even in the nonlinear regime.

We employed this method in \citetalias{frontiere2022simulating} to validate the non-radiative \CRKHACC\ hydrodynamic solver. The subgrid models introduced in this paper incorporate a multitude of physical scales, making them incompatible with a scale-free framework. However, \citet{owen1998cooling} demonstrated that self-similar evolution can still be preserved under radiative cooling, provided that the cooling function follows a power law whose slope is a specific function of the spectral index of density perturbations. While this cooling function is not physically realistic, it serves as a useful idealization for validating the radiative solver in the nonlinear regime. 

In what follows, we assume an Einstein--de Sitter (EdS) cosmology with $(\Omega_c, \Omega_b, h) = (0.8, 0.2, 0.5)$ and an initial power spectrum $P(k) = A_0 k^{n_s}$, where the spectral index is $n_s = -2$ (similar to $\Lambda$CDM on the scales considered), and the normalization $A_0$ is set by requiring $\sigma_8 = 0.5$. The cooling function takes the form
\begin{equation}
\frac{\Lambda(u)}{n_H^2} = A_1 u^\beta,
\end{equation}
with self-similarity enforced by requiring that the cooling time of a halo collapsing at the nonlinear scale is a fixed fraction, $f_*$, of the Hubble time. This condition constrains the exponent $\beta$ to depend on the spectral index as
\begin{equation}
\beta = \frac{3}{2} \cdot \frac{3 + n_s}{1 - n_s} + 1,
\end{equation}
as derived in \citet{owen1998cooling}. For our choice of \mbox{$n_s = -2$}, self-similarity requires a power-law slope of $\beta = 1.5$. 

The normalization
\begin{equation}
A_1 = \frac{1}{f_*} \frac{H_0}{\rho_*} \left[ \frac{T_* k_B}{\mu m_H (\gamma-1)} \right]^{1-\beta}\!\!\!\!\!\!\!,
\end{equation}
is set by the free parameter $f_*$. We use asterisks to denote quantities evaluated at the nonlinear collapse scale, which --- as shown in \citetalias{frontiere2022simulating} --- is characterized by the following mass, density, radius, temperature, and entropy:
\begin{align}
M_* &= 4.05\times 10^{11} a^6\ h^{-1}\,M_\odot, \nonumber\\
\rho_* &= 5.55\times 10^{13} a^{-3}\ h^2\, M_\odot \,{\rm Mpc}^{-3},\nonumber\\
R_* &= 0.12 a^3\ h^{-1}\,{\rm Mpc},\nonumber\\
T_* &= 5.19\times 10^5 a^3\ \,{\rm K},\nonumber\\
S_* &= -7.94 + 5{\rm ln}(a).
\label{eq:ss-scalingrelations}
\end{align}

We run three simulations, each containing $N = 2 \times 512^3$ dark matter and baryonic particles in a box of width $L_\mathrm{box} = \unit{40}{\Mpch}$, with mass resolutions of $m_{\rm dm} = \unit{1.06\times10^8}{\massh}$ and $m_{\rm b} = \unit{2.65\times10^7}{\massh}$.  The three runs differ only in the normalization of the cooling law, with the cooling time set to $f_* = {1, 10, 100}$ times the Hubble time at the nonlinear collapse scale. Each simulation evolves from $z = 200$ to $0$ with a gravitational softening length of $\epsilon_{\rm soft} = \unit{4.88}{\kpch}$. These tests include only radiative gas cooling, with all prescriptions for star formation, winds, and AGN feedback disabled.

We identify spherical overdensity (SO) halos using a threshold of $\Delta_c = 200$ relative to the critical density and compare halos across redshift to assess temporal self-similarity. Only halos with mass $M_{200c} \geq \unit{3\times10^{11}}{\massh}$ are included, as this threshold was shown by \citetalias{frontiere2022simulating} to avoid artificial mass segregation effects. To minimize scatter from major mergers, we further restrict the sample to relaxed halos where the offset between the potential minimum and the center of mass is less than $7\%$ of the SO radius, $R_{200c}$ \citep{child2018}.

The results from the three simulations are summarized in Figure~\ref{fig:selfsimilar}, which shows halo baryon density, temperature, and entropy as functions of scaled mass across two decades, for redshifts $0 \leq z \leq 1.5$. The density refers specifically to the baryonic component, calculated as $\rho_b = 3 M_{200c,b}/(4\pi R_{200c}^3)$, where $M_{200c,b}$ is the total baryonic mass within radius $R_{200c}$. For temperature, we compute the mass-weighted average over all baryonic particles in each halo. The two quantities are combined into the entropy $S = {\rm ln}(T/\rho_b^{2/3})$. Each marker in the figure denotes the mean value for halos within a given $M/M_*$ (where $M$ denotes $M_{200c}$) and redshift bin. 

To assess self-similarity, we compare halos at fixed $M/M_*$ across different redshifts. For example, at $M/M_* = 20$, the sample includes five redshifts between $z = 0$ and $z = 0.67$. Since $M_* \propto a^6$, the corresponding physical halo masses in this bin differ by more than an order of magnitude. Despite this variation, the density, temperature, and entropy agree to within a few percent across redshifts in this bin. This low level of scatter is consistent across the full scaled mass range for all three radiative simulations, indicating a strong degree of self-similarity. As expected, the simulation with more efficient cooling (i.e., lower $f_*$) exhibits higher baryon density, lower temperature, and correspondingly lower entropy.

\section{Cluster Code Comparison}\label{sec:nifty}

\begin{figure*}[htp]
\centering
\includegraphics[width=0.32\textwidth]{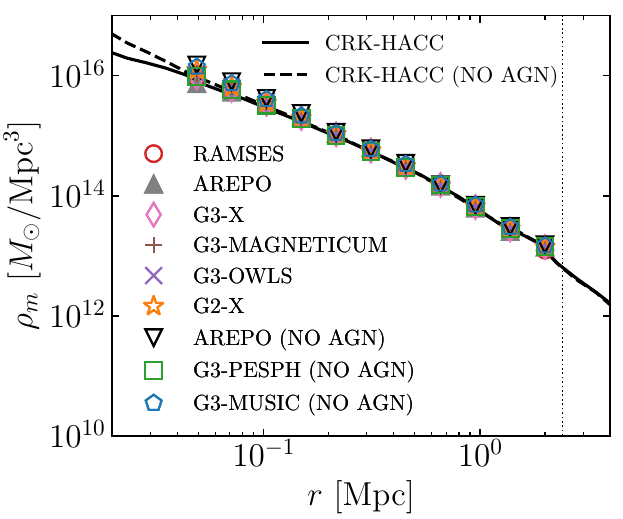}
\includegraphics[width=0.32\textwidth]{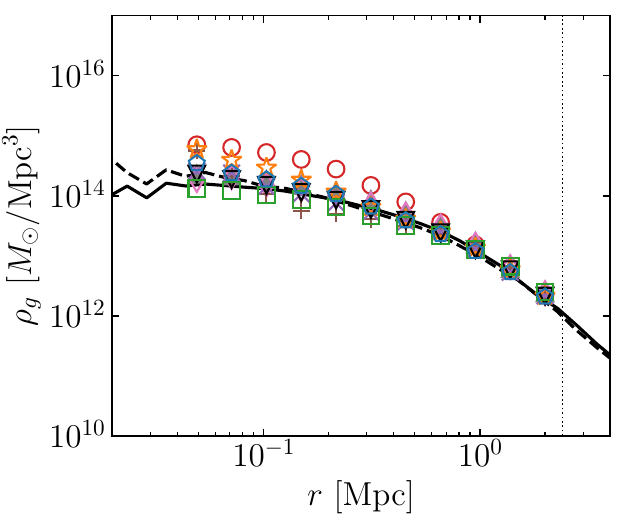}
\includegraphics[width=0.32\textwidth]{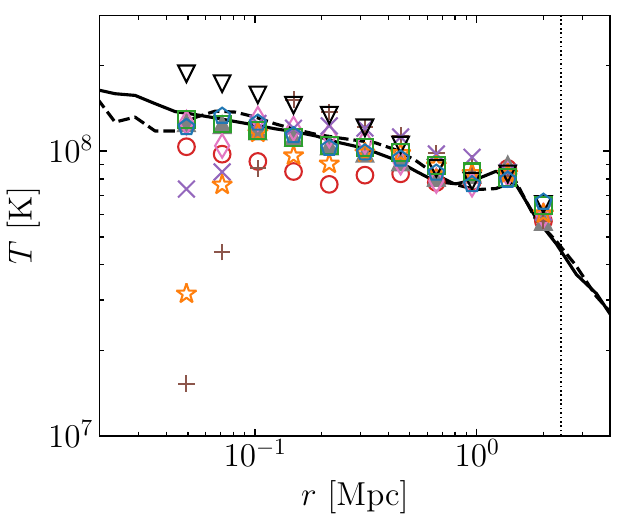}
\vspace{0.5cm}
\includegraphics[width=0.32\textwidth]{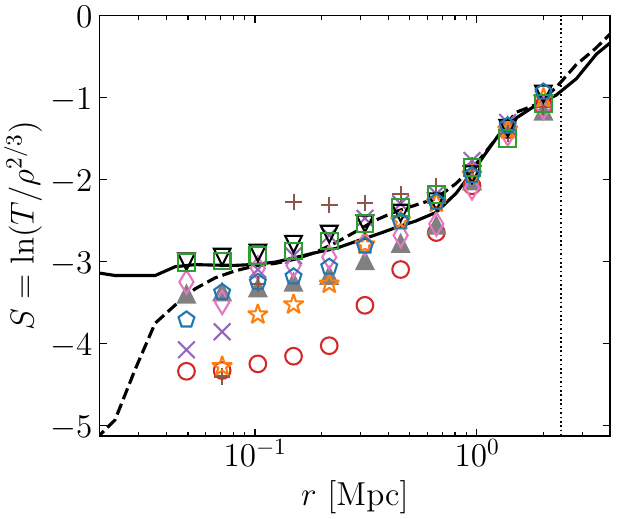}
\includegraphics[width=0.32\textwidth]{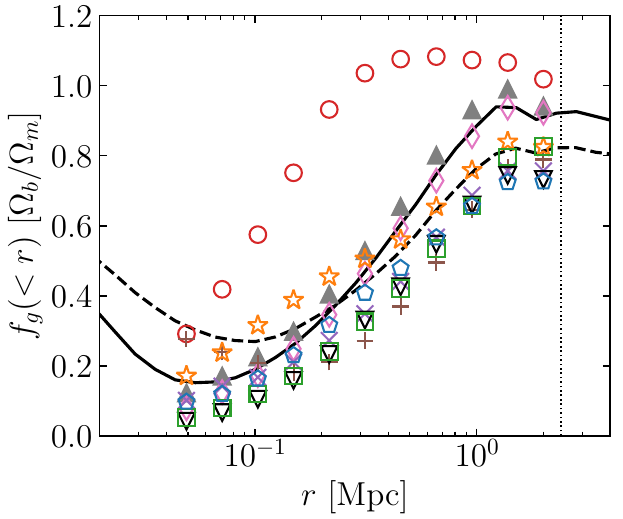}
\includegraphics[width=0.32\textwidth]{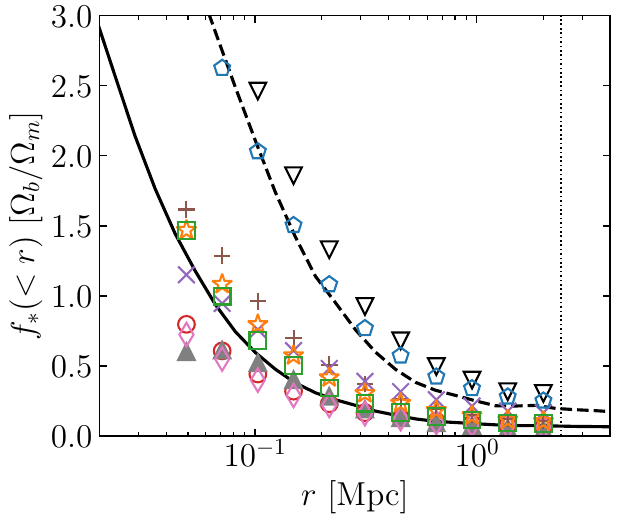}
\caption{Clockwise from top left: Radial profiles of the total matter density, gas density, temperature, stellar fraction, gas fraction, and entropy for the nIFTy cluster at $z = 0$. Solid and dashed lines represent the \CRKHACC subgrid runs with and without AGN feedback, respectively. Symbols show results from the simulations presented in \citet{sembolini2016niftyradiative}. The vertical dotted line in each panel marks $R_{200c} = \unit{2.41}{Mpc}$, the cluster radius in the \CRKHACC run with full subgrid physics. \\}
\label{fig:nifty}
\end{figure*}

We present results from \CRKHACC subgrid runs initialized with the nIFTy dataset \citep{sembolini2016nifty}. The nIFTy initial conditions were generated using a zoom-in technique \citep{klypin2001} to extract a massive galaxy cluster within the large-volume Multidark simulation \citep{prada2012}. The adopted cosmology is WMAP-7+BAO+SNI \citep{Komatsu_2011}, with parameters ($\Omega_m$, $\Omega_b$, $\Omega_\Lambda$, $n_s$, $\sigma_8$, $h$) = (0.27, 0.0469, 0.73, 0.95, 0.82, 0.7). In the zoomed region, the highest-resolution particle masses are ${m_{\rm dm} = \unit{9.01 \times 10^8}{\massh}}$ for dark matter and ${m_{\rm b} = \unit{1.9 \times 10^8}{\massh}}$ for baryons. 

The nIFTy initial conditions were first used by \citet{sembolini2016nifty} to compare various cosmological codes employing non-radiative hydrodynamic solvers. We extended this comparison to the non-radiative \CRKHACC solver in \citetalias{frontiere2022simulating}, where we found excellent agreement with both Eulerian and modern SPH codes. This agreement is particularly notable since modern SPH methods typically require artificial conductivity to match Eulerian codes, but no such treatment is necessary within the \CRKSPH framework. 

Here, we further extend our comparison to the radiative hydrodynamic simulations presented in \citet{sembolini2016niftyradiative}, which include a broad range of methods: grid-based (\smaller{RAMSES}; \citealt{teyssier2002}), hybrid moving-mesh (\smaller{AREPO}; \citealt{Springel2010}), modern SPH (\smaller{G3-X}; \citealt{beck2016}; \smaller{G3-PESPH}; \citealt{huang2019}; \smaller{G3-MAGNETICUM}; \citealt{hirschmann2014}), and traditional SPH codes (\smaller{G3-MUSIC}; \citealt{sembolini2013}; \smaller{G3-OWLS}; \citealt{schaye2010owls}; \smaller{G2-X}; \citealt{pike2014}). In addition, these codes implement a wide variety of subgrid models for cooling, star formation, wind, and AGN feedback. To compare against this suite, we perform two \CRKHACC runs using the nIFTy initial conditions: one including the full set of subgrid models described in this work, and one omitting AGN feedback. Conveniently, the mass resolution in the nIFTy zoom-in region closely matches that used for calibrating our subgrid models (Section~\ref{subsec:calib}), allowing us to directly apply our fiducial parameters.

We find the main cluster to have a mass $M_{200c}$ = $1.59\times10^{15}\,M_\odot$ ($1.61\times10^{15}\,M_\odot$) and radius $R_{200c}$ = $\unit{2.41}{Mpc}$ ($\unit{2.42}{Mpc}$) at $z = 0$ in the full subgrid (non-AGN) simulation. These values are consistent with the range reported by \citet{sembolini2016niftyradiative}, with the slight increase in mass in the non-AGN run also observed in the two \smaller{AREPO} simulations. Radial profiles of the main cluster from our two runs are compared to the full suite of radiative nIFTy simulations in Figure~\ref{fig:nifty}.

As noted by \citet{sembolini2016niftyradiative}, systematic differences between Eulerian and traditional SPH codes are largely washed out once subgrid physics is included, particularly for this massive, unrelaxed cluster. Instead, all codes consistently produce entropy cores, regardless of the presence of AGN feedback; a trend reproduced in our two \CRKHACC runs. The primary effect of AGN feedback is the suppression of star formation on small scales, along with a slightly steeper gas fraction profile. Overall, \CRKHACC produces nIFTy cluster profiles in broad agreement with those from current state-of-the-art hydrodynamic codes.

\section{Generated Cloudy Tables}\label{sec:cloudy_tab}
The inputs to \CLOUDY specify a gas of hydrogen number density $n_{\rm H}~[\mathrm{cm}^{-3}]$, temperature $T~[\text{K}]$, helium abundance $n_{\rm He}/n_{\rm H}$, individual metal abundances $n_{Z_i}/n_{\rm H}$, and redshift $z$. We provide \CLOUDY with the input radiation field from either \cite{faucher2020} or optionally \cite{haardt2012} (hereafter \citetalias{faucher2020} and \citetalias{haardt2012}), modeling a uniform time-dependent UV background. As was done in \cite{vogelsberger2013model}, to incorporate self-shielding effects for high-density gas --- where the optically thin approximation is invalid --- we attenuate the background radiation field using a density- and redshift-dependent fit from Appendix A in \cite{rahmati2013}. Namely, the correction amplitude $A_\text{corr}$, follows:
\begin{equation}\label{eqn:SS}
\resizebox{0.85\linewidth}{!}{$
    A_\text{corr}=(1-f) \left[1+\left(\frac{n_{\rm H}}{n_{\rm 0}}\right)^\beta\right]^{\alpha_1} \!\!\!\!\!\! + f\left[1+\left(\frac{n_{\rm H}}{n_{\rm 0}}\right)\right]^{\alpha_2}
$}
\end{equation}
where coefficients $(f,n_{\rm 0},\alpha_1,\alpha_2,\beta)$ are redshift-dependent (specified in Table A1 in \citealp{rahmati2013}), and linearly interpolated. \Cref{eqn:SS} asymptotically tends to unity for low-density gas, i.e. no correction, and rapidly decreases toward zero at high density. We do not apply any attenuation for redshift $z > 6$.

Although this approach does not rigorously simulate radiative transfer, it self-consistently modifies our cooling treatment using an approximated attenuation that was measured by post-processing cosmological simulations using the radiative transfer code \smaller{TRAPHIC} \citep{pawlik2008traphic,pawlik2011traphic}.

To tabulate radiative cooling rates, we ran a suite of \CLOUDY simulations across a broad parameter space encompassing conditions relevant to cosmological gas. Specifically, we sampled hydrogen number density $n_{\rm H}$ over the range $\log_{10}(n_{\rm H}\, [\mathrm{cm}^{-3}]) \in [-8, +5]$ using 200 points, and temperature $\log_{10}(T\, [\mathrm{K}]) \in [0, 9]$ with 226 points. Metallicity was parameterized using the solar-scaled factor $R$ (defined in \cref{eqn:nzi}), which was sampled linearly over the range $R \in [0, 10]$ with 50 values. The helium composition was parameterized using the helium mass fraction assuming zero metallicity $Y_{\rm Z=0}$, sampled linearly at 5 values in the range $Y_{\rm Z=0} \in [0.25, 0.37]$. This quantity is mapped to the helium abundance required by \CLOUDY via the relation:
\begin{equation}
\frac{n_{\rm He}}{n_{\rm H}} = \frac{Y_{\rm Z=0}}{1 - Y_{\rm Z=0}} \cdot \frac{m_{\rm H}}{m_{\rm He}},
\label{eq:nhe}
\end{equation}
as derived from \cref{eqn:Yfrac}. The advantage of parameterizing the table by $Y_{\rm Z=0}$ is that it is metallicity independent, as opposed to directly using the total helium fraction $Y$. 

We generated separate tables for redshifts linearly sampled in scale factor using 50 values between $z = 0$ and $z = 15$ for the \citetalias{haardt2012} model, and capped at $z = 7.7$ for the \citetalias{faucher2020} model due to discontinuities in the cooling rates at higher redshift. At runtime, for any particle with properties $n_{{\rm H},i},\,T_i,X_i,\,Y_i,\,Z_i,\,z_i$, the corresponding $R_i$ and $Y_{{\rm Z=0},i}$ values are computed using \cref{eqn:R} and the transformation
\[
\frac{Y_i}{X_i} = \frac{Y_{{\rm Z=0},i}}{1 - Y_{{\rm Z=0},i}}.
\]
The simulation then identifies the two tabular redshifts $z_1$ and $z_2$ that bracket $z_i$, and performs linear interpolation across all five parameters --- $n_{\rm H}$, $T$, $R$, $Y_{\rm Z=0}$, and $z$ --- to compute the local cooling function $\Lambda_i$.

In practice, simulation particles evolve internal energy $u\propto T/\mu$, rather than temperature. Thus, for consistency we tabulate and interpolate \CLOUDY cooling rates as a function of $T/\mu$.  The mean molecular weight (per particle) is defined as,
\begin{align}
    \mu = \frac{\bar{m}}{m_{\rm H}} &= \frac{1}{m_{\rm H}}\frac{n_{\rm H} m_{\rm H} + n_{\rm He} m_{\rm He} + \sum_i n_{Z_i}m_{Z_i}+n_{\rm e}m_{\rm e}}{n_{\rm H} + n_{\rm He} + \sum_i n_{Z_i} + n_{\rm e} } \nonumber\\
    &\approx \frac{1}{X + Y\frac{m_{\rm H}}{m_{\rm He}} + Z\sum_i\frac{m_{\rm H}}{m_{Z_i}}+X\frac{n_{\rm e}}{n_{\rm H}}}
\end{align}
where $\bar{m}$ is the mean mass of the gas, and we ignore the electron mass contribution ($n_{\rm e}m_{\rm e}$).

In addition to storing the tabulated cooling rates $\Lambda$, we also save several \CLOUDY output quantities that are used in analysis, including values relevant for X-ray and Lyman-$\alpha$ observables. These include the neutral hydrogen abundance $n_{\rm HI}$, the free electron number density $n_{\rm e}$, the mean molecular weight $\mu$, and the volumetric emissivity $J_{E_{\rm min}-E_{\rm max}}$ [${\rm erg\, cm^{-3} \,s^{-1}}$], which is used to compute luminosity measurements. The high-bandwidth memory and parallel compute capacity of the GPU enable efficient tabular lookups and interpolation across all \CLOUDY outputs, making the five-dimensional tables well-suited for on-the-fly evaluation of cooling and emission properties.

Following \citet{braspenning2024}, the emissivity is calculated by integrating the \CLOUDY-provided continuous emission coefficient $\epsilon_\nu$ over a spectral energy interval \( E_{\rm min} \) to \( E_{\rm max} \):
\begin{equation}
    J_{E_{\rm min}-E_{\rm max}} = \int_{E_{\rm min}}^{E_{\rm max}} \frac{\epsilon_\nu}{E_\nu}\, dE_\nu,
\label{eq:JE}
\end{equation}
where \( E_\nu \) is the midpoint spectral energy of each frequency bin. The integrated energy bands we store include the bolometric range (0.5-10\,keV), the ROSAT band (0.5-2.0\,keV), and the low (0.2-2.3\,keV) and high (2.3-8.0\,keV) eROSITA bands. These bands are defined in the rest frame at redshift zero and are scaled by a factor of $1+z$ in the integral, with \( E_{\rm min}(z) = (1+z)\,E_{\rm min}^{z=0} \) and \( E_{\rm max}(z) = (1+z)\,E_{\rm max}^{z=0} \). 

\begin{table*}[ht!]
\centering
\caption{Example \CLOUDY parameter file}
\label{tab:cloudy}
\begin{tabular}{p{0.40\textwidth} p{0.55\textwidth}}
\hline
\hline
\textbf{Input Command} & \textbf{Notes} \\
\hline
\texttt{hden LOGNH}  & Sets the log hydrogen number density, \texttt{LOGNH}, sampled at 200 points in $\log_{10}(n_{\rm H}\, [\mathrm{cm}^{-3}]) \in [-8, +5]$. \\
\texttt{element abundance helium NHE linear} & Sets the helium abundance, \texttt{NHE}, computed using \cref{eq:nhe}, and sampled at 5 points in $Y_{\rm Z=0} \in [0.25, 0.37]$. \\
\texttt{metals R linear} & Sets the metal abundance with \texttt{R} following equation \cref{eqn:nzi}, sampled at 50 points in $R \in [0, 10]$. \\
\texttt{constant temperature 0.0 vary} \newline \texttt{grid 0.0, 9.0, 0.04 ncpus 1} & Sets the temperature grid from 226 points in $\log_{10}(T\, [\mathrm{K}]) \in [0, 9]$. \\
\texttt{CMB redshift Z} & Sets the CMB at redshift \texttt{Z}, sampled at 50 points in $a = 1/(1+z)$ from $z=0$ to $7.7\ (15)$ for the \citetalias{faucher2020} (\citetalias{haardt2012}) UV model.\\
\texttt{table HM12 redshift Z factor LOGA} & Sets the UV background at redshift \texttt{Z}, with the log attenuation factor, \texttt{LOGA}, computed using \cref{eqn:SS}. The \citetalias{faucher2020} model is applied by replacing the \texttt{HM12} source file, \texttt{hm12\_galaxy.ascii}, with the \citetalias{faucher2020} data.\\
\texttt{stop zone 1} \newline \texttt{iterate to convergence} & 
Sets the stopping condition and iterative convergence for single-zone models. \\
\texttt{set continuum resolution 1.0} & Sets the frequency resolution to its default value.\\
\texttt{save grid last "output\_grid"} & Saves the failure status of the run.\\
\texttt{save cooling last "output\_cooling"} & Saves the cooling rate.\\
\texttt{save overview last "output\_overview"} & Saves the heating rate, neutral fraction, and free electron abundance.\\
\texttt{save diffuse continuum last "output\_diffuse"} & Saves the continuous emission coefficient used in \cref{eq:JE}. \\
\texttt{save xspec atable lines last "output\_lines"} & Saves the XSPEC line emission additive table. \\
\texttt{save xspec atable reflected lines last "output\_reflect\_lines"} & Saves the XSPEC reflected line emission additive table. \\
\texttt{save xspec atable diffuse continuum last "output\_contin\_diffuse"} & Saves the XSPEC diffuse continuum emission additive table. \\
\texttt{save xspec atable reflected diffuse continuum last "output\_reflect\_diffuse"} & Saves the XSPEC reflected diffuse continuum emission additive table. \\
\texttt{save xspec mtable last "output\_mtable"} & Saves the XSPEC multiplicative table. \\
\texttt{save radius last "output\_rad"} & Saves the cloud radius. \\
\hline \\[6pt]
\end{tabular}
\end{table*}

\cref{tab:cloudy} provides an example \CLOUDY parameter file used to sample a fixed $n_{{\rm H},i},\,X_i,\,Y_i,\,Z_i,\,z_i$ for all temperature values. Occasionally, \CLOUDY fails to converge on an answer, typically at low redshift and high density, in which case we linearly interpolate an answer from the nearest converged sample points.

For completeness, we also store individual components of the emergent emission spectrum, which outputs direct and reflected contributions to both the diffuse continuum and line emission in XSPEC-compatible FITS format.\footnote{XSPEC is a spectral fitting software package widely used in X-ray astronomy; see \url{https://heasarc.gsfc.nasa.gov/xanadu/xspec/}.} These tables report photon fluxes in units of \([{\rm photons\, s^{-1}\, cm^{-2}\, bin^{-1}}]\), and offer an alternative, photon-based format suitable for XSPEC-style modeling and mock observation pipelines. The tables are additive, and one can reconstruct the total volume emissivity by summing all components, dividing by the \CLOUDY radius output, and multiplying by the bin-centered photon energy and bin width.

\section{Stellar Enrichment Integration}\label{app:Enrich}

We present here formulae for the integrated mass loss from supernovae and stellar winds, based on the rates and yields provided in \citetalias{hopkins2023fire}. These integrated expressions are convenient for two reasons. First, they ensure that cumulative mass loss is independent of the timestep size. Second, in cosmological simulations with relatively coarse timesteps, on the order of $\unit{1\text{--}10}{\Myr}$, they avoid significant errors in cumulative mass loss that would otherwise result from evaluating the rates instantaneously at the start and end of each timestep. In what follows, we measure time in Myr and compute the mass loss over time $t_0$ to $t_1 = t_0 + \Delta t$. Rates are expressed in Gyr$^{-1}$ per unit solar mass.

\subsection{Supernovae Mass Loss}\label{app:SNloss}

We begin with the SN mass loss obtained by integrating \cref{eqn:snint}. For Type Ia SN, the ejecta mass is fixed at $M^{\rm Ia} = 1.4\,M_\odot$ and is composed entirely of metals. The SN Ia rate follows
\begin{equation}
    R^{\rm Ia}(t) = \begin{cases}
        0, & t < t_r;\\
        a (t/t_r)^c, & t \geq t_r;
    \end{cases}
\end{equation}
where $a = 8.3\times10^{-3}$, $c = -1.1$, and $t_r = \unit{44}{\Myr}$. The integrated mass loss evaluates to
\begin{equation}\label{eqn:mSNIA}
\frac{\Delta M_{T,Z}^{\rm Ia}}{M_*} = M^{\rm Ia} \left(\frac{a}{c+1}\left. \frac{t^{c+1}}{t_r^c} \right|^{t_1}_{{\rm max}(t_0,t_r)} \right),
\end{equation}
for $t_1 \geq t_r$ and 0 otherwise. As stated above, $\Delta M_Y^{\rm Ia} = 0$.

For CC SN, the ejecta mass has time dependence:
\begin{equation}
    M_{\rm CC}(t) = \begin{cases}
        M_0(t/t_m)^{c_{m_1}}, & t \leq t_m;\\
        M_0(t/t_m)^{c_{m_2}}, & t > t_m;
    \end{cases}
\label{eqn:mcc}
\end{equation}
where $M_0 = \unit{10}{M_\odot}$, $t_m = 6.5$, and ($c_{m_1}$, $c_{m_2}$) = ($-2.22$, $-0.267$). The CC SN rate follows:
\begin{equation}
    R_{\rm CC}(t) = \begin{cases}
        0, & t \leq t_{r_1}\ {\rm or}\ t > t_{r_3};\\
        a_{r_1} (t/t_{r_1})^{c_{r_1}}, & t_{r_1} < t \leq t_{r_2};\\
        a_{r_2} (t/t_{r_2})^{c_{r_2}}, & t_{r_2} < t \leq t_{r_3};
    \end{cases}
\label{eqn:rcc}
\end{equation}
where ($a_{r_1}$, $a_{r_2}$, $a_{r_3}$) = (0.39, 0.51, 0.18), ($t_{r_1}$, $t_{r_2}$, $t_{r_3}$) = (3.7, 7, 44), and $c_{r_j} \equiv {\rm ln}(a_{r_{j+1}}/a_{r_j})/{\rm ln}(t_{r_{j+1}}/t_{r_j})$. The integrated total mass loss becomes
\begin{align}
\frac{\Delta M_T^{\rm CC}}{M_*} =  &\ F_{11}^T(t_{r_1},t_m) + F_{21}^T(t_m, t_{r_2}) \notag \\
+ &\ F_{22}^T(t_{r_2}, t_{r_3}),
\end{align}
where we define the function
\begin{equation}
F_{ij}^T(t_l,t_u) \equiv \frac{M_0 a_{r_j}}{\alpha_{ij}+1} \left.\frac{t^{\alpha_{ij}+1}}{t_m^{c_{m_i}}t_{r_j}^{c_{r_j}}}\right|_{[t_l,t_0,t_u]}^{[t_l,t_1,t_u]},
\end{equation}
with coefficient $\alpha_{ij} \equiv c_{m_i}+c_{r_j}$ and the notation $[t_l, t, t_u] \equiv {\rm min}[{\rm max}(t_l,t),t_u]$ conveys that $t$ is bounded within $[t_l, t_u]$.

The \FIIIRE helium yield is modeled using piecewise power-laws:
\begin{equation}
    f_Y(t) = \begin{cases}
        0, & t \leq t_{f_1}\ {\rm or}\ t > t_{f_5};\\
        a_{f_1}^Y (t/t_{f_1})^{c_{f_1}^Y}, & t_{f_1} < t \leq t_{f_2};\\
        \dots \\
        a_{f_4}^Y (t/t_{f_4})^{c_{f_4}^Y}, & t_{f_4} < t \leq t_{f_5};
    \end{cases}
\label{eqn:fcc}
\end{equation}
where ($a_{f_1}^Y$, $a_{f_2}^Y$, $a_{f_3}^Y$, $a_{f_4}^Y$, $a_{f_5}^Y$) = (0.461, 0.33, 0.358, 0.365, 0.359), ($t_{f_1}$, $t_{f_2}$, $t_{f_3}$, $t_{f_4}$, $t_{f_5}$) = (3.7, 8, 18, 30, 44), and $c_{f_k}^Y \equiv {\rm ln}(a_{f_{k+1}}^Y/a_{f_k}^Y)/{\rm ln}(t_{f_{k+1}}/t_{f_k})$. The integrated helium mass loss evaluates to
\begin{align}
\Delta M_Y^{\rm CC}/M_* = &\ F_{111}^Y(t_{r_1},t_m) + F_{211}^Y(t_m, t_{r_2}) \notag \\
+ &\ F_{221}^Y(t_{r_2}, t_{f_2}) + F_{222}^Y(t_{f_2}, t_{f_3}) \notag \\
+ &\ F_{223}^Y(t_{f_3}, t_{f_4}) + F_{224}^Y(t_{f_4}, t_{f_5}),
\label{eqn:mycc}
\end{align}
where
\begin{equation}
F_{ijk}^Y(t_l,t_u) \equiv \frac{M_0 a_{r_j} a_{f_k}^Y}{\alpha_{ijk}^Y+1} \left.\frac{t^{\alpha_{ijk}^Y+1}}{t_m^{c_{m_i}}t_{r_j}^{c_{r_j}}t_{f_k}^{c_{f_k}^Y}}\right|_{[t_l,t_0,t_u]}^{[t_l,t_1,t_u]},
\label{eqn:gijky}
\end{equation}
and the coefficient $\alpha_{ijk}^Y \equiv c_{m_i}+c_{r_j}+c_{f_k}^Y$.

Table 1 of \citetalias{hopkins2023fire} provides the coefficients $a_{f_k}^{Z_i}$ for nine metal elements whose mass fraction yields, $f_{Z_i}(t)$, are modeled in the same manner as helium in \cref{eqn:fcc}. Since we do not track individual metal elements in \CRKHACC, we compute the total metal mass fraction yield, $f_Z(t) = \sum_{i=1}^9 f_{Z_i}(t)$, and fit the same functional form given in \cref{eqn:fcc} to $f_Z(t)$. Our fitting procedure gives ($a_{f_1}^Z$, $a_{f_2}^Z$, $a_{f_3}^Z$, $a_{f_4}^Z$, $a_{f_5}^Z$) = (0.2925, 0.164, 0.166, 0.0422, 0.0237) with $c_{f_k}^Z$ following the same definition as above. The integrated metal mass loss, $\Delta M_Z^{\rm CC}$, follows identically to \cref{eqn:mycc,eqn:gijky} with the coefficients $a_{f_k}^Y$ replaced by $a_{f_k}^Z$.

\subsection{Stellar Outflows}\label{app:stellaroutflow}

Next we compute the stellar outflow mass loss by integrating \cref{eqn:swind}. In \FIIIRE, the mass loss rate is written as the sum of two terms, $f_w(t) = f_{w}^1(t) + f_w^2(t)$, which model wind loss from OB and AGB stars, respectively. The OB term is modeled as a piecewise power law:
\begin{equation}
    f_w^1(t) = \begin{cases}
        a_{w_0}(t/t_{w_0})^{c_{w_0}}, & t \leq t_{w_1};\\
        a_{w_1}(t/t_{w_1})^{c_{w_1}}, & t_{w_1} < t \leq t_{w_2};\\
        a_{w_2}(t/t_{w_2})^{c_{w_2}}, & t_{w_2} < t \leq t_{w_3};\\
        a_{w_3}(t/t_{w_3})^{c_{w_3}}, & t > t_{w_3};
    \end{cases}
\end{equation}
where ($a_{w_1}$, $a_{w_2}$, $a_{w_3}$) = (3$z_{\rm Fe}^{0.87}$, 20$z_{\rm Fe}^{0.45}$ , 0.6$z_{\rm Fe}$) with $a_{w_0}$ = $a_{w_1}$, ($t_{w_1}$, $t_{w_2}$, $t_{w_3}$) = (1.7, 4, 20) with $t_{w_0}$ = $t_{w_1}$, and $c_{w_i} = {\rm ln}(a_{w_{i+1}}/a_{w_i})/{\rm ln}(t_{w_{i+1}}/t_{w_i})$ with $c_{w_0}$ = $0$ and $c_{w_3}$ = $-3.1$. We also have that $z_{\rm Fe} \equiv 10^{[{\rm Fe}/{\rm H}]}$ which reduces to $R$ in our model. We separate the AGB component into two terms, $f_w^2(t) = f_w^{2a}(t) + f_{w}^{2b}(t)$, with
\begin{align}
f_w^{2a}(t) = &\ a_{w_4}(t_{w_4}/t)^{1.6} e^{-(t_{w_4}/t)^6} \notag \\ 
f_w^{2b}(t) = &\ a_{w_4}(t_{w_4}/t)^{1.6} \left[a_{w_5}^{-1} + (t_{w_4}/t)^2 \right]^{-1},
\end{align}
where ($a_{w_4}$, $a_{w_5}$) = (0.11, 0.01) and $t_{w_4} = 800$. Combining all of this, the total integrated mass loss is
\begin{align}
\frac{\Delta M_T^{\rm w}}{M_*} = &\ F_0^T(0, t_{w_1}) + F_1^T(t_{w_1},t_{w_2}) \notag \\
+ &\ F_2^T(t_{w_2},t_{w_3}) + F_3^T(t_{w_3},\infty) \notag \\
+ &\ \int_{t_0}^{t_1} f_w^{2a}(t)\ dt + \int_{t_0}^{t_1} f_w^{2b}(t)\ dt,
\label{eqn:mtf3}
\end{align}
where
\begin{equation}
F_i^T(t_l, t_u) \equiv \frac{a_{w_i}}{c_{w_i}+1} \left. \frac{t^{c_{w_i}+1}}{t_{w_i}^{c_{w_i}}} \right|^{[t_l,t_1, t_u]}_{[t_l,t_0, t_u]}.
\end{equation}
The last two integrals in \cref{eqn:mtf3} cannot be solved analytically. Instead, we numerically evaluate each integral and construct a fitting function made from a piecewise combination of straight line segments and tanh functions in log space. The fits are summarized in \cref{tab:integrals}.

The helium yield is modeled by taking into account the formation of helium from nuclear hydrogen burning as well as the loss of helium to heavier elements during nuclear helium burning. More specifically, we have
\begin{equation}
f_Y(t) = X_*y_{\rm HHe}(t) + Y_*[1-y_{\rm HeC}(t)],
\end{equation}
where $y_{\rm HHe}$ ($y_{\rm HeC}$) describes the formation (loss) of helium to hydrogen (helium) burning, and $X_*$ and $Y_*$ are the hydrogen and helium fractions of the SSP. The time dependence of the formation and loss channels is modeled in \FIIIRE using piecewise power laws:
\begin{equation}
    y_j(t) = \begin{cases}
        a_{j_0}(t/t_{j_0})^{c_{j_0}}, & t \leq t_{j_1};\\
        a_{j_1}(t/t_{j_1})^{c_{j_1}}, & t_{j_1} < t \leq t_{j_2};\\
        \dots  & \dots \\
        a_{j_n}(t/t_{j_n})^{c_{j_n}}, & t_{j_n} < t \leq t_{j_{n+1}};
    \end{cases}
\end{equation}
where $c_{j_n} \equiv {\rm ln}(a_{j_{n+1}}/a_{j_n})/{\rm ln}(t_{j_{n+1}}/t_{j_n})$, $a_{j_0} = a_{j_1}$, and $t_{j_0} = t_{j_1}$. For $y_{\rm HHe}$, we have ($a_{{\rm HHe}_1}$, $a_{{\rm HHe}_2}$, $a_{{\rm HHe}_3}$, $a_{{\rm HHe}_4}$, $a_{{\rm HHe}_5}$) = \{$0.4{\rm min}[(z_{\rm CNO}+0.001)^{0.6},2]$, $0.08$, $0.07$, $0.042$, $0.042$\}, ($t_{{\rm HHe}_1}$, $t_{{\rm HHe}_2}$, $t_{{\rm HHe}_3}$, $t_{{\rm HHe}_4}$, $t_{{\rm HHe}_5}$) = (2.8, 10, 2300, 3000, $10^5$), and $c_{{\rm HHe}_0} = 3$. We also have $z_{\rm CNO} \equiv (Z_{*,{\rm C}}+Z_{*,{\rm N}}+Z_{*,{\rm O}})/(Z_{\rm C}+Z_{\rm N}+Z_{\rm O})_\odot$ which reduces to $\Tilde{Z}_* \equiv Z_*/Z_\odot$ in our model. For $y_{\rm HeC}$, we have ($a_{{\rm HeC}_1}$, $a_{{\rm HeC}_2}$, $a_{{\rm HeC}_3}$, $a_{{\rm HeC}_4}$) = ($10^{-6}$, $0.001$, $0.005$, $0.005$), ($t_{{\rm HeC}_1}$, $t_{{\rm HeC}_2}$, $t_{{\rm HeC}_3}$, $t_{{\rm HeC}_4}$) = (5, 40, $10^4$, $10^5$), and $c_{{\rm HeC}_0} = 3$. The total integrated helium mass loss evaluates to
\begin{align}
\frac{\Delta M_Y^{\rm w}}{M_*} = &\ Y_*\left(\frac{\Delta M_T^{\rm w}}{M_*} - \int_{t_0}^{t_1} f_w(t)y_{\rm HeC}(t)\ dt \right)\notag \\
+ &\ X_*\int_{t_0}^{t_1} f_w(t)y_{\rm HHe}(t)\ dt.
\label{eqn:myf3}
\end{align}
As before, we separate the $f_w$ term into its (integrable) OB term, $f_w^1$, and its (non-integrable) AGB terms, $f_w^{2a}$ and $f_w^{2b}$. The first integral in \cref{eqn:myf3} evaluates to
\begin{align}
\int_{t_0}^{t_1} f_w &\ y_{\rm HeC} dt\ =\ F_{00}^{\rm HeC}(0,t_{w_1}) + F_{10}^{\rm HeC}(t_{w_1},t_{w_2})\notag \\
+ &\ F_{20}^{\rm HeC}(t_{w_2},t_{{\rm HeC}_1}) + F_{21}^{\rm HeC}(t_{{\rm HeC}_1}, t_{w_3}) \notag \\
+ &\ F_{31}^{\rm HeC}(t_{w_3},t_{{\rm HeC}_2}) + F_{32}^{\rm HeC}(t_{{\rm HeC}_2}, t_{{\rm HeC}_3}) \notag \\
+ &\ F_{33}^{\rm HeC}(t_{{\rm HeC}_3},t_{{\rm HeC}_4}) \notag \\
+ &\ \int_{t_0}^{t_1} f_w^{2a}y_{\rm HeC}\ dt + \int_{t_0}^{t_1} f_w^{2b}y_{\rm HeC}\ dt,
\label{eqn:fwyhec}
\end{align}
where 
\begin{equation}
F_{ij}^{\rm s}(t_l,t_u) \equiv \frac{a_{w_i}a_{{\rm s}_j}}{\alpha_{ij}+1} \left. \frac{t^{\alpha_{ij}+1}}{t_{w_i}^{c_{w_i}} t_{{\rm s}_j}^{c_{{\rm s}_j}}} \right|^{[t_l,t_1,t_u]}_{[t_l,t_0,t_u]},
\label{eqn:fijs}
\end{equation}
with $\alpha_{ij} \equiv c_{w_i}+ c_{{\rm s}_j}$. As before, \cref{tab:integrals} provides fitting functions for the $f_w^2$ integrals in \cref{eqn:fwyhec}. The second integral in \cref{eqn:myf3} evaluates to
\begin{align}
\int_{t_0}^{t_1} f_w &\  y_{\rm HHe} dt\ =\ F_{00}^{\rm HHe}(0,t_{w_1}) + F_{10}^{\rm HHe}(t_{w_1},t_{{\rm HHe}_1})\notag \\
+ &\ F_{11}^{\rm HHe}(t_{{\rm HHe}_1},t_{w_2}) + F_{21}^{\rm HHe}(t_{w_2}, t_{{\rm HHe}_2}) \notag \\
+ &\ F_{22}^{\rm HHe}(t_{{\rm HHe}_2},t_{w_3}) + F_{32}^{\rm HHe}(t_{w_3},t_{{\rm HHe}_3}) \notag \\
+ &\ F_{33}^{\rm HHe}(t_{{\rm HHe}_3},t_{{\rm HHe}_4}) + F_{34}^{\rm HHe}(t_{{\rm HHe}_4},t_{{\rm HHe}_5})\notag \\
+ &\ \left. G(\tilde{Z}_*) \frac{t}{t_{{\rm HHe}_2}} \right|^{[0,t_1,t_{{\rm HHe}_2}]}_{[0,t_0,t_{{\rm HHe}_2}]} \notag \\
+ &\ \int_{{\rm max}(t_0,t_{{\rm HHe}_2})}^{{\rm max}(t_1,t_{{\rm HHe}_2})} f_w^{2a}y_{\rm HHe}\ dt \notag \\ 
+ &\ \int_{{\rm max}(t_0,t_{{\rm HHe}_2})}^{{\rm max}(t_1,t_{{\rm HHe}_2})} f_w^{2b}y_{\rm HHe}\ dt,
\label{eqn:fwyhhe}
\end{align}
where $F_{ij}^{\rm HHe}$ follows \cref{eqn:fijs}. The final three terms in this expression arise from the  $\int f_w^2 y_{\rm HHe}$ integral which has metallicity dependence at early times, $t \leq t_{{\rm HHe}_2} = \unit{10}{\rm Myr}$, due to the presence of the $z_{\rm CNO} = \tilde{Z}_*$ term in $a_{{\rm HHe}_1}$. To capture this metallicity dependence, we fit the function, 
\begin{equation}
G(\tilde{Z}_*) = \begin{cases}
        {\rm exp}(g_l), & \tilde{Z}_* \leq \tilde{z}_l;\\
        {\rm exp} \Big[g_1 + g_2({\rm ln}\tilde{Z}_*-g_4) \\
        \qquad + g_3({\rm ln}\tilde{Z}_*-g_4)^2
    \Big], & \tilde{z}_l < \tilde{Z}_* < \tilde{z}_u;\\
        {\rm exp}(g_u), & \tilde{Z}_* \geq \tilde{z}_u;
    \end{cases}
\label{eqn:gz}
\end{equation}
to the numerical integral of $\int f_w^2 y_{\rm HHe}$ evaluated from $t = 0$ to $t_{{\rm HHe}_2}$ for different $\tilde{Z}_*$ values. We find the best-fit coefficients ($g_l$, $g_1$, $g_2$, $g_3$, $g_4$, $g_u$) = ($-5.420649$, $-4.819271$, $0.207927$, $0.017928$, $-3.40815$, $-3.468966$) with $(\tilde{z}_l,\tilde{z}_u)$ = ($1.555676\times10^{-4}$, $3.199267$). We make the assumption in \cref{eqn:fwyhhe} that the metallicity dependence of the $f_w^2$ integral grows linearly with time up to $t_{{\rm HHe}_2}$ and then use the fitting functions provided in \cref{tab:integrals} to approximate the $f_w^2$ integrals at later times.

The \FIIIRE metal yield is modeled as
\begin{equation}
f_Z(t) = Z_* + (1-Z_*)y_{\rm HeC}(t) - X_* y_{\rm HHe}(t) y_{\rm HeC}(t),
\end{equation}
with the corresponding integrated metal mass loss
\begin{align}
\frac{\Delta M_Z^w}{M_*} = &\ Z_* \frac{\Delta M_T^w}{M_*} + (1-Z_*) \int_{t_0}^{t_1} f_w(t) y_{\rm HeC}(t)\ dt \notag \\
&\ - X_* \int_{t_0}^{t_1} f_w(t) y_{\rm HHe}(t) y_{\rm HeC}(t)\ dt.
\end{align}
The first integral was solved previously in \cref{eqn:fwyhec} while the second integral evaluates to
\begin{align}
\int_{t_0}^{t_1} f_w &\ y_{\rm HHe} y_{\rm HeC} dt\ =\ F_{000}(0,t_{w_1})\notag \\
+ &\ F_{100}(t_{w_1},t_{{\rm HHe}_1}) + F_{110}(t_{{\rm HHe}_1}, t_{w_2})\notag \\
+ &\ F_{210}(t_{w_2},t_{{\rm HeC}_1}) + F_{211}(t_{{\rm HeC}_1},t_{{\rm HHe}_2}) \notag \\
+ &\ F_{221}(t_{{\rm HHe}_2},t_{w_3}) + F_{321}(t_{w_3},t_{{\rm HeC}_2}) \notag \\
+ &\ F_{322}(t_{{\rm HeC}_2}, t_{{\rm HHe}_3}) + F_{332}(t_{{\rm HHe}_3},t_{{\rm HHe}_4}) \notag \\
+ &\ F_{342}(t_{{\rm HHe}_4}, t_{{\rm HeC}_3}) + F_{343}(t_{{\rm HeC}_3}, t_{{\rm HeC}_4}) \notag \\
+ &\ \left. G(\tilde{Z}_*) \frac{t}{t_{{\rm HHe}_2}} \right|^{[0,t_1,t_{{\rm HHe}_2}]}_{[0,t_0,t_{{\rm HHe}_2}]} \notag \\
+ &\ \int_{{\rm max}(t_0,t_{{\rm HHe}_2})}^{{\rm max}(t_1,t_{{\rm HHe}_2})} f_w^{2a}y_{\rm HHe}y_{\rm HeC}\ dt \notag \\ 
+ &\ \int_{{\rm max}(t_0,t_{{\rm HHe}_2})}^{{\rm max}(t_1,t_{{\rm HHe}_2})} f_w^{2b}y_{\rm HHe}y_{\rm HeC}\ dt,
\label{eqn:fwyhhehec}
\end{align}
where
\begin{align}
F_{ijk}(t_l,t_u) = &\ \frac{a_{w_i}a_{{\rm HHe}_j}a_{{\rm HeC}_k}}{\alpha_{ijk}+1} \times \notag \\
&\ \left. \frac{t^{\alpha_{ijk}+1}}{t_{w_i}^{c_{w_i}}t_{{\rm HHe}_j}^{c_{{\rm HHe}_j}}t_{{\rm HeC}_k}^{c_{{\rm HeC}_k}}} \right|^{[t_l,t_1,t_u]}_{[t_l,t_0,t_u]},
\end{align}
and $\alpha_{ijk} \equiv c_{w_i}+c_{{\rm HHe}_j}+c_{{\rm HeC}_k}$. We model the metallicity dependence in the same manner as before, with the best-fit coefficients to $G(\tilde{Z}_*)$ being ($g_l$, $g_1$, $g_2$, $g_3$, $g_4$, $g_u$) = ($-17.61843$, $-17.61708$, $0.007352$, $0.007056$, $-9.173174$, $-16.78266$) and $(\tilde{z}_l,\tilde{z}_u)$ = ($9.545485\times10^{-5}$, $3.199267$). Fitting functions for the remaining two $f_w^2$ integrals are provided in \cref{tab:integrals}.

\subsection{Initial Versus Evolved Mass}

We note that the integrals provided above assumed that the stellar mass, $M_*$, in \cref{eqn:snint,eqn:swind} is constant and equal to its initial value at time $t = 0$. In \citetalias{hopkins2023fire}, the rates are instead normalized so that the mass loss equations can be evaluated using the time-evolving $M_*(t)$. To account for this, we introduce a correction factor,
\begin{equation}
\alpha^{\rm enrich} = M_*(t_0) \frac{1-{\rm exp}[-\Delta M_T^{\rm enrich}/M_*(t_0)]}{\Delta M_T^{\rm enrich}},
\end{equation}
where $M_*(t_0)$ is the SSP mass at the start of the timestep, and $\Delta M_T^{\rm enrich}$ is the total mass lost due to supernovae and stellar winds computed using the formulae provided above for $M_*(t=0)$. The correction factor is derived by separating variables in \cref{eqn:snint,eqn:swind} and using $\int dM/M = {\rm ln}\ M$. The mass loss for each individual component, $\Delta M_c^{\rm enrich}$, is corrected by multiplication with $\alpha^{\rm enrich}$. This correction is analytically exact in the case of the total mass and we find that it is accurate to within $2\%$ for the helium and metal components.

\setlength{\LTcapwidth}{\textwidth}
\begin{deluxetable*}{@{\extracolsep{\fill}}ccccccc}
\tablewidth{\textwidth}
\setlength{\tabcolsep}{18pt}   
\tablecaption{
Fitting coefficients of the numerical integrals used in the \FIIIRE\ AGB stellar wind mass loss. 
Each integral is fit in log-space with a piecewise series of straight line segments, 
${\rm exp}[a + b\,{\rm ln}t]$, and tanh functions, 
${\rm exp}[a + b\,{\rm tanh}(c({\rm ln}t-d))]$. 
Rows with empty entries in the $c$ and $d$ columns indicate a straight-line fit; otherwise, the tanh function applies. 
The lower and upper time bounds of each piecewise fit, $t_l$ and $t_u$, are provided in Myr.
\label{tab:integrals}}

\tablehead{
  \colhead{Integral} & \colhead{$a$} & \colhead{$b$} &
  \colhead{$c$} & \colhead{$d$} & \colhead{$t_l$} & \colhead{$t_u$}
}
\startdata
\multirow{8}{*}{$\int_0^t f_w^{2a}(t) \, dt$} 
& $-431.99$   & $66.81093$   &            &            & $0$    & $600$   \\ 
& $-96.7544$  & $100.0$      & $2.43306$  & $5.73977$  & $600$  & $900$   \\ 
& $-95.70486$ & $100.0$      & $1.254149$ & $5.00497$  & $900$  & $1800$  \\ 
& $-2.458194$ & $0.849421$   &            &            & $1800$ & $2400$  \\ 
& $-0.530956$ & $0.601807$   &            &            & $2400$ & $3200$  \\ 
& $0.992478$  & $0.413051$   &            &            & $3200$ & $4800$  \\ 
& $2.20617$   & $0.269867$   &            &            & $4800$ & $8000$  \\ 
& $2.9763$    & $0.184177$   &            &            & $8000$ & $\infty$ \\ \hline
\multirow{3}{*}{$\int_0^t f_w^{2b}(t) \, dt$} 
& $-5.2309$   & $1.395576$   &            &            & $0$    & $20$    \\ 
& $-4.877508$ & $1.277652$   &            &            & $20$   & $50$    \\ 
& $-2.980048$ & $4.863875$   & $0.421655$ & $2.123926$ & $50$   & $\infty$ \\ \hline
\multirow{8}{*}{$\int_0^t f_w^{2a}(t)y_{\rm HeC}(t) \, dt$} 
& $-390.0083$ & $59.29077$   &            &            & $0$    & $600$   \\ 
& $-102.7369$ & $100.0$      & $2.403543$ & $5.7355$   & $600$  & $900$   \\ 
& $-101.5443$ & $100.0$      & $1.182235$ & $4.931659$ & $900$  & $1800$  \\ 
& $-9.3368$   & $0.977616$   &            &            & $1800$ & $2400$  \\ 
& $-7.38478$  & $0.726819$   &            &            & $2400$ & $3200$  \\ 
& $-5.7873$   & $0.528893$   &            &            & $3200$ & $4800$  \\ 
& $-4.4839$   & $0.375126$   &            &            & $4800$ & $8000$  \\ 
& $-3.600583$ & $0.276840$   &            &            & $8000$ & $\infty$ \\ \hline
\multirow{5}{*}{$\int_0^t f_w^{2b}(t)y_{\rm HeC}(t) \, dt$} 
& $-25.49524$ & $4.672025$   &            &            & $0$     & $22$  \\ 
& $-9.886332$ & $3.693667$   & $1.291699$ & $3.344415$ & $22$    & $65$  \\ 
& $-39.50205$ & $34.87625$   & $0.402449$ & $0$        & $65$    & $700$ \\
& $-6.56253$  & $0.241310$   &            &            & $700$   & $2000$  \\ 
& $-5.69778$  & $0.127541$   &            &            & $2000$  & $\infty$ \\ \hline
\multirow{8}{*}{$\int_{10}^t f_w^{2a}(t)y_{\rm HHe}(t) \, dt$} 
& $-434.459$  & $66.78649$   &            &            & $10$   & $600$   \\ 
& $-99.392$   & $100.0$      & $2.435569$ & $5.740131$ & $600$  & $900$   \\ 
& $-98.35381$ & $100.0$      & $1.260357$ & $5.010987$ & $900$  & $2000$  \\ 
& $-4.663074$ & $0.791674$   &            &            & $2000$ & $2500$  \\ 
& $-1.685144$ & $0.411062$   &            &            & $2500$ & $3200$  \\ 
& $-0.50251$  & $0.264532$   &            &            & $3200$ & $4800$  \\ 
& $0.19617$   & $0.182106$   &            &            & $4800$ & $8000$  \\ 
& $0.678$     & $0.128502$   &            &            & $8000$ & $\infty$ \\ \hline
\multirow{2}{*}{$\int_{10}^t f_w^{2b}(t)y_{\rm HHe}(t) \, dt$} 
& $-103.755$  & $100.0$      & $1.611936$ & $1.072965$ & $10$   & $18$  \\ 
& $-6.93124$  & $6.186073$   & $0.427509$ & $1.83905$  & $18$   & $\infty$  \\ \hline
\multirow{8}{*}{$\int_{10}^t f_w^{2a}(t)y_{\rm HHe}(t)y_{\rm HeC}(t) \, dt$} 
& $-334.945$  & $50.2729$    &            &            & $10$   & $600$   \\ 
& $-105.3747$ & $100.0$      & $2.406011$ & $5.735859$ & $600$  & $900$   \\ 
& $-104.1948$ & $100.0$      & $1.188165$ & $4.937966$ & $900$  & $2000$  \\ 
& $-11.53772$ & $0.919162$   &            &            & $2000$ & $2500$  \\ 
& $-8.25936$  & $0.500152$   &            &            & $2500$ & $3200$  \\ 
& $-7.0058$   & $0.344835$   &            &            & $3200$ & $4800$  \\ 
& $-6.29761$  & $0.261287$   &            &            & $4800$ & $8000$  \\ 
& $-5.759975$ & $0.201465$   &            &            & $8000$ & $\infty$ \\ \hline
\multirow{6}{*}{$\int_{10}^t f_w^{2b}(t)y_{\rm HHe}(t)y_{\rm HeC}(t) \, dt$} 
& $-59.9234$  & $17.13235$   &            &            & $10$   & $12.5$   \\ 
& $-30.05034$ & $5.304919$   &            &            & $12.5$ & $22.5$   \\ 
& $-12.57509$ & $3.888042$   & $1.229671$ & $3.318168$ & $22.5$ & $75$  \\ 
& $-42.61491$ & $35.40138$   & $0.404669$ & $0.0$      & $75$   & $1000$  \\ 
& $-8.81622$  & $0.193901$   &            &            & $1000$ & $2500$  \\ 
& $-7.87374$  & $0.073443$   &            &            & $2500$ & $\infty$  \\ 
\enddata
\end{deluxetable*}
\clearpage
\bibliography{bib,jdbib}
\bibliographystyle{aasjournal}

\end{document}